\documentclass[11pt,a4paper]{article}
\usepackage[T1]{fontenc}
\usepackage[utf8]{inputenc}
\usepackage{lmodern}
\usepackage[margin=1in]{geometry}
\usepackage{amsmath,amssymb,amsfonts,amsthm}
\usepackage[affil-it]{authblk}
\usepackage{booktabs,array,microtype}
\usepackage{flafter}
\usepackage{needspace}
\usepackage{tikz}
\usetikzlibrary{arrows.meta,positioning,calc,decorations.pathreplacing}
\usepackage[section]{placeins}
\usepackage{cite}
\usepackage{xurl}
\usepackage[nottoc]{tocbibind}
\usepackage[pdfencoding=auto,hidelinks]{hyperref}
\makeatletter
\renewcommand*\l@subsubsection{\@dottedtocline{3}{3.8em}{4.1em}}
\makeatother
\newcommand{\dd}{\mathrm d}
\newcommand{\pd}{\partial}
\newcommand{\Mloc}{\mathcal M}
\newcommand{\Boxeta}{\Box_\eta}
\newcommand{\GN}{G_0}
\newcommand{\doi}[1]{\href{https://doi.org/#1}{doi:\nolinkurl{#1}}}
\newtheorem{proposition}{Proposition}
\title{\textbf{A Machian wave effect in conformal, scalar--tensor gravitational theory}}
\author{Jos\'e J.\ A.\ Rodal}
\affil{Space Studies Institute, Research Triangle Park, NC, USA\\
\texttt{jrodal@alum.mit.edu}}
\date{Revised manuscript --- 28 October 2026}
\hypersetup{pdftitle={A Machian Wave Effect in Conformal Scalar-Tensor Gravitational Theory},pdfauthor={Jose J. A. Rodal},pdfsubject={Complete revised analysis of Mach-effect propulsion in GR and Hoyle-Narlikar theory}}
\begin{document}
\maketitle

\begin{abstract}
Woodward's simultaneous energy-density identities, imposed exactly with a constant physical speed of light and nonzero proper density, fix the potential and exclude the independent potential derivatives used in his derivation. A background interpretation instead requires the perturbation terms omitted by that derivation. In general relativity (GR), the second time derivative isolated from the exact Landau--Lifshitz equations belongs to the gravitational operator and supplies no additional material mass source. In the Hoyle--Narlikar (HN) derivation of Woodward and Fearn, the linear second time derivative cancels from the time--time component; the covariant trace cancels the squared-gradient terms retained in Fearn's diagonal sum. Their propulsion estimates contain further errors in mechanical power, momentum transfer and constitutive dimensions. The mechanical analysis retains damping, complex harmonic response and the dissipative energy and entropy balances. For a smooth strictly positive scalar, an admissible Weyl transformation, universal material coupling and consistently mapped matter, driving and boundary data, the complete continuum conformal HN action becomes the Einstein action. Under these hypotheses, the additional measured force relative to the corresponding Einstein result is identically zero for the same physical matter model. Compatible global source-integral and absorber conditions restrict the admitted solutions and preserve this equality. Woodward and Fearn's mass-fluctuation and propulsion formulas do not follow from the stated GR or continuum conformal HN equations. Their additional scalar propulsion force is excluded in the local continuum sector under the stated hypotheses.
\end{abstract}

\clearpage
\setcounter{tocdepth}{3}
\tableofcontents
\clearpage

\section{Introduction}
Woodward's Mach-effect experiments use electrically driven piezoelectric stacks coupled to unequal reaction masses, with reported force signals of fractions of a microNewton to a few microNewtons~\cite{Fearn2013,Fearn2014,Fearn2015I,Fearn2015II}. Independent investigations were reported by Tajmar and Buldrini at the Estes Park workshop and by K\"o{\ss}ling et al. in the SpaceDrive programme~\cite{Tajmar2016,Buldrini2016,Koessling2018}. The theoretical question is whether the proposed mass fluctuation and propulsion follow from Einstein's equations or from the conformal theory of Hoyle and Narlikar. I examine the equations used to claim these results.

Woodward~\cite{Woodward2004,WoodwardStarship} associates a second time derivative of internal energy with a transient inertial-mass change and correlates that change with an acceleration to obtain a mean force. The cosmic-potential normalization in Refs.~\cite{Woodward2004,WoodwardStarship} originates in Sciama's account of inertia~\cite{Sciama1953}. Tajmar~\cite{Tajmar2017} subsequently attributed the fluctuation specifically to mechanical power and claimed agreement with the experimental forces. Both proposals require a derivation of the mass source and of the momentum transferred to the device. The broader interpretations of Mach's principle~\cite{Einstein1956,CiufoliniWheeler1995,BondiSamuel1997,Barbour1995} do not specify either relation.

\paragraph{Woodward's equation and its defining identities.}
Equation~(7) of Ref.~\cite{Woodward2004}, repeated as Eq.~(3) of Ref.~\cite{Woodward2020Reply}, is
\begin{align}
 \nabla^2\varphi-\frac{\ddot\varphi}{c^2}
 ={}&4\pi G\rho_0
 +\frac{\varphi}{\rho_0c^4}\ddot E_0
 -\left(\frac{\varphi}{\rho_0c^4}\right)^2\dot E_0^{\,2}
 -\frac{\dot\varphi^{\,2}}{c^4}.
 \label{eq:intro-woodward-wave}
\end{align}
Here $\rho_0$ and $E_0$ are proper mass and energy densities, dots denote partial time derivatives, and $\varphi$ denotes Woodward's positive total potential. I distinguish it from the local perturbation $\phi$ and cosmic potential $\Phi$ used below. Woodward~\cite{Woodward2004,Woodward2020Reply} identifies the additional terms on the right as transient gravitational sources.

The text following Eq.~(4) of Ref.~\cite{Woodward2004} identifies the potential with $c^2$; Eqs.~(A11) and~(A13) use $\rho_0=E_0/c^2$ and $E_0=\rho_0\varphi$. The same exact normalization is used to obtain Eq.~(3) of Ref.~\cite{Woodward2020Reply}. For the same densities, spacetime-constant physical $c$ and $\rho_0\ne0$,
\begin{equation}
 E_0=\rho_0c^2,\qquad E_0=\rho_0\varphi
 \quad\Longrightarrow\quad
 \varphi=\frac{E_0}{\rho_0}=c^2.
 \label{eq:intro-exact-constraint}
\end{equation}
Imposed throughout the differentiation region, these identities require
\begin{equation}
 \partial_a\varphi=0,\qquad
 \dot\varphi=\ddot\varphi=0,\qquad \nabla\varphi=0.
 \label{eq:intro-potential-fixed}
\end{equation}
The product rule gives
\begin{align}
 \dot E_0&=\varphi\dot\rho_0+\rho_0\dot\varphi,
 \nonumber\\
 \ddot E_0&=\varphi\ddot\rho_0
       +2\dot\rho_0\dot\varphi+\rho_0\ddot\varphi.
 \label{eq:intro-product-rule}
\end{align}
On the constraint, $\dot E_0=c^2\dot\rho_0$ and $\ddot E_0=c^2\ddot\rho_0$. Neither density is required to be constant. The inconsistency is the simultaneous use of the exact constraint and an independently varying $\varphi$.

Equation~(A12) of Ref.~\cite{Woodward2004} is
\begin{equation}
 \nabla^2\varphi-\frac{\ddot E_0}{\rho_0c^2}
 +\frac{\dot E_0^{\,2}}{\rho_0^2c^4}=4\pi G\rho_0.
 \label{eq:intro-A12}
\end{equation}
Substitution of Eq.~\eqref{eq:intro-product-rule} into Eq.~\eqref{eq:intro-A12}, before imposing $\varphi/c^2=1$, gives
\begin{align}
 \nabla^2\varphi-\frac{\ddot\varphi}{c^2}
 ={}&4\pi G\rho_0+\frac{\varphi\ddot\rho_0}{\rho_0c^2}
 -\left(\frac{\varphi\dot\rho_0}{\rho_0c^2}\right)^2
 -\frac{\dot\varphi^{\,2}}{c^4}+\mathcal R_{\rm mix},
 \nonumber\\
 \mathcal R_{\rm mix}
 ={}&\frac{2}{\rho_0c^2}
       \left(1-\frac{\varphi}{c^2}\right)
       \dot\varphi\dot\rho_0.
 \label{eq:intro-mixed-remainder}
\end{align}
The cancellation between Eqs.~(A15) and~(A16) of Ref.~\cite{Woodward2004} sets $\varphi/c^2=1$. The displayed product expansions are correct. They do not release $\varphi$ from Eq.~\eqref{eq:intro-potential-fixed}.

\paragraph{A background equality is a different assumption.}
Expand both relations about constant $\overline\rho>0$ and $\overline\varphi=c^2$:
\begin{align}
 \rho_0&=\overline\rho+\rho_{(1)}+\rho_{(2)}+O(\delta^3),\nonumber\\
 \varphi&=c^2+\varphi_{(1)}+\varphi_{(2)}+O(\delta^3),\nonumber\\
 E_0&=\overline\rho c^2+E_{(1)}+E_{(2)}+O(\delta^3).
 \label{eq:intro-background-split}
\end{align}
The subscript $(j)$ denotes order $\delta^j$, without a factorial. The derivatives through second order obey the same amplitude ordering: the characteristic time and length scales do not vary with $\delta$. The two identities separately require
\begin{align}
 E_0=\rho_0c^2:\qquad
 E_{(1)}&=c^2\rho_{(1)},&
 E_{(2)}&=c^2\rho_{(2)},\nonumber\\
 E_0=\rho_0\varphi:\qquad
 E_{(1)}&=c^2\rho_{(1)}+\overline\rho\varphi_{(1)},&
 E_{(2)}&=c^2\rho_{(2)}+\overline\rho\varphi_{(2)}
                    +\rho_{(1)}\varphi_{(1)}.
 \label{eq:intro-two-identities-expanded}
\end{align}
Thus imposing both through second order gives $\varphi_{(1)}=\varphi_{(2)}=0$. If they hold only through first order, introduce their independent second-order defects,
\begin{align}
 E_0-\rho_0c^2&=\mathcal R_c^{(2)}+O(\delta^3),\nonumber\\
 E_0-\rho_0\varphi&=\mathcal R_\varphi^{(2)}+O(\delta^3),\nonumber\\
 \varphi_{(1)}&=0,\qquad
 \overline\rho\varphi_{(2)}=\mathcal R_c^{(2)}-\mathcal R_\varphi^{(2)}.
 \label{eq:intro-first-order-defects}
\end{align}
The first-order identities do not determine either defect. Differentiation of the second relation gives, through second order,
\begin{align}
 \dot E_0={}&c^2(\dot\rho_{(1)}+\dot\rho_{(2)})
 +\overline\rho(\dot\varphi_{(1)}+\dot\varphi_{(2)})
 +\varphi_{(1)}\dot\rho_{(1)}+\rho_{(1)}\dot\varphi_{(1)}
 +\dot{\mathcal R}_\varphi^{(2)}+O(\delta^3),\nonumber\\
 \ddot E_0={}&c^2(\ddot\rho_{(1)}+\ddot\rho_{(2)})
 +\overline\rho(\ddot\varphi_{(1)}+\ddot\varphi_{(2)})
 +\varphi_{(1)}\ddot\rho_{(1)}+2\dot\rho_{(1)}\dot\varphi_{(1)}
 +\rho_{(1)}\ddot\varphi_{(1)}
 +\ddot{\mathcal R}_\varphi^{(2)}+O(\delta^3).
 \label{eq:intro-product-second-order}
\end{align}
These formulas also describe the background-only case with an independent $\varphi_{(1)}$ when $E_0=\rho_0\varphi$ is retained exactly and $\mathcal R_\varphi^{(2)}=0$. In that case,
\begin{equation}
 \mathcal R_{\rm mix}
 =-\frac{2\varphi_{(1)}\dot\varphi_{(1)}\dot\rho_{(1)}}{\overline\rho c^4}
   +O(\delta^4)=O(\delta^3).
 \label{eq:intro-background-order}
\end{equation}
If both identities hold through first order, Eq.~\eqref{eq:intro-first-order-defects} instead gives $\mathcal R_{\rm mix}=O(\delta^5)$. This does not remove the second-order defect: substitution in Eq.~\eqref{eq:intro-A12} adds $\ddot{\mathcal R}_\varphi^{(2)}/(\overline\rho c^2)$ to the right side of Eq.~\eqref{eq:intro-mixed-remainder} through second order.

Before the substitution $\dot\rho_0=\dot E_0/c^2$, the temporal expression is
\begin{align}
 \mathcal D_E
 &=\frac1{c^2}\partial_t\left(\frac{\dot E_0}{\rho_0}\right)
 =\frac{\ddot E_0}{\rho_0c^2}
  -\frac{\dot E_0\dot\rho_0}{\rho_0^2c^2},\nonumber\\
 \mathcal D_{E,\rm W}
 &=\frac{\ddot E_0}{\rho_0c^2}
  -\frac{\dot E_0^{\,2}}{\rho_0^2c^4}.
 \label{eq:intro-quotient}
\end{align}
For the background-only interpretation with exact $E_0=\rho_0\varphi$, the second-order expansion is
\begin{align}
 \mathcal D_E={}&\frac{\ddot\varphi_{(1)}+\ddot\varphi_{(2)}}{c^2}
 +\frac{\ddot\rho_{(1)}+\ddot\rho_{(2)}}{\overline\rho}
 -\frac{\rho_{(1)}\ddot\rho_{(1)}+\dot\rho_{(1)}^{\,2}}{\overline\rho^2}
 +\frac{\varphi_{(1)}\ddot\rho_{(1)}+\dot\rho_{(1)}\dot\varphi_{(1)}}{\overline\rho c^2}
 +O(\delta^3),\nonumber\\
 \mathcal D_E-\mathcal D_{E,\rm W}
 ={}&\frac{\dot\rho_{(1)}\dot\varphi_{(1)}}{\overline\rho c^2}
  +\frac{\dot\varphi_{(1)}^{\,2}}{c^4}+O(\delta^3).
 \label{eq:intro-independent-potential}
\end{align}
The discrepancy is second order although $\mathcal R_{\rm mix}$ is third order. Woodward's product rule is exact if $E_0=\rho_0\varphi$ is exact, and is valid through second order if its defect and its first two derivatives are $O(\delta^3)$. The replacement $\dot\rho_0=\dot E_0/c^2$ requires the separate energy--mass relation to the required differentiated order. If both relations are imposed through second order, the potential has no perturbation through that order. If both hold only through first order, the defect terms in Eq.~\eqref{eq:intro-product-second-order} must be retained; the quotient discrepancy then starts at third order. Equality at one event supplies no derivative identity.

The three time-dependent terms on the right of Eq.~\eqref{eq:intro-woodward-wave} have the expansion $\mathcal W_{(1)}+\mathcal W_{(2)}+O(\delta^3)$, where
\begin{align}
 \mathcal W_{(1)}={}&\frac{\ddot E_{(1)}}{\overline\rho c^2},\nonumber\\
 \mathcal W_{(2)}={}&\frac{\ddot E_{(2)}}{\overline\rho c^2}
 +\left(\frac{\varphi_{(1)}}{\overline\rho c^4}
        -\frac{\rho_{(1)}}{\overline\rho^2c^2}\right)\ddot E_{(1)}
 -\frac{\dot E_{(1)}^{\,2}}{\overline\rho^2c^4}
 -\frac{\dot\varphi_{(1)}^{\,2}}{c^4}.
 \label{eq:intro-source-orders}
\end{align}
The second energy derivative contributes at first and second order; the squared derivatives begin at second order. If both identities hold through first order, Eq.~\eqref{eq:intro-first-order-defects} gives
\begin{align}
 \mathcal W_{(1)}={}&\frac{\ddot\rho_{(1)}}{\overline\rho},\nonumber\\
 \mathcal W_{(2)}={}&\frac{\ddot\rho_{(2)}}{\overline\rho}
 -\frac{\rho_{(1)}\ddot\rho_{(1)}+\dot\rho_{(1)}^{\,2}}{\overline\rho^2}
 +\frac{\ddot{\mathcal R}_c^{(2)}}{\overline\rho c^2}.
 \label{eq:intro-source-defects}
\end{align}
Density derivatives remain, but no independent first-order potential survives. If the identities are imposed through second order, the defect in Eq.~\eqref{eq:intro-source-defects} also vanishes. If only their background values are equated, the independent potential and the quotient discrepancy in Eq.~\eqref{eq:intro-independent-potential} must be retained. These expansions evaluate the proposed equation; they do not establish its time-dependent terms as gravitational mass sources.

\paragraph{The initial fields do not agree for variable \texorpdfstring{$c$}{c}.}
The first two definitions in Woodward's handwritten notes~\cite{WoodwardHandwritten} are
\begin{equation}
 \mathcal F_\rho^\mu=-\left(c\frac{\dot\rho_0}{\rho_0},\mathbf f\right),
 \qquad
 \mathcal F_E^\mu=-\left(\frac{\dot E_0}{\rho_0c},\mathbf f\right).
 \label{eq:intro-initial-fields}
\end{equation}
Differentiating $\rho_0=E_0/c^2$ gives
\begin{align}
 \dot\rho_0&=\frac{\dot E_0}{c^2}
             -\frac{2E_0\dot c}{c^3},\nonumber\\
 \frac{\dot E_0}{\rho_0c}
 &=c\frac{\dot\rho_0}{\rho_0}+2\dot c,
 \qquad
 \mathcal F_E^0-\mathcal F_\rho^0=-2\dot c.
 \label{eq:intro-variable-c}
\end{align}
Here $\mathbf f$ denotes the common spatial field in the two definitions. The fields coincide only for $\dot c=0$. Differentiating the energy-density expression while allowing $c$ to vary therefore does not preserve the initial density-defined field.

Woodward~\cite[pp.~4,7]{Woodward2020Reply} invokes local invariance to permit nonzero derivatives, as in Ref.~\cite[pp.~70--71]{WoodwardStarship}. In a diagonal chart with $x^0=ct$, $g_{00}>0$ and $g_{11}<0$, a null ray with $\dd x^2=\dd x^3=0$ gives
\[
 v_{\rm coord}\equiv\left|\frac{\dd x^1}{\dd t}\right|
 =c\sqrt{\frac{g_{00}}{-g_{11}}},\qquad
 \frac{\dd\ell}{\dd\tau_{\rm obs}}
 =\frac{\sqrt{-g_{11}}\,v_{\rm coord}}{\sqrt{g_{00}}}=c.
\]
Here $\dd\ell$ and $\dd\tau_{\rm obs}$ are local proper lengths and clock intervals for coordinate-stationary observers. Derivatives of $v_{\rm coord}$ need not vanish; they are not derivatives of the invariant $c$ in $E_0=\rho_0c^2$. Substituting a coordinate speed in that proper-density relation changes its meaning. A coordinate reformulation must transform the metric factors and full field operator consistently. No global inertial chart or constant coordinate light speed is assumed. A varying physical $c$ in fixed units requires a different dynamical theory~\cite{Ellis2007}.

\paragraph{The four-acceleration contraction.}
With signature $(+---)$ and constant invariant speed $c$, a geometric four-acceleration satisfies
\begin{align}
 U^\mu U_\mu&=c^2,\qquad A^\mu=U^\nu\nabla_\nu U^\mu,\nonumber\\
 U_\mu A^\mu&=\frac12\frac{\dd}{\dd\tau}(U^\mu U_\mu)=0,
 \qquad U^{\hat\mu}=(c,\mathbf0)
 \ \Longrightarrow\ A^{\hat0}=0.
 \label{eq:intro-acceleration-contraction}
\end{align}
If the time component of $\mathcal F_E^\mu$ in Eq.~\eqref{eq:intro-initial-fields} is identified with a geometric four-acceleration in the instantaneous rest frame, it gives
\begin{equation}
 a^{\hat0}=-c\frac{\dot E_0}{E_0},\qquad
 U_{\hat\mu}a^{\hat\mu}=-c^2\frac{\dot E_0}{E_0}\ne0
 \quad(\dot E_0\ne0).
 \label{eq:intro-proposed-contraction}
\end{equation}
A hat on a spacetime index denotes an orthonormal-frame component; the hats on whole fields introduced below denote Weyl-invariant representatives. Here the dot is the local rest-frame time derivative. For $\dot E_0\ne0$, this identification contradicts $U_\mu A^\mu=0$. This conclusion does not assume constant rest mass. For a timelike four-momentum $p^\mu=M_{\rm rest}U^\mu$,
\begin{equation}
 \mathcal Q^\mu=\frac{Dp^\mu}{\dd\tau}
 =M_{\rm rest}A^\mu+\frac{\dd M_{\rm rest}}{\dd\tau}U^\mu,
 \qquad U_\mu \mathcal Q^\mu=c^2\frac{\dd M_{\rm rest}}{\dd\tau}.
 \label{eq:intro-four-momentum-transfer}
\end{equation}
The longitudinal component describes rest-energy transfer, not acceleration;
$\mathcal Q^\mu$ includes the momentum transfers of the chosen system. For a conserved
particle current and a local proper density, write
\begin{equation}
 \nabla_\mu(NU^\mu)=0,\qquad
 \rho_0=\mu_\rho N,\qquad
 \frac{D}{D\tau}=U^\mu\nabla_\mu.
 \label{eq:intro-local-continuity}
\end{equation}
Here $N$ is proper particle-number density and $\mu_\rho=\rho_0/N$ is
rest-energy per conserved particle divided by $c^2$; it may include stored
internal energy and is not identified with the HN scalar. The product rule gives
\begin{equation}
 \frac{D\ln(\rho_0/\rho_*)}{D\tau}
 =\frac1{\mu_\rho}\frac{D\mu_\rho}{D\tau}
  -\nabla_\mu U^\mu.
 \label{eq:intro-density-compression}
\end{equation}
This is pointwise and requires no spatially uniform density. For constant
$\mu_\rho$, compression alone changes $\rho_0$. In the nonrelativistic
continuum limit, with $D_t=\partial_t+\mathbf v\cdot\boldsymbol\nabla$,
\begin{equation}
 \partial_t\ln(\rho_0/\rho_*)
 =\frac{D_t\mu_\rho}{\mu_\rho}
  -\boldsymbol\nabla\cdot\mathbf v
  -\mathbf v\cdot\boldsymbol\nabla\ln(\rho_0/\rho_*).
 \label{eq:intro-eulerian-density}
\end{equation}
Here $\rho_*>0$ is a fixed reference density. The partial time derivatives
in Woodward's field equation are not material derivatives; their conversion
must retain advection. These identities establish neither a gravitational
source law nor a relation between compression and the HN mass field.

\paragraph{The total-density and background-subtracted equations.}
With $E_0=\rho_0c^2$, Eq.~\eqref{eq:intro-A12} gives
\begin{equation}
 \nabla^2\varphi-\partial_t^2\ln(\rho_0/\rho_*)=4\pi G\rho_0.
 \label{eq:intro-density-log-form}
\end{equation}
(If $E_0=\rho_0\varphi$ is also imposed exactly, then $\varphi=c^2$ and
\begin{equation}
 -\partial_t^2\ln(\rho_0/\rho_*)=4\pi G\rho_0.
 \label{eq:intro-periodic-density-ode}
\end{equation}
For a $C^2$, strictly positive periodic density with period $\mathcal P$, integration gives
\begin{equation}
 0=-\left[\partial_t\ln(\rho_0/\rho_*)\right]_0^{\mathcal P}
   =4\pi G\int_0^{\mathcal P}\rho_0\,\dd t>0.
 \label{eq:intro-periodic-density-contradiction}
\end{equation}
This contradiction is restricted to the un-subtracted equation and the simultaneous exact identities.)

Let $(\overline\varphi,\overline\rho)$ instead be a consistent background solution of Eq.~\eqref{eq:intro-density-log-form}, with $\overline\rho>0$, and define $y_\rho=\ln(\rho_0/\overline\rho)$. Subtraction gives
\begin{equation}
 \nabla^2(\varphi-\overline\varphi)-\ddot y_\rho
   =4\pi G(\rho_0-\overline\rho).
 \label{eq:intro-density-subtracted}
\end{equation}
When $\dot y_\rho$ is periodic and the averaging commutes with the spatial derivatives,
\begin{equation}
 \nabla^2\langle\varphi-\overline\varphi\rangle
   =4\pi G\langle\rho_0-\overline\rho\rangle.
 \label{eq:intro-density-subtracted-average}
\end{equation}
The positive-total-density contradiction does not apply to this signed source. Indeed, the separately postulated temporal-only equation with constant $\overline\rho>0$,
\begin{equation}
 \ddot y_\rho+4\pi G\overline\rho(e^{y_\rho}-1)=0,
 \qquad
 \frac12\dot y_\rho^{\,2}
 +4\pi G\overline\rho(e^{y_\rho}-y_\rho-1)=\mathcal E_\rho,
 \label{eq:intro-density-subtracted-first-integral}
\end{equation}
has periodic solutions for $\mathcal E_\rho>0$, and their mean density is $\overline\rho$. Here $\mathcal E_\rho$ is an integration constant, not a material energy. This modified equation is not obtained by subtracting a constant positive-density background with $\overline\varphi=c^2$ from Eq.~\eqref{eq:intro-density-log-form}: that background does not solve the un-subtracted equation. No universal obstruction to background-subtracted periodic density follows from Eq.~\eqref{eq:intro-periodic-density-contradiction}. A perturbative mass estimate is not itself a derivation of the background and scalar profile required by Eq.~\eqref{eq:intro-density-subtracted}.

Corrections to the earlier analyses~\cite{Rodal2019,Rodal2025v1,Rodal2016Electroelastic} are collected in Appendix~\ref{app:prior-corrections}. Section~\ref{sec:GR} derives the relevant GR operator. Section~\ref{sec:HN} examines the HN equations, the derivations of Fearn and Woodward, and the proposed mechanical force. The HN analysis culminates in the transformation of the action, matter and boundary data to the corresponding Einstein problem.

Figure~\ref{fig:mach-phys-rigorous} of this article distinguishes the local metric perturbation from the proposed cosmic normalization.
\begin{figure}[htbp]
\centering
\begin{tikzpicture}[
 font=\small, >={Latex},
 massdot/.style={circle,draw=black,fill=gray!65,inner sep=0pt,minimum size=2.4pt},
 lab/.style={fill=white,inner sep=3pt,align=center,font=\footnotesize}]
\def\R{3.0}
\draw[thick,gray!55] (0,0) circle (\R);
\foreach \ang in {0,18,...,342}{
 \node[massdot] at ({\R*cos(\ang)},{\R*sin(\ang)}) {};}
\node[lab] at (0,4.05) {\textbf{Cosmological matter and boundary data}};
\node[lab] at (0,3.45) {Sciama's heuristic scale: $|\Phi|/c^2\sim1$};
\foreach \ang in {45,135,225,315}{
 \draw[->,thick,gray!65]
 ({0.88*\R*cos(\ang)},{0.88*\R*sin(\ang)})
 -- ({0.7*cos(\ang)},{0.7*sin(\ang)});}
\node[lab] at (1.7,1.75) {Distance-weighted\\influence: $\sum_i GM_i/R_i$};
\fill[black] (0,0) circle (0.12);
\node[lab] at (0,0.43) {$\mathcal M(t)$};
\foreach \r in {.65,1.05,1.45}{\draw[densely dashed,gray!65] (0,0) circle (\r);}
\node[lab] at (-1.7,0) {\textbf{Local device}\\$|\phi|/c^2\ll1$};
\draw[decorate,decoration={brace,amplitude=5pt}]
 (1.58,.9) -- (1.58,-.9)
 node[midway,xshift=.55cm,rotate=90,font=\scriptsize] {Local region};
\node[lab] at (0,-3.55) {Cosmic normalization and local perturbation\\must enter the same covariant problem.};
\end{tikzpicture}
\caption{Cosmic normalization and local metric perturbation. The shell schematically represents the influence of distant masses $M_i$ at distances $R_i$ in Sciama's model~\cite{Sciama1953}. For the nonlinear GR term in Eq.~\eqref{eq:gr-time-density}, the coefficient relative to the flat temporal operator is $O(U/c^2)$, with $U=|\phi|$. Replacing it by the cosmic normalization changes the derived field equation.}
\label{fig:mach-phys-rigorous}
\end{figure}

\section{Second time derivatives of mass--energy in general relativity}
\label{sec:GR}

Woodward's derivation~\cite{Woodward2004,WoodwardStarship}, based on Sciama's inertial-induction model~\cite{Sciama1953}, introduces a gravitational source with the characteristic dependence
\begin{equation}
 \mathcal S_{\rm W}\sim\frac{\Phi}{c^2}\frac{\ddot{\Mloc}}{\Mloc},
 \label{eq:gr-woodward-form}
\end{equation}
where $\Phi/c^2\simeq-1$ is the proposed cosmic normalization and $\Mloc$ is a local mass--energy variable. Davidson~\cite{Davidson1957} showed that some cosmological solutions of Einstein's equations are consistent with Sciama's interpretation of inertia. That result does not derive Eq.~\eqref{eq:gr-woodward-form}. I obtain the corresponding second-derivative term directly from Einstein's equations and determine its coefficient and its place in the gravitational operator.

\subsection{Conventions and the exact Landau--Lifshitz equation}
We retain the timelike signature $(+---)$ and use coordinates $x^\mu=(ct,\mathbf x)$, so that
\begin{equation}
 \eta_{\mu\nu}=\eta^{\mu\nu}=\operatorname{diag}(1,-1,-1,-1),\qquad
 \Boxeta=\eta^{\alpha\beta}\pd_\alpha\pd_\beta
 =\frac1{c^2}\pd_t^2-\nabla^2.
 \label{eq:gr-conventions}
\end{equation}
Hoyle and Narlikar~\cite[\mbox{p.~20}, Eqs.~(42)--(43)]{Hoyle1974} define $R^i{}_{mlk}$ and contract $R_{mk}=R^i{}_{mki}$ on the last lower index. Renaming their indices gives the convention retained in Sect.~3.1.1 of Ref.~\cite{Rodal2019}:
\begin{align}
 R^a{}_{bcd}
 &=\partial_c\Gamma^a{}_{bd}-\partial_d\Gamma^a{}_{bc}
   +\Gamma^a{}_{ec}\Gamma^e{}_{bd}-\Gamma^a{}_{ed}\Gamma^e{}_{bc},\nonumber\\
 R_{ab}&=\sum_rR^r{}_{abr},\qquad R=g^{ab}R_{ab},\qquad
 G_{ab}=R_{ab}-\tfrac12g_{ab}R.
 \label{eq:curvature-definition}
\end{align}
With $T_{ab}$ in mass-density units, the Einstein equation is
\begin{equation}
 G_{ab}+\Lambda g_{ab}=-\frac{8\pi\GN}{c^2}T_{ab}.
 \label{eq:gr-hn-einstein}
\end{equation}
Here $\rho$ is the Newtonian mass density, $T^{00}=\rho+O(\rho v^2/c^2)$, and $\Lambda>0$ denotes positive vacuum energy. The negative material coupling agrees with Eq.~(42) of Ref.~\cite{HoyleNarlikar1964}. The local relaxed equations below use $\Lambda=0$; a nonzero $\Lambda$ can be included algebraically through $T^{ab}\mapsto T^{ab}+\Lambda c^2g^{ab}/(8\pi\GN)$, with the boundary conditions specified separately.

Metric signature and Ricci contraction must be translated together when comparing other conventions~\cite{MTW1973,Einstein1956,Eddington1975,Weinberg1972}. Here all curvature signs follow Eq.~\eqref{eq:curvature-definition}; at fixed metric, the opposite contraction $\sum_rR^r{}_{arb}$ reverses $R_{ab}$, $R$ and $G_{ab}$.

Let $g=\det(g_{\mu\nu})$. Define the gothic inverse metric and its difference from the auxiliary Minkowski metric by
\begin{equation}
 \mathfrak g^{\mu\nu}=\sqrt{-g}\,g^{\mu\nu},\qquad
 H^{\mu\nu}=\mathfrak g^{\mu\nu}-\eta^{\mu\nu}=2\mathfrak h^{\mu\nu}.
 \label{eq:gr-gothic-definition}
\end{equation}
Here $\mathfrak h^{\mu\nu}$ denotes the variable called $h^{\mu\nu}$ in Ref.~\cite{Rodal2019}; the distinct typeface separates it from the induced boundary metric $h_{ab}$ below. It is not the covariant metric perturbation. Equation~\eqref{eq:gr-gothic-definition} is exact and does not assume a weak field. Harmonic coordinates satisfy
\begin{equation}
 \pd_\nu\mathfrak g^{\mu\nu}=0
 \quad\Longleftrightarrow\quad\pd_\nu H^{\mu\nu}=0.
 \label{eq:gr-harmonic}
\end{equation}
On a harmonic coordinate chart, the Landau--Lifshitz superpotential~\cite{LandauLifshitz1975,PoissonWill2014} satisfies
\begin{align}
 \mathcal U^{\mu\alpha\nu\beta}
 &=\mathfrak g^{\mu\nu}\mathfrak g^{\alpha\beta}
   -\mathfrak g^{\mu\beta}\mathfrak g^{\alpha\nu},\nonumber\\
 \pd_\alpha\pd_\beta\mathcal U^{\mu\alpha\nu\beta}
 &=\frac{16\pi\GN}{c^2}(-g)
   \big(T^{\mu\nu}+t_{\rm LL}^{\mu\nu}\big).
 \label{eq:gr-superpotential}
\end{align}
The Landau--Lifshitz quantity $t_{\rm LL}^{\mu\nu}$ is expressed in the same mass-density units as $T^{\mu\nu}$. It contains products of first derivatives of the metric, with metric-dependent coefficients. Applying the product rule in harmonic coordinates gives
\begin{equation}
 \pd_\alpha\pd_\beta\mathcal U^{\mu\alpha\nu\beta}
 =\mathfrak g^{\alpha\beta}\pd_\alpha\pd_\beta H^{\mu\nu}
  -(\pd_\beta H^{\mu\alpha})(\pd_\alpha H^{\nu\beta}).
 \label{eq:gr-superpotential-expanded}
\end{equation}
Consequently the exact relaxed equation is
\begin{align}
 \Boxeta H^{\mu\nu}
 ={}&\frac{16\pi\GN}{c^2}(-g)
       \big(T^{\mu\nu}+t_{\rm LL}^{\mu\nu}\big)
       +\mathcal N^{\mu\nu},\nonumber\\
 \mathcal N^{\mu\nu}
 ={}&(\pd_\beta H^{\mu\alpha})(\pd_\alpha H^{\nu\beta})
       -H^{\alpha\beta}\pd_\alpha\pd_\beta H^{\mu\nu}.
 \label{eq:gr-exact-relaxed}
\end{align}
Substituting $H^{\mu\nu}=2\mathfrak h^{\mu\nu}$ yields the normalization employed in Eq.~(3) of Ref.~\cite{Rodal2019}:
\begin{align}
 \Boxeta\mathfrak h^{\mu\nu}=\frac{8\pi\GN}{c^2}\Big\{
 &(-g)(T^{\mu\nu}+t_{\rm LL}^{\mu\nu})\nonumber\\
 &+\frac{c^2}{4\pi\GN}
 \big[(\pd_\beta\mathfrak h^{\mu\alpha})(\pd_\alpha\mathfrak h^{\nu\beta})
      -\mathfrak h^{\alpha\beta}\pd_\alpha\pd_\beta\mathfrak h^{\mu\nu}\big]\Big\}.
 \label{eq:gr-old-normalization}
\end{align}
The factor $1/\GN$ in this expression results from dividing a metric nonlinearity by the gravitational coupling. Define
\begin{equation}
 \tau^{\mu\nu}=(-g)(T^{\mu\nu}+t_{\rm LL}^{\mu\nu})
       +\frac{c^2}{16\pi\GN}\mathcal N^{\mu\nu},\qquad
 \Boxeta H^{\mu\nu}=\frac{16\pi\GN}{c^2}\tau^{\mu\nu}.
 \label{eq:relaxed}
\end{equation}
$\tau^{\mu\nu}$ depends on $H^{\mu\nu}$. The equation therefore remains nonlinear; $\mathcal N^{\mu\nu}$ is determined by the metric, rather than by an independent material source. Neither $t_{\rm LL}^{\mu\nu}$ nor its individual nonlinear companions are generally covariant local stress tensors.

\subsection{Metric perturbations, the trace, and the Newtonian limit}
To distinguish the metric perturbation from the gothic variable, write
\begin{equation}
 g_{\mu\nu}=\eta_{\mu\nu}+\psi_{\mu\nu}+O(\psi^2),\qquad
 \psi=\eta^{\mu\nu}\psi_{\mu\nu},\qquad
 \overline\psi^{\mu\nu}=\psi^{\mu\nu}-\tfrac12\eta^{\mu\nu}\psi.
 \label{eq:gr-linear-metric}
\end{equation}
Raising indices with $\eta^{\mu\nu}$ to this order gives
\begin{equation}
 g^{\mu\nu}=\eta^{\mu\nu}-\psi^{\mu\nu}+O(\psi^2),\qquad
 \sqrt{-g}=1+\tfrac12\psi+O(\psi^2),\qquad
 H^{\mu\nu}=-\overline\psi^{\mu\nu}+O(\psi^2).
 \label{eq:gr-linear-gothic}
\end{equation}
Consequently $\eta_{\mu\nu}H^{\mu\nu}=\psi+O(\psi^2)$, and the trace of $\mathfrak h^{\mu\nu}$ defined in Eq.~\eqref{eq:gr-gothic-definition} is $\psi/2$. Substituting Eq.~\eqref{eq:gr-linear-gothic} into Eq.~\eqref{eq:gr-harmonic} gives
\begin{equation}
 \pd_\nu\overline\psi^{\mu\nu}=0,
 \label{eq:gr-linear-harmonic}
\end{equation}
the trace-reversed harmonic condition~\cite{Schutz1985,Carroll2013}. In the present curvature convention,
\begin{equation}
 R^{(1)}=\Boxeta\psi-\pd_\mu\pd_\nu\psi^{\mu\nu}
       =\tfrac12\Boxeta\psi
       \quad\hbox{in harmonic coordinates}.
 \label{eq:gr-linear-ricci}
\end{equation}
For $\Lambda=0$, contraction of Eq.~\eqref{eq:gr-hn-einstein} then gives
\begin{equation}
 \Boxeta\psi=\frac{16\pi\GN}{c^2}T
 \label{eq:gr-linear-trace}
\end{equation}
to linear order. The factor $16\pi$ follows from the trace-reversed harmonic condition.

Let $U>0$ denote the positive Newtonian potential magnitude, with $\phi=-U$ and $\nabla^2U=-4\pi\GN\rho$. The leading isotropic, weak-field GR metric is
\begin{equation}
 g_{00}=1-\frac{2U}{c^2}+O(c^{-4}),\qquad
 g_{ij}=-\left(1+\frac{2U}{c^2}\right)\delta_{ij}+O(c^{-4}).
 \label{eq:gr-newtonian-metric}
\end{equation}
Terms $g_{0i}=O(c^{-3})$ and the higher-order pieces are determined by the source motion. Equations~\eqref{eq:gr-linear-gothic} and~\eqref{eq:gr-newtonian-metric} imply
\begin{equation}
 \psi=\frac{4U}{c^2}+O(c^{-4}),\qquad
 H^{00}=\frac{4U}{c^2}+O(c^{-4}),\qquad
 H^{ij}=O(c^{-4}).
 \label{eq:gr-H00}
\end{equation}
The spatial metric perturbation contributes to the trace at the same order as the time--time perturbation. Setting $\psi_{ij}=0$ while retaining the harmonic trace equation therefore discards terms of the retained order. For a slow test particle $g_{00}$ determines the leading acceleration, but the trace still requires $g_{ij}$. In the static limit, Eq.~\eqref{eq:gr-linear-trace} gives $R^{(1)}=-2\nabla^2U/c^2=8\pi\GN\rho/c^2$.

\subsection{The second time derivative is part of the gravitational operator}
The term of interest in Eq.~\eqref{eq:gr-exact-relaxed} is
\begin{equation}
 \mathcal N^{\mu\nu}\supset-H^{00}\pd_0^2H^{\mu\nu}
 =-\frac{H^{00}}{c^2}\pd_t^2H^{\mu\nu}.
 \label{eq:principal}
\end{equation}
Moving $-H^{\alpha\beta}\pd_\alpha\pd_\beta H^{\mu\nu}$ to the left of Eq.~\eqref{eq:gr-exact-relaxed} gives, exactly,
\begin{equation}
 \mathfrak g^{\alpha\beta}\pd_\alpha\pd_\beta H^{\mu\nu}
 =\frac{16\pi\GN}{c^2}(-g)(T^{\mu\nu}+t_{\rm LL}^{\mu\nu})
 +(\pd_\beta H^{\mu\alpha})(\pd_\alpha H^{\nu\beta}).
 \label{eq:gr-principal-restored}
\end{equation}
Division by $\sqrt{-g}$ restores $g^{\alpha\beta}\pd_\alpha\pd_\beta$, the principal operator of the curved-spacetime equation. The second time derivative belongs to this operator. Its appearance on the right of the flat-wave equation does not create an additional material source.

Substitution of Eq.~\eqref{eq:gr-H00} into the selected time--time term gives
\begin{equation}
 \left.\tau^{00}\right|_{H^{00}\pd_t^2H^{00}}
 =-\frac{1}{16\pi\GN}H^{00}\pd_t^2H^{00}
 \simeq-\frac{U\ddot U}{\pi\GN c^4}.
 \label{eq:gr-time-density}
\end{equation}
If one formally inserts $U=\GN\Mloc(t)/r$ at fixed $r$, this selected expression becomes
\begin{equation}
 \left.\tau^{00}\right|_{H^{00}\pd_t^2H^{00}}
 \simeq-\frac{U}{\pi c^4r}\ddot{\Mloc}.
 \label{eq:gr-formal-mass-derivative}
\end{equation}
The coefficient contains the local potential $U$, not the cosmic potential $\Phi$. At fixed derivative scales, $H^{00}=O(\epsilon_g)$ with weak-field amplitude $\epsilon_g=U/c^2$ makes $H^{00}\partial_0^2H^{00}=O(\epsilon_g^2)$. Replacing its local $O(\epsilon_g)$ coefficient by $|\Phi|/c^2=O(1)$ promotes the time-dependent contribution to $O(\epsilon_g)$ in the same expansion. This changes the retained equation, not an omitted higher-order remainder. Prescribing $\Mloc(t)$ also requires the energy fluxes and stresses that change it, as follows from the conservation equations below.

\subsection{The complete quadratic source and weak-field ordering}
Equation~\eqref{eq:relaxed} must be expanded as a whole. Write
$H^{\mu\nu}=H_{(1)}^{\mu\nu}+H_{(2)}^{\mu\nu}+\cdots$ and
$T^{\mu\nu}=T_{(1)}^{\mu\nu}+T_{(2)}^{\mu\nu}+\cdots$, where the order counts powers of the weak field and of $16\pi\GN L^2T^{\mu\nu}/c^2$. This is not a velocity expansion. With $H_{\rm tr}=\eta_{\mu\nu}H^{\mu\nu}$,
\begin{align}
 -g&=1+H_{\rm tr}
  +\tfrac12\big(H_{\rm tr}^2-H^{\mu\nu}H_{\mu\nu}\big)+O(H^3),\nonumber\\
 \tau_{(1)}^{\mu\nu}&=T_{(1)}^{\mu\nu},\nonumber\\
 \tau_{(2)}^{\mu\nu}&=T_{(2)}^{\mu\nu}
  +H_{{\rm tr}(1)}T_{(1)}^{\mu\nu}
  +\frac{c^2}{16\pi\GN}\mathcal Q_{\rm grav}^{\mu\nu}[H_{(1)}].
 \label{eq:gr-complete-source-expansion}
\end{align}
The material correction $T_{(2)}^{\mu\nu}$ follows from the same matter equations and boundary data as $T_{(1)}^{\mu\nu}$. It cannot be omitted while assigning a metric-dependent mass to the source.

The complete quadratic gravitational polynomial is~\cite[Eq.~(26)]{Blanchet2014}
\begin{align}
 \mathcal Q_{\rm grav}^{\mu\nu}[H]={}&
 -H^{\alpha\beta}\pd_\alpha\pd_\beta H^{\mu\nu}
 +\tfrac12\pd^\mu H_{\alpha\beta}\pd^\nu H^{\alpha\beta}
 -\tfrac14\pd^\mu H_{\rm tr}\pd^\nu H_{\rm tr}\nonumber\\
 &+\pd_\beta H^{\mu\alpha}
       \big(\pd^\beta H_\alpha{}^\nu+\pd_\alpha H^{\nu\beta}\big)
 -2\pd^{(\mu}H_{\alpha\beta}\pd^\alpha H^{\nu)\beta}
 \nonumber\\
 &+\eta^{\mu\nu}\left[
 -\tfrac14\pd_\gamma H_{\alpha\beta}\pd^\gamma H^{\alpha\beta}
 +\tfrac18\pd_\alpha H_{\rm tr}\pd^\alpha H_{\rm tr}
 +\tfrac12\pd_\alpha H_{\beta\gamma}\pd^\beta H^{\alpha\gamma}
 \right].
 \label{eq:gr-complete-quadratic}
\end{align}
All indices in Eq.~\eqref{eq:gr-complete-quadratic} are raised and lowered with $\eta_{\mu\nu}$, and every $H$ on its right is a first-order field satisfying $\pd_\nu H^{\mu\nu}=0$. Thus
\begin{equation}
 \mathcal Q_{\rm grav}^{\mu\nu}[H_{(1)}]
 =\mathcal N_{(2)}^{\mu\nu}
 +\frac{16\pi\GN}{c^2}t_{{\rm LL}(2)}^{\mu\nu}.
 \label{eq:gr-quadratic-parts}
\end{equation}
Changing both the metric signature and the sign of the gothic perturbation in Ref.~\cite{Blanchet2014} leaves this quadratic polynomial unchanged. It agrees with the conventions of Eq.~\eqref{eq:gr-gothic-definition}.

For a static leading Newtonian field, $H_{(1)}^{00}=4U/c^2$ and $H_{(1)}^{0i}=H_{(1)}^{ij}=0$, Eq.~\eqref{eq:gr-complete-quadratic} gives
\begin{equation}
 \mathcal Q_{\rm grav}^{00}=-\frac{14}{c^4}|\nabla U|^2,
 \qquad
 \tau_{{\rm grav}(2)}^{00}
 =-\frac{7}{8\pi\GN c^2}|\nabla U|^2.
 \label{eq:gr-static-quadratic}
\end{equation}
If $U$ varies with time, that component truncation violates harmonic gauge:
$\pd_0H^{00}+\pd_iH^{0i}=0$ requires the accompanying current field. Equation~\eqref{eq:gr-time-density} is consequently one selected term, not the complete quadratic source of a time-dependent apparatus.

Let a perturbation vary on spatial scale $L$ and temporal scale $\omega^{-1}$, where $\omega=2\pi f$ and $f$ is the ordinary frequency. The independent dimensionless parameters are
\begin{equation}
 \epsilon_g=\frac{U}{c^2},\qquad
 \epsilon_\omega=\frac{\omega L}{c}.
 \label{eq:gr-ordering-parameters}
\end{equation}
For component amplitudes of characteristic size $O(\epsilon_g)$, all quadratic terms in Eq.~\eqref{eq:gr-complete-quadratic} fall into the three derivative classes
\begin{equation}
 \mathcal Q_{{\rm grav},ss}=O\!\left(\frac{\epsilon_g^2}{L^2}\right),\qquad
 \mathcal Q_{{\rm grav},st}=O\!\left(\frac{\epsilon_g^2\epsilon_\omega}{L^2}\right),\qquad
 \mathcal Q_{{\rm grav},tt}=O\!\left(\frac{\epsilon_g^2\epsilon_\omega^2}{L^2}\right).
 \label{eq:gr-quadratic-classes}
\end{equation}
Here $ss$, $st$ and $tt$ label two spatial derivatives, one derivative of each kind, and two time derivatives. Both $H\pd\pd H$ and $(\pd H)(\pd H)$ are included. These are characteristic amplitude envelopes obtained by assigning $O(\epsilon_g)$ to every component, not component-by-component post-Newtonian estimates or pointwise bounds. For the near-zone field of slowly moving, weakly stressed matter, assume compatible source conservation, no independently imposed homogeneous gravitational field, and total current scaling $T^{0i}/T^{00}=O(v/c)$. The source equations then give
\begin{align}
 H^{00}&=O(\epsilon_g),\qquad
 H^{0i}=O(\epsilon_g v/c),\nonumber\\
 H^{ij}&=O\!\left[\epsilon_g\left(\frac{v^2}{c^2}
                    +\frac{|\sigma|}{\rho c^2}\right)\right]+O(\epsilon_g^2).
 \label{eq:gr-component-ordering}
\end{align}
Here $v$ is the characteristic material speed, $\rho>0$ a characteristic mass density and $|\sigma|$ the non-kinetic material and electromagnetic stress scale. An electrical or thermal energy flux that does not obey this mass-current scaling must be retained independently in $T^{0i}$. The harmonic conditions constrain divergences; they do not, by themselves, fix these amplitudes. The orders $H^{0i}=O(\epsilon_g\epsilon_\omega)$ and $H^{ij}=O(\epsilon_g\epsilon_\omega^2)$ require $v=O(\omega L)$, $|\sigma|/\rho=O(v^2)$ and consistent ordering of the nonlinear contribution~\cite{Blanchet2014,PoissonWill2014}. For a vibrating stack, $v\sim\omega A$ depends on displacement amplitude $A$, which is generally much smaller than $L$. Its currents and stresses require their own ordering. The determinant correction in Eq.~\eqref{eq:gr-complete-source-expansion} is $O(\epsilon_gT_{(1)}^{\mu\nu})$; the material correction requires the current, stress and driving solution.

In particular,
\begin{equation}
 \frac{|H^{00}c^{-2}\pd_t^2H^{\mu\nu}|}
      {|\nabla^2H^{\mu\nu}|}_{\rm char}
 =O(\epsilon_g\epsilon_\omega^2).
 \label{eq:near-gr}
\end{equation}
This compares characteristic terms; it is not a pointwise bound at a zero of the Laplacian. Different background and perturbation scales must be kept separately.

For the 2013 configuration, $L=0.019$ m, $f=32.35$ kHz, and stack mass $0.046$ kg~\cite{Fearn2013} give
\begin{equation}
 \epsilon_\omega=1.29\times10^{-5},\qquad
 \epsilon_\omega^2=1.66\times10^{-10},\qquad
 \frac{\GN(0.046\ {\rm kg})}{Lc^2}=1.80\times10^{-27}.
 \label{eq:gr-stack-scales}
\end{equation}
The corresponding characteristic quadratic-to-linear ratios are
\begin{equation}
 \frac{(|\mathcal Q_{{\rm grav},ss}|,|\mathcal Q_{{\rm grav},st}|,|\mathcal Q_{{\rm grav},tt}|)}
      {\epsilon_g/L^2}
 =O\!\left(1.80\times10^{-27},\,
           2.32\times10^{-32},\,
           2.98\times10^{-37}\right).
 \label{eq:gr-quadratic-lab-ratios}
\end{equation}
These numbers are the uniform-component envelopes of Eq.~\eqref{eq:gr-quadratic-classes}. Equation~\eqref{eq:gr-component-ordering} further suppresses contributions containing current or spatial components. It leaves the selected $H^{00}\partial_t^2H^{00}$ estimate unchanged and does not eliminate other contributions at the same resulting order. They do not specify the individual contractions or a measured force: the three quoted apparatus parameters determine neither the current and stress fields nor the oscillating fraction of $U$. The largest purely spatial cubic scale divided by the quadratic temporal scale is $\epsilon_g/\epsilon_\omega^2=1.08\times10^{-17}$ for this self-field. External fields require their own amplitudes and length scales.

For a self-gravitating system in the usual virial counting, $\epsilon_{\rm PN}\sim v^2/c^2\sim U/c^2$ and $\omega L\sim v$, so that
\begin{equation}
 H^{00}\pd_0^2H^{00}=O(\epsilon_{\rm PN}^3/L^2),\qquad
 \nabla^2H^{00}=O(\epsilon_{\rm PN}/L^2).
 \label{eq:gr-pn-count}
\end{equation}
The selected term is 2PN relative to the leading Newtonian time--time equation. Its absolute order is $c^{-6}$. A complete calculation at that PN order also retains the other velocity, stress and nonlinear terms of the same order~\cite{PoissonWill2014,Blanchet2014}. For an electrically driven resonance, gravitational virial counting does not apply; Eqs.~\eqref{eq:gr-ordering-parameters}--\eqref{eq:gr-quadratic-lab-ratios} keep the independent scales explicit.

\subsection{Source conservation and the meaning of a fluctuating mass}
Taking the partial divergence of Eq.~\eqref{eq:relaxed} and using Eq.~\eqref{eq:gr-harmonic} gives
\begin{equation}
 \pd_\nu\tau^{\mu\nu}=0.
 \label{eq:gr-total-conservation}
\end{equation}
Integration over a fixed spatial region $\mathcal V_0$ gives
\begin{align}
 \frac{\dd}{\dd t}\int_{\mathcal V_0}\tau^{00}\,\dd^3x
   &=-c\int_{\partial\mathcal V_0}\tau^{0j}n_j\,\dd S,\nonumber\\
 \frac{\dd}{\dd t}\left(c\int_{\mathcal V_0}\tau^{i0}\,\dd^3x\right)
   &=-c^2\int_{\partial\mathcal V_0}\tau^{ij}n_j\,\dd S.
 \label{eq:gr-integrated-conservation}
\end{align}
Here $n_j$ is the outward Euclidean unit normal to the fixed spatial boundary and $\dd S$ is its area element. The volume energy and momentum are conserved when the corresponding boundary fluxes vanish. A powered laboratory subsystem has surface terms from its conductors, supports, electromagnetic fields and radiation. Dropping these terms while prescribing a changing subsystem energy contradicts Eq.~\eqref{eq:gr-integrated-conservation}.

At linear order, with suitable isolated retarded boundary conditions, the solution is
\begin{equation}
 H^{\mu\nu}(t,\mathbf x)
 =\frac{4\GN}{c^2}\int
 \frac{T^{\mu\nu}(t-|\mathbf x-\mathbf x'|/c,\mathbf x')}
 {|\mathbf x-\mathbf x'|}\,\dd^3x'.
 \label{eq:gr-retarded-linear}
\end{equation}
The leading source in this expression must satisfy $\pd_\nu T^{\mu\nu}=0$. An arbitrary time-dependent monopole of a closed source violates that condition. Electrical, elastic and kinetic energies may exchange while their sum, including the supply, remains constant; their spatial distribution and higher moments may vary.

The relation to stresses follows directly by differentiating a moment of the conserved source twice. For convergent moments with vanishing relevant surface terms,
\begin{equation}
 \frac{\dd^2}{\dd t^2}
 \int\tau^{00}x^ix^j\,\dd^3x
 =2c^2\int\tau^{ij}\,\dd^3x.
 \label{eq:gr-moment-conservation}
\end{equation}
The leading radiation expansion uses the corresponding conserved material source~\cite{Thorne1980,Thorne1983,Maggiore2008}. Equation~\eqref{eq:gr-moment-conservation} relates second derivatives of mass moments to stresses. It supplies no independent monopole mass fluctuation proportional to a prescribed second derivative of stored energy.

\subsection{Local fields and the proposed cosmic potential}
Equation~\eqref{eq:gr-principal-restored} has principal operator
$g^{\alpha\beta}\partial_\alpha\partial_\beta$ before any weak-field expansion.
A constant background normalization illustrates why a coordinate coefficient
cannot be identified with a new material source. For
$\dd s^2=N_{\rm bg}^2c^2\dd t^2-A_{\rm bg}^2\dd\mathbf x^2$,
with constant $N_{\rm bg},A_{\rm bg}>0$,
\begin{equation}
 \left(\frac{\partial_t^2}{N_{\rm bg}^2c^2}
       -\frac{\nabla_{\mathbf x}^2}{A_{\rm bg}^2}\right)
 =\frac{\partial_{t_{\rm loc}}^2}{c^2}
       -\nabla_{\mathbf x_{\rm loc}}^2,
 \qquad t_{\rm loc}=N_{\rm bg}t,\quad
 \mathbf x_{\rm loc}=A_{\rm bg}\mathbf x.
 \label{eq:gr-background-normalization}
\end{equation}
An order-unity coefficient in the first representation is possible. It is
absorbed into the same physical clocks and rods that define the local
acceleration, energy and drive frequency. Replacing only the coefficient in
Eq.~\eqref{eq:gr-formal-mass-derivative} by $|\Phi|/c^2$ does not perform this
transformation.

The slowly varying cosmological case can be calculated locally. For a
spatially flat Friedmann metric~\cite{Friedmann1922,EllisMaartensMacCallum2012},
set $H_{\rm cos}=\dot a_{\rm cos}/a_{\rm cos}$ and
$\mathbf r=a_{\rm cos}(t_{\rm c})\mathbf x_{\rm c}$:
\begin{align}
 \dd s^2
 &=c^2\dd t_{\rm c}^2-a_{\rm cos}^2\dd\mathbf x_{\rm c}^2\nonumber\\
 &=\big(c^2-H_{\rm cos}^2r^2\big)\dd t_{\rm c}^2
       +2H_{\rm cos}\mathbf r\cdot\dd\mathbf r\,\dd t_{\rm c}
       -\dd\mathbf r^2.
 \label{eq:gr-flrw-physical-coordinates}
\end{align}
Let $t_{\rm loc}=t_{\rm c}+H_{\rm cos}(t_{\rm c})r^2/(2c^2)$.
Here $r=|\mathbf r|$ is the physical radial distance on a constant-$t_{\rm c}$
hypersurface and the areal radius. Its defining relation remains
$\mathbf r=a_{\rm cos}(t_{\rm c})\mathbf x_{\rm c}$. Exactly,
\[
 \dd t_{\rm loc}
 =\left(1+\frac{\dot H_{\rm cos}r^2}{2c^2}\right)\dd t_{\rm c}
  +\frac{H_{\rm cos}}{c^2}\mathbf r\cdot\dd\mathbf r.
\]
The time and mixed terms in Eq.~\eqref{eq:gr-flrw-physical-coordinates} give
\[
 \begin{split}
 &(c^2-H_{\rm cos}^2r^2)\dd t_{\rm c}^2
       +2H_{\rm cos}\mathbf r\cdot\dd\mathbf r\,\dd t_{\rm c}\\
 &\quad=\big[c^2-(\dot H_{\rm cos}+H_{\rm cos}^2)r^2\big]
       \dd t_{\rm loc}^2
       -\frac{H_{\rm cos}^2}{c^2}(\mathbf r\cdot\dd\mathbf r)^2
       +O(r^3).
 \end{split}
\]
Through second order in the distance from the central comoving geodesic,
\begin{align}
 \dd s^2={}&\left[1-
       \frac{(\dot H_{\rm cos}+H_{\rm cos}^2)r^2}{c^2}\right]
       c^2\dd t_{\rm loc}^2
       -\left(\delta_{ij}
          +\frac{H_{\rm cos}^2r_i r_j}{c^2}\right)\dd r^i\dd r^j
       +O(r^3),
 \label{eq:gr-flrw-local-chart}
\end{align}
where $O(r^3)$ denotes the Taylor remainder of the metric coefficients, and
$H_{\rm cos}$ and its dot are evaluated at $t_{\rm loc}$ to the displayed
order. This chart is locally inertial on the central geodesic; it is not
asserted to be an exact Fermi chart away from that geodesic. Define the
cosmological curvature scale
\begin{equation}
 \mathcal C_{\rm cos}
 =\frac{\max\{|\dot H_{\rm cos}+H_{\rm cos}^2|,H_{\rm cos}^2\}}{c^2}.
 \label{eq:gr-cosmological-curvature-scale}
\end{equation}
For $\mathcal C_{\rm cos}L^2\ll1$ and a background varying slowly during a
drive cycle, the background metric corrections are
$O(\mathcal C_{\rm cos}L^2)$, its connections are
$O(\mathcal C_{\rm cos}L)$, and its curvature is
$O(\mathcal C_{\rm cos})$. Linearization about this background therefore
retains the local wave operator, background-curvature terms and their
couplings to the laboratory perturbation. In particular, the cosmological
change in the coefficient of the local second time derivative is
$O(\mathcal C_{\rm cos}L^2)$, not an independent $|\Phi|/c^2\simeq1$
material source. A nonzero spatial curvature adds its usual
$|k|/a_{\rm cos}^2$ contribution to the curvature scale. An accelerated
laboratory frame also retains its acceleration and rotation terms.

Any smooth matching to an asymptotically flat or cosmological exterior
must satisfy the local Einstein equations and $\nabla_aT^{ab}=0$.
The chart in Eq.~\eqref{eq:gr-flrw-local-chart} need not be harmonic;
$\partial_\nu\tau^{\mu\nu}=0$ in
Eq.~\eqref{eq:gr-total-conservation} belongs to the harmonic representation
and is not being imposed as a tensor equation in this new chart.
With no surface layer, the induced metric and extrinsic curvature are
continuous across the matching surface; a surface layer requires its own
stress and momentum transfer. Matching can determine tidal fields,
incoming radiation and boundary fluxes. It does not convert an isolated
part of the principal operator into the proposed mass source.
The calculation above is a local expansion of a cosmological background,
not a construction of a global apparatus solution or a bound on all
external-field forces. The source law used by Woodward~\cite{Woodward2004,WoodwardStarship}
still requires a derivation from the conserved stress and the full metric.
Brans' local inertial-induction analysis~\cite{Brans1977} and the distinct
cosmological examples of G\"odel and Ozsv\'ath--Sch\"ucking~\cite{Godel1949,OzsvathSchucking1962}
also preclude identifying a general statement of Mach's principle with a
unique laboratory force law.

\subsection{The source substitution in Tajmar's thruster model}
\label{sec:tajmar-gravity}

Equation~(3) of Ref.~\cite{Tajmar2017} moves the time derivative in a scalar gravitational wave equation to the source side. Equation~(4) of Ref.~\cite{Tajmar2017} then defines
\begin{equation}
 \delta\rho_{\rm geom}=\frac{1}{4\pi Gc^2}\pd_t^2\phi_g.
 \label{eq:tajmar-effective-source}
\end{equation}
Equation~(4) of Ref.~\cite{Tajmar2017} also substitutes $\phi_g=-G\mathcal M/r$. At fixed $r$,
\begin{equation}
 \pd_t^2\phi_g=\frac{\phi_g}{\mathcal M}\ddot{\mathcal M},
 \qquad
 \delta\rho_{\rm geom}
 =\frac{\phi_g}{4\pi Gc^2\mathcal M}\ddot{\mathcal M}.
 \label{eq:tajmar-local-potential}
\end{equation}
Here $\phi_g$ is the potential of the local mass $\mathcal M$. Equation~(5) of Ref.~\cite{Tajmar2017} assigns this coefficient the cosmic value $\phi_g/c^2=-1$. That replacement does not follow from Eq.~(4) of Ref.~\cite{Tajmar2017}: the derivative of the local point-mass potential has been calculated, but its coefficient has been replaced by a different potential. Adding a constant cosmic potential leaves $\pd_t^2\phi_g$ unchanged. Making the cosmic potential vary proportionally to the local mass would require an additional dynamical relation absent from the derivation.

There is a second failure at the starting step. Moving a derivative from one side of a wave equation to the other does not convert it into a material mass fluctuation. Equations~\eqref{eq:gr-exact-relaxed} and~\eqref{eq:gr-principal-restored} display this distinction explicitly in GR. Tajmar's scalar rearrangement and potential substitution~\cite{Tajmar2017} therefore do not derive his mass-fluctuation law from Einstein's equations. The mechanical calculation in Ref.~\cite{Tajmar2017} is examined in Sect.~\ref{sec:tajmar-mechanics}.

\section{Second time derivatives of mass--energy in Hoyle--Narlikar's conformal gravitational theory}
\label{sec:HN}
Fearn and Woodward invoke Hoyle--Narlikar's conformal field equations to justify the mass fluctuations and propulsion claimed for their devices~\cite{Fearn2015I,Fearn2015II,Fearn2016}. The time--time component cancels the linear second time derivative retained in Refs.~\cite{Fearn2015I,Fearn2015II}, and the tensor trace cancels the squared-gradient terms retained by the diagonal sum in Ref.~\cite{Fearn2016}. Both derivations are mathematically incorrect.

Hoyle and Narlikar developed their theory to connect inertia with the distribution of matter~\cite{HoyleNarlikar1964,HoyleNarlikar1966,Hoyle1974}, drawing on advanced and retarded interactions in absorber electrodynamics~\cite{WheelerFeynman1945,FeynmanWheeler1949}. Cosmological applications include variable-mass hypotheses and quasi-steady-state radiation~\cite{NarlikarDas1980,NarlikarArp1997,NarlikarEtAl2015}. The action below specifies the smooth conformal scalar--tensor sector used here. An additional creation field is excluded. Its Weyl symmetry distinguishes the scalar from an independent Brans--Dicke field~\cite{BransDicke1961}. After examining the proposed derivations, I establish the corresponding Einstein description in Sect.~\ref{sec:equivalence}.

\subsection{Hoyle--Narlikar's conformal scalar--tensor theory}
\label{sec:action}
\label{sec:HN-review}
\subsubsection{Definitions, notation and conventions}
I use Latin spacetime indices for comparison with HN, retaining the signature and curvature convention of Sect.~\ref{sec:GR}. Covariant differentiation is denoted by $\nabla_a$, or by a semicolon. The scalar wave operator is $\Box=g^{ab}\nabla_a\nabla_b$. Its local Minkowski form is $c^{-2}\partial_t^2-\nabla^2$. Equation~\eqref{eq:curvature-definition} contracts the upper index with the third (last) lower index: $R_{ab}=\sum_rR^r{}_{abr}$. Contracting the same four-index tensor on its second lower index gives $\sum_rR^r{}_{arb}=-R_{ab}$. The dummy index $r$ is distinct from the free indices $a,b$.

The invariant trace of the material tensor is $T=g^{ab}T_{ab}$. When using SI units, $T_{ab}=\mathsf T_{ab}/c^2$ denotes the tensor in mass-density units, with $\mathsf T_{ab}$ in energy-density units. In the action and its variational equations below we set $c=1$; SI factors are restored explicitly in the weak-field comparison and apparatus calculations. 

The HN scalar $m(x)$ sets the inertial scale. In the SI normalization of Ref.~\cite{Rodal2019}, its dimensions are $\sqrt{\mathrm{kg}/\mathrm m}$, not kg. A particle mass may be written $M_{\rm p}(x)=\lambda m(x)$, where the fixed coupling $\lambda$ supplies the remaining dimensions. For dust, $\rho_{\rm rest}=\lambda mN$, where $N$ is proper particle-number density. The macroscopic mass--energy variable $\mathcal M$ used for a device is a different quantity, obtained from its complete material and field energy. Identifying the scalar $m$ directly with the mass or density of one capacitor confuses these definitions.

\subsubsection{The complete action and universal material coupling}
With $c=1$, let
\begin{equation}
 \kappa=8\pi\GN,\qquad m_0^2=\frac6\kappa,\qquad
 \Omega=\frac{m}{m_0}>0,\qquad \widehat g_{ab}=\Omega^2g_{ab}.
 \label{eq:invariantmetric}
\end{equation}
The constant $m_0$ fixes the units and the Einstein-frame normalization. In the SI mass-density normalization of Ref.~\cite{Rodal2019}, $m_0^2=3c^2/(4\pi\GN)$ has units of kg/m. Restoring those units gives an overall factor $c$ in the $m$-normalized gravitational action below and the particle action $-\lambda c\int m\,\dd s_g$, with $x^0=ct$. The admissibility hypotheses for the equivalence result are stated in Sect.~\ref{sec:equivalence}.

The defining matter hypothesis is
\begin{equation}
 S_{\rm mat}[g,m,\Psi]=\widehat S_{\rm mat}[\widehat g,\widehat\Psi].
 \label{eq:universal}
\end{equation}
Here $\widehat\Psi$ denotes invariant matter variables. Equation~\eqref{eq:universal} includes the dielectric, elastic body, electromagnetic field, electrical supply, heat reservoirs, supports and cosmological matter. Their constitutive parameters must transform with the corresponding material standards.

For a spacetime domain $\mathcal X$ with a smooth non-null boundary, let $n^a$ be the outward unit normal, $n_an^a=\epsilon_n=\pm1$, $h_{ab}=g_{ab}-\epsilon_n n_an_b$ and $K=h^{ab}\nabla_an_b$. Here $h_{ab}$ is the induced metric, $K_{ab}=h_a{}^ch_b{}^d\nabla_cn_d$ is its extrinsic curvature, $K=h^{ab}K_{ab}$ is its trace and $\epsilon_n$ records the causal character of the unit normal. In boundary coordinates, $h=\det(h_{ij})$. Consider
\begin{align}
 S={}&\frac1{2\kappa}\int_{\mathcal X}\sqrt{-g}\,
 \big[\Omega^2R-6\nabla_a\Omega\nabla^a\Omega-2\Lambda\Omega^4\big]\,\dd^4x
 \nonumber\\
 &-\frac1\kappa\int_{\partial\mathcal X}\epsilon_n\sqrt{|h|}\,\Omega^2K\,\dd^3y
 +\widehat S_{\rm mat}[\widehat g,\widehat\Psi].
 \label{eq:fullaction}
\end{align}
The contraction in Eq.~\eqref{eq:curvature-definition} gives
\begin{equation}
 g^{ab}\delta R_{ab}=\nabla_r
 \left(g^{ar}\delta\Gamma^b{}_{ab}-g^{ab}\delta\Gamma^r{}_{ab}\right).
 \label{eq:HN-palatini-sign}
\end{equation}
In outward Gaussian normal coordinates, $\dd s^2=\epsilon_n\dd z^2+h_{ij}\dd y^i\dd y^j$, $n^a=\delta^a_z$ and $n_a=g_{ab}n^b=\epsilon_n\delta_a^z$. Thus
\begin{equation}
 K_{ij}=-\Gamma^z{}_{ij}n_z
 =-\left(-\frac{\epsilon_n}{2}\partial_z h_{ij}\right)\epsilon_n
 =\frac12\partial_z h_{ij}.
 \label{eq:GNC-extrinsic-curvature}
\end{equation}
At fixed $\Omega$ and induced metric on the boundary, the parts of the variations containing a normal coordinate derivative of $\delta h_{ij}$ are
\begin{equation}
 \left.\delta S_{\rm bulk}\right|_{\partial_z\delta h}
 =\frac1{2\kappa}\int_{\partial\mathcal X}
 \epsilon_n\sqrt{|h|}\,\Omega^2 h^{ij}\partial_z\delta h_{ij}\,\dd^3y
 =-\left.\delta S_{\rm GHY}\right|_{\partial_z\delta h}.
 \label{eq:boundary-variation-sign}
\end{equation}
Thus the negative boundary sign cancels the bulk variation in the stated HN convention. In the normalization of the scalar field $m$, the gravitational bulk Lagrangian is
\begin{equation}
 \sqrt{-g}\left[\frac{m^2R}{12}-\frac12(\nabla m)^2-\alpha m^4\right],
 \qquad \alpha=\frac{\Lambda}{6m_0^2},
 \label{eq:maction}
\end{equation}
and its boundary term is $-\frac16\int_{\partial\mathcal X}\epsilon_n\sqrt{|h|}m^2K\,\dd^3y$. The quartic term is required for a cosmological constant in the invariant metric. An observationally relevant positive $\Lambda$ cannot simply be added to one representation while omitted from the other.

Equation~\eqref{eq:fullaction} contains the fixed Weyl-compensator combination~\cite{BlaschkeDabrowski2012,Anastasiou2024}. An independent scalar kinetic term, nonuniversal matter coupling, additional curvature invariant or quantum anomaly would define a different theory.

\subsubsection{The tensor equation and its scalar contraction}
At fixed invariant matter variables define
\begin{equation}
 T_{ab}=\frac2{\sqrt{-g}}\left.\frac{\delta S_{\rm mat}}{\delta g^{ab}}\right|_{m,\widehat\Psi},\qquad
 \mathcal J=-\frac1{\sqrt{-g}}\left.\frac{\delta S_{\rm mat}}{\delta m}\right|_{g,\widehat\Psi},\qquad T=g^{ab}T_{ab}.
\end{equation}
These stress conventions give a positive dust energy density with signature $(+---)$. Variation with respect to the scalar gives
\begin{equation}
 \left(\Box+\frac R6\right)m-4\alpha m^3=\mathcal J,
 \label{eq:scalar}
\end{equation}
while variation with respect to the metric gives
\begin{align}
 \frac12m^2G_{ab}={}&-3T_{ab}
 +m(g_{ab}\Box m-\nabla_a\nabla_bm)\nonumber\\
 &+2\left(\nabla_am\nabla_bm-\frac14g_{ab}(\nabla m)^2\right)
 -3\alpha m^4g_{ab}.
 \label{eq:metriceq}
\end{align}
At fixed invariant matter variables, universality implies the off-shell Ward identity and stress transformation
\begin{equation}
 T=m\mathcal J,\qquad T_{ab}=\Omega^2\widehat T_{ab},\qquad
 T=\Omega^4\widehat T.
 \label{eq:ward}
\end{equation}
With noninvariant matter fields as independent variables, their Euler--Lagrange terms enter the Ward identity, as displayed in Eq.~\eqref{eq:raw-matter-Ward}; the relation $T=m\mathcal J$ with the scalar derivative then taken at fixed noninvariant fields holds on the matter equations. Thus the scalar source and material stress tensor cannot be prescribed independently.

For particles with action $-\lambda\int m\,\dd s_g$, the dust source is $\mathcal J=\lambda N$. A strained, polarized or heated medium must instead be varied from its own matter action. Although its stress correction may be small compared with its rest-energy density, it cannot be discarded when it is of the same order as the proposed signal. The use of a dust scalar source and an independently chosen stressed-medium tensor is not an exact completion of the theory.

Taking the trace of Eq.~\eqref{eq:metriceq} gives
\begin{equation}
 R=\frac{6T}{m^2}-6\frac{\Box m}{m}+24\alpha m^2.
 \label{eq:trace}
\end{equation}

The contraction uses $g^{ab}G_{ab}=-R$, while the contraction of
$g_{ab}\Box m-\nabla_a\nabla_bm$ is $3\Box m$. The quadratic derivative
combination is traceless. Collecting these terms gives Eq.~\eqref{eq:trace},
and combining it with the scalar equation gives the source identity
Eq.~\eqref{eq:ward}. The scalar and metric sources must therefore arise from
one material model.

\subsubsection{The conformal scalar equation}
For $\alpha=0$ and $\mathcal J=0$, Eq.~\eqref{eq:scalar} has the form of a conformally coupled massless Klein--Gordon equation. Its metric constraint, Eq.~\eqref{eq:trace}, remains. The isolated scalar equation therefore cannot establish an independent dilaton or monopole-radiation channel. The Weyl transformation of the action in Sect.~\ref{sec:equivalence} determines the physical degrees of freedom.

\subsection{Decomposition of the scalar field}
\label{sec:HN-perturbation}
For constant $m=m_0$, contraction of Eq.~\eqref{eq:gr-hn-einstein} gives
\begin{equation}
 -R+4\Lambda=-\frac{6T}{m_0^2},\qquad
 R-4\Lambda=\frac{6T}{m_0^2}.
 \label{eq:constant-m-trace}
\end{equation}
Here $\Lambda>0$ denotes positive vacuum energy. In SI mass-density units, $6/m_0^2=8\pi G_0/c^2$ fixes the Einstein coupling; it does not introduce an independent gravitational-coupling field in addition to $g_{ab}$ and $m$.

Let the scalar be composed of a constant part and a small spacetime-dependent part,
\begin{equation}
 m=m_0+m_s=m_0(1+s),\qquad |s|\ll1,\qquad \Box m=\Box m_s.
 \label{eq:HN-decomposition}
\end{equation}
It follows that
\begin{equation}
 \frac{6}{m^2}=\frac{6}{m_0^2}(1-2s+O(s^2)),\qquad
 \frac{\Box m}{m}=\frac{\Box m_s}{m_0}(1-s+\cdots).
 \label{eq:HN-coupling-expansion}
\end{equation}
Thus the expansion must be applied to the coefficient multiplying the material tensor as well as to the scalar derivative terms. What is retained depends on the background and on the order assigned to the material perturbation.

Write $T=T_0+\delta T$, $R=R_0+\delta R$, about a background solution with constant $m_0$. Linearizing Eq.~\eqref{eq:trace}, in units $c=1$, gives
\begin{equation}
 \delta R=\frac{6}{m_0^2}\delta T
 -\frac{12T_0}{m_0^2}s-6\Box_0s+48\alpha m_0^2s.
 \label{eq:HN-linear-trace}
\end{equation}
The coefficient correction $-12T_0s/m_0^2$ is first order when $T_0\ne0$. If instead the whole material source is assigned first perturbative order about a vacuum reference, $sT$ is second order and can consistently be omitted. The background source therefore determines whether freezing the coupling is consistent at first order.

Define the algebraic coefficient
\begin{equation}
 G_{\rm alg}(x)\equiv\frac{3c^2}{4\pi m^2(x)},\qquad
 G_{\rm alg}=G_0+G_s.
 \label{eq:Galgebraic}
\end{equation}
Thus $G_s/G_0=-2s+O(s^2)$. The coefficient $G_{\rm alg}$ changes under a Weyl transformation; its measurement cannot be specified independently of the transformation of the material standards.

The scalar equation must be linearized at the same time. Define the conformal scalar operator $\mathcal O_g=\Box_g+R/6$ and the combined source $\mathcal J_{\rm tot}=\mathcal J+4\alpha m^3$. For a constant background $m_0$, $\delta m=m_s$ and $\mathcal O_0=\Box_0+R_0/6$, so that
\begin{align}
 \mathcal O_0\delta m
 &=\delta\mathcal J_{\rm tot}-\left(\delta\Box+\frac{\delta R}{6}\right)m_0,
 \label{eq:linearized}\\
 \left(\Box_0+\frac{R_0}{6}-12\alpha m_0^2\right)\delta m
 +\frac{m_0}{6}\delta R&=\delta\mathcal J.
 \label{eq:HN-coupled-linear}
\end{align}
For constant $m_0$, $\delta\Box\,m_0=0$, whereas $m_0\delta R/6$ remains first order. Small $|m_s|/m_0$ therefore does not decouple the scalar and metric perturbations, including about a vacuum background: vanishing background material sources do not imply $\delta R=0$. Equations~\eqref{eq:HN-linear-trace} and~\eqref{eq:HN-coupled-linear}
must be imposed together. Eliminating $\delta R$, with
$R_0=6T_0/m_0^2+24\alpha m_0^2$, gives
\begin{equation}
 \delta T=m_0\delta\mathcal J+T_0s,
 \label{eq:HN-linear-Ward}
\end{equation}
which is the linearized Ward identity, not a separate propagation equation
for $s$.

\subsection{Fluctuation of the scalar field as an effective source}
\label{sec:HN-ms-source}
Rearranging Eq.~\eqref{eq:trace}, restoring SI units and taking $\Lambda=0$ for the local weak-field comparison, define
\begin{equation}
 T_{\rm field}\equiv\frac{3c^2}{4\pi G_0}\frac{\Box m_s}{m_0}
 =-\frac{3c^2}{8\pi G_0}\frac{\Box G_s}{G_0}+O(s^2).
 \label{eq:T_field_def}
\end{equation}
Here the order symbol presupposes a fixed derivative scale. To the indicated order, the scalar equation for curvature is
\begin{equation}
 R=\frac{8\pi G_0}{c^2}\left[(1-2s)T-T_{\rm field}\right]+\cdots.
 \label{eq:HN-effective-trace}
\end{equation}
The term $T_{\rm field}$ is the scalar derivative term transferred from the geometrical side of Eq.~\eqref{eq:trace}. This rearrangement introduces no additional material source; the coefficient correction $-2sT$ remains part of the same expansion.

\subsubsection{The material trace and the scalar source}
For a rest-frame medium, let $\varepsilon$ be its total energy density and let $\Pi_{ij}$ denote the spatial momentum-flux stress, positive for an isotropic positive pressure. Here $\Pi^i{}_i\equiv\delta^{ij}\Pi_{ij}$ denotes the Euclidean spatial trace in a local orthonormal rest frame. The physical trace in SI mass-density units is
\begin{equation}
 T=\frac{\varepsilon-\Pi^i{}_i}{c^2}.
 \label{eq:material-trace}
\end{equation}
For a perfect fluid this becomes $T=\varepsilon/c^2-3p/c^2$. If one defines the engineering Cauchy stress as positive in tension, its sign relative to $\Pi_{ij}$ must be reversed. The electromagnetic vacuum stress is traceless, but a dielectric, its polarization and the mechanical stresses supporting its field energy are not described by the vacuum tensor alone.

For particles with a universal scalar mass scale, $\rho_{\rm rest}=\lambda mN$. In the dust approximation the action gives $\mathcal J=\lambda N$ and $T=m\mathcal J$. For a general medium, however, $T=m\mathcal J$ remains the action identity, while the source $\mathcal J$ changes with the material action. The approximation of negligible pressure compared with rest energy does not justify dropping its contribution at an order at which a proposed electromechanical gravitational signal is being calculated.

\subsubsection{Particle-number variation is not inertial-mass variation}
Applying the covariant wave operator to the dust relation $\rho_{\rm rest}=\lambda mN$ gives
\begin{align}
 \Box\rho_{\rm rest}
 &=\lambda\left(N\Box m+m\Box N+
 2\nabla_am\nabla^aN\right),\\
 \frac{\Box\rho_{\rm rest}}{\rho_{\rm rest}}
 &=\frac{\Box m}{m}+\frac{\Box N}{N}
 +\frac{2\nabla_am\,\nabla^aN}{mN}.
 \label{eq:densityidentity}
\end{align}
Even the dust model therefore does not imply $\Box m/m=\Box\rho_{\rm rest}/\rho_{\rm rest}$.

With the unit particle tangent $u^a=U^a/c$, $u^au_a=1$, particle conservation is $\nabla_a(Nu^a)=0$, not $\nabla_aN=0$. In a nonrelativistic solid it gives
\begin{equation}
 \partial_tN+\nabla\cdot(N\mathbf v)=0,\qquad
 \frac{\Delta N}{N}=-\nabla\cdot\boldsymbol\xi
 \label{eq:numberstrain}
\end{equation}
to first order in the Lagrangian displacement $\boldsymbol\xi$; $\Delta$ denotes the material perturbation. An Eulerian perturbation additionally includes advection of a nonuniform reference density. The electrically driven compression of a piezoelectric stack therefore changes $N$ even when its particle number and individual rest masses are constant.

Equations~\eqref{eq:densityidentity} and~\eqref{eq:numberstrain} separate compression from variation of the inertial scale. A local proportionality between $m$ and the density is an additional assumption, not an interface condition or a consequence of particle conservation. It must satisfy the scalar and metric equations with the material boundary conditions.

Figure~\ref{fig:interface_fallacy} contrasts the additional tracking assumption with the admissible Einstein-gauge description. The rest-density and number-density variables in its right panel are evaluated in that representation; the conformal change of variables is derived in Sect.~\ref{sec:equivalence}.
\begin{figure}[tbp]
\centering
\begin{tikzpicture}[font=\footnotesize,>={Latex}]
\begin{scope}[xshift=-3.65cm]
 \node[align=center,text width=6.7cm] at (2,2.95)
 {\textbf{A. Additional tracking ansatz}\\$\delta m/m_0=\delta\rho_{\rm rest}/\overline\rho_{\rm rest}$};
 \fill[gray!15] (1,-.65) rectangle (3,1.75);
 \draw[->] (-.2,0)--(4.7,0) node[right] {$x$};
 \draw[->] (0,-.8)--(0,1.95) node[above] {Perturbation};
 \draw[red!75!black,thick] plot[smooth] coordinates{(0,0)(.6,.2)(1.1,1)(2,1.1)(2.9,1)(3.4,.2)(4,0)};
 \draw[blue!75!black,dashed,thick] plot[smooth] coordinates{(0,0)(.6,.2)(1.1,1)(2,1.1)(2.9,1)(3.4,.2)(4,0)};
 \node[red!75!black] at (2,1.5) {$\delta m/m_0$};
 \node[blue!75!black] at (2,.62) {$\delta\rho_{\rm rest}/\overline\rho_{\rm rest}$};
 \node at (2,-1.05) {Material region};
 \node[align=center,text width=6.6cm] at (2,-1.75)
 {Coincident profiles require a derivation;\\they are not an interface condition.};
\end{scope}
\begin{scope}[xshift=3.65cm]
 \node[align=center,text width=6.7cm] at (2,2.95)
 {\textbf{B. Einstein-gauge representation}\\$m=m_0$,\quad $\rho_{\rm rest}=\lambda m_0N$};
 \fill[gray!15] (1,-.65) rectangle (3,1.75);
 \draw[->] (-.2,0)--(4.7,0) node[right] {$x$};
 \draw[->] (0,-.8)--(0,1.95) node[above] {Perturbation};
 \draw[blue!75!black,dashed,thick] plot[smooth] coordinates{(0,0)(.6,.2)(1.1,1)(2,1.1)(2.9,1)(3.4,.2)(4,0)};
 \draw[red!75!black,very thick] (0,0)--(4,0);
 \node[red!75!black] at (2,-.4) {$\delta m=0$};
 \node[blue!75!black] at (2,1.5) {$\delta\rho_{\rm rest}/\overline\rho_{\rm rest}=\delta N/N_0$};
 \node at (2,-1.05) {Material region};
 \node[align=center,text width=6.6cm] at (2,-1.75)
 {Number conservation permits local\\density changes through deformation.};
\end{scope}
\end{tikzpicture}
\caption{Density variation and the HN scalar. Panel A imposes an additional proportionality between the scalar and rest density. Panel B uses the Einstein-gauge representation, in which $m=m_0$ while conserved particles produce a changing density through deformation. Here $\overline\rho_{\rm rest}=\lambda m_0N_0$ and $N_0$ are the reference rest density and number density. The profiles are schematic; the constant scalar is a representation choice.}
\label{fig:interface_fallacy}
\end{figure}

\subsection{Comparison with Fearn's analysis of the fluctuating scalar field}
\label{sec:HN-fearn-compare}

Fearn et al. derive a mass fluctuation from the $00$ component of the HN tensor equation~\cite{Fearn2015I,Fearn2015II}. Fearn subsequently obtains a similar expression by a diagonal sum~\cite[pp.~104--105, Eqs.~(90)--(93)]{Fearn2016}. The two calculations fail at different operations.

\subsubsection{The time--time component and the proposed mass fluctuation}

Equations~(42)--(43) of Ref.~\cite{Fearn2015I} and Eqs.~(4)--(5) of Ref.~\cite{Fearn2015II} identify a mass fluctuation from the time--time component of the derivative tensor. Setting $c=1$, that tensor before division by $m^2/2$ is
\begin{equation}
 \mathcal B_{ab}
 =m\left(g_{ab}\Box m-\nabla_a\nabla_bm\right)
  +2\left(\nabla_am\nabla_bm
               -\frac14g_{ab}\nabla_cm\nabla^cm\right).
 \label{eq:fearn-B-definition}
\end{equation}
Refs.~\cite[p.~1523, following Eq.~(42)]{Fearn2015I} and
\cite[p.~1870, following Eq.~(4)]{Fearn2015II} use
$g_{ab}=\operatorname{diag}(-1,1,1,1)$ for this component calculation.
The field-equation indices $\alpha,\beta$ in those equations are both
covariant. Their signature differs from ours, but the cancellation does not:
for $g_{00}=\varepsilon_{\rm sig}$, $g^{00}=\varepsilon_{\rm sig}$ and
$g^{ij}=-\varepsilon_{\rm sig}\delta^{ij}$, where
$\varepsilon_{\rm sig}=\pm1$,
\begin{equation}
 g_{00}g^{ab}m_{,ab}-m_{,00}
 =\varepsilon_{\rm sig}^{\,2}
   \left(m_{,00}-\sum_{i=1}^3m_{,ii}\right)-m_{,00}
 =-\sum_{i=1}^3m_{,ii}.
 \label{eq:fearn-two-signatures}
\end{equation}
The same signature reversal leaves $\mathcal B_{ab}$ and
$\mathcal D_{ab}=2\mathcal B_{ab}/m^2$ unchanged. An overall sign change
from reversing the Ricci convention multiplies both sides of the complete
field equation; it cannot change this zero coefficient into a nonzero one.
Let $e_{\hat0}^a=u_{\rm obs}^a$ be any unit timelike observer vector, $u_{\rm obs}^au_{{\rm obs}\,a}=1$, and
$e_{\hat i}^a$ an orthonormal spatial triad. Contracting
Eq.~\eqref{eq:fearn-B-definition} twice with $u_{\rm obs}^a$ gives
\begin{align}
 u_{\rm obs}^au_{\rm obs}^b\mathcal B_{ab}
 &=-m\sum_{i=1}^3m_{;\hat i\hat i}
   +\frac32(u_{\rm obs}^a\nabla_am)^2
   +\frac12\sum_{i=1}^3(e_{\hat i}^a\nabla_am)^2,
 \label{eq:fearn-general-frame}\\
 m_{;\hat i\hat i}
 &=e_{\hat i}(e_{\hat i}m)
   -(\nabla_{e_{\hat i}}e_{\hat i})^a\nabla_am
   \qquad\hbox{(no sum on $i$)}.
 \nonumber
\end{align}
The $m_{;\hat0\hat0}$ terms cancel exactly. The frame need not be inertial;
the connection terms in the spatial Hessian remain. In local inertial
coordinates at the evaluation point this gives
\begin{equation}
 g_{00}\Box m-\partial_t^2m=-\sum_{i=1}^3\partial_i^2m,
 \label{eq:fearn-00-cancellation}
\end{equation}
and hence
\begin{equation}
 \mathcal B_{00}
 =-m\nabla^2m+\frac32\dot m^2
                 +\frac12|\boldsymbol\nabla m|^2.
 \label{eq:fearn-B00-correct}
\end{equation}
The cancellation in the linear Hessian part also holds with signature
$(-+++)$ when the metric and wave-operator signs are changed together.

There is no $m\ddot m$ term. Equation~(43) of Ref.~\cite{Fearn2015I}, and likewise
Eq.~(5) of Ref.~\cite{Fearn2015II}, does not follow from the preceding tensor
equation in the respective source. Applying the normalization of
Refs.~\cite{Fearn2015I,Fearn2015II} to the actual component gives
\begin{equation}
 -\frac{2\mathcal B_{00}}{3m^2}
 =\frac{2\nabla^2m}{3m}
  -\frac{\dot m^2}{m^2}
  -\frac{|\boldsymbol\nabla m|^2}{3m^2}.
 \label{eq:fearn-normalized-00}
\end{equation}
The linear second time derivative required by Woodward's formula~\cite{Woodward2004} is absent. For spatially uniform $m=m_0+m_s(t)$, the remaining $\dot m^2$ term begins at second order. For a local source, the spatial derivatives in Eq.~\eqref{eq:fearn-normalized-00} must also be retained. These are the corrected geometric terms, not an independent material mass fluctuation.

Following Eq.~(6) of Ref.~\cite[p.~1871]{Fearn2015II}, Fearn et al. propose
\begin{equation}
 v_{\rm PZT}m_{;i}=m_{;0},\qquad
 v_{\rm PZT}^{\,2}m_{;ii}=m_{;00}
 \qquad (i=1,2,3;\ \hbox{no sum}),
 \label{eq:fearn-profile-assumptions}
\end{equation}
where $v_{\rm PZT}$ denotes the deformation speed in their $c=1$ units.
These are additional conditions on the scalar field. They do not follow
from the deformation velocity of the material. A conversion of its spatial
Hessian into time derivatives requires a scalar profile satisfying the
coupled metric, scalar and material equations and their boundary data.
Ref.~\cite{Fearn2015II} supplies no such solution. Fearn subsequently
identified the earlier incorporation of the spatial derivatives into time
derivatives as a mistake~\cite[p.~105, after Eq.~(93)]{Fearn2016}.

For a coordinate basis with $x^0=ct$, $g_{0i}=0$ and arbitrary $g_{00}>0$,
including $g_{00}=1-2U/c^2+O(U^2/c^4)$,
\begin{align}
 g_{00}\Box m-m_{;00}
 &=g_{00}\sum_{i,j=1}^3g^{ij}m_{;ij},
 \nonumber\\
 \mathcal B_{00}
 &=m g_{00}\sum_{i,j=1}^3g^{ij}m_{;ij}
   +\frac32(\partial_0m)^2
   -\frac12g_{00}\sum_{i,j=1}^3g^{ij}\partial_im\partial_jm.
 \label{eq:fearn-coordinate-00}
\end{align}
The first term is the contracted covariant spatial Hessian. In local inertial
coordinates at the evaluation point, $g_{00}=1$, $g^{ij}=-\delta^{ij}$ and
$\Gamma^a{}_{bc}=0$ there, so it reduces to $-m\nabla^2m$ as in
Eq.~\eqref{eq:fearn-00-cancellation}. Merely choosing Minkowski metric
components at a point does not set the connection to zero. Thus the nonunit
lapse does not restore $m\partial_0^2m$; the measured component is
$\mathcal B_{00}/g_{00}$. If $g_{0i}\ne0$, the coefficient of
$m_{;00}$ in $g_{00}\Box m-m_{;00}$ is instead
\begin{equation}
 g_{00}g^{00}-1=-\sum_{i=1}^3g_{0i}g^{i0}.
 \label{eq:fearn-shift-coefficient}
\end{equation}
For a weak metric with $g_{0i}=O(U/c^2)$ this coefficient is
$O(U^2/c^4)$, and cannot produce a term linear in $m_s$ in the joint
weak-field expansion about constant $m_0$. An arbitrary coordinate shift
can mix temporal and spatial derivatives in a coordinate component;
Eq.~\eqref{eq:fearn-general-frame} gives the exact observer projection.
Neither the orthonormal calculation nor its coordinate form sets the
curvature to zero.
The connection in these equations is that of the spacetime metric
$g_{ab}$ in the HN action. An effective acoustic metric, when a material wave equation admits
one, is a different constitutive construction; it cannot replace $g_{ab}$ in
the HN equation. Material acceleration can produce nonzero coordinate
connection coefficients even in flat spacetime. None has been discarded in
Eq.~\eqref{eq:fearn-general-frame}, and their magnitude is not needed for
its exact cancellation of the temporal Hessian.

\subsubsection{Contraction of the tensor equation}

Equation~(90) of Ref.~\cite{Fearn2016} has two covariant indices, $\delta T_{\alpha\beta}$. On p.~105, immediately before Eqs.~(91)--(92), Fearn explicitly chooses the flat metric $(+1,-1,-1,-1)$. Equation~(93) then adds covariant diagonal components and calls the result a trace. The signature and index positions are therefore the same as those used here. The derivative tensor in Eq.~(90), after division by $m^2/2$, is
\begin{equation}
 \mathcal D_{ab}
 =\frac2m\left(g_{ab}\Box m-\nabla_a\nabla_bm\right)
  +\frac4{m^2}\left(\nabla_am\nabla_bm
                  -\frac14g_{ab}\nabla_cm\nabla^cm\right).
 \label{eq:fearn-D-definition}
\end{equation}
Contracting with $g^{ab}$ and using $g^{ab}g_{ab}=4$ gives
\begin{align}
 g^{ab}\mathcal D_{ab}
 &=\frac2m(4\Box m-\Box m)
   +\frac4{m^2}\left((\nabla m)^2-(\nabla m)^2\right)\nonumber\\
 &=\frac6m\Box m.
 \label{eq:fearn-D-trace}
\end{align}
The squared-gradient terms cancel identically, before any approximation or metric choice.

For signature $(+---)$ in a local orthonormal basis, this contraction is
\begin{equation}
 \mathcal D^a{}_a
 =\mathcal D_{00}-\mathcal D_{11}
                  -\mathcal D_{22}-\mathcal D_{33},
 \label{eq:fearn-signed-trace}
\end{equation}
not the sum with four positive signs. In a general coordinate basis the trace is $g^{ab}\mathcal D_{ab}$, with the metric factors retained. With $x^0=ct$, write
$m_{,i}=\partial_i m$ and $m_{,ii}=\partial_i^2m$. Direct evaluation of
Eq.~\eqref{eq:fearn-D-definition} in flat coordinates gives
\begin{align}
 \mathcal D_{00}
 &=-\frac2m\sum_{j=1}^3m_{,jj}
   +\frac1{m^2}\left(\frac{3\dot m^2}{c^2}
                       +\sum_{j=1}^3m_{,j}^{\,2}\right),
 \nonumber\\
 \mathcal D_{ii}
 &=\frac2m\left(-\frac{\ddot m}{c^2}
                  +\sum_{j=1}^3m_{,jj}-m_{,ii}\right)
   +\frac1{m^2}\left(4m_{,i}^{\,2}+\frac{\dot m^2}{c^2}
                       -\sum_{j=1}^3m_{,j}^{\,2}\right)
       \quad\hbox{(no sum on $i$)},
 \nonumber\\
 \sum_{i=1}^3\mathcal D_{ii}
 &=-\frac6{mc^2}\ddot m+\frac4m\sum_{j=1}^3m_{,jj}
   +\frac1{m^2}\left(\frac{3\dot m^2}{c^2}
                       +\sum_{j=1}^3m_{,j}^{\,2}\right).
 \label{eq:fearn-components}
\end{align}
Consequently
\begin{equation}
 \mathcal D_{00}-\sum_{i=1}^3\mathcal D_{ii}
 =\frac6m\left(\frac{\ddot m}{c^2}
                         -\sum_{j=1}^3m_{,jj}\right)
 =\frac6m\Box m.
 \label{eq:fearn-explicit-contraction}
\end{equation}
No squared-gradient term survives.

In the energy-density normalization of Ref.~\cite{Fearn2016},
$-(8\pi G/c^4)\delta\mathsf T^{\rm eff}_{ab}=\mathcal D_{ab}$, where $\mathsf T_{ab}$ is in energy-density units as defined above.
Its actual scalar trace is consequently
\begin{equation}
 \frac{g^{ab}\delta\mathsf T^{\rm eff}_{ab}}{c^2}
 =-\frac{3c^2}{4\pi G}\frac{\Box m}{m}.
 \label{eq:fearn-effective-trace}
\end{equation}
For $m=m(t)$ this is $-3\ddot m/(4\pi Gm)$, with no $\dot m^2/m^2$ term.
For the same spatially homogeneous test field $m=m(t)$, the three
expressions are
\begin{align}
 \frac{g^{ab}\delta\mathsf T^{\rm eff}_{ab}}{c^2}
 &=-\frac{3}{4\pi G}\frac{\ddot m}{m},\nonumber\\
 \frac{\delta\mathsf T^{\rm eff}_{00}
              +\sum_{i=1}^3\delta\mathsf T^{\rm eff}_{ii}}{c^2}
 &=\frac{3}{4\pi G}
       \left(\frac{\ddot m}{m}-\frac{\dot m^{\,2}}{m^2}\right),\nonumber\\
 \left.\frac{\operatorname{Tr}(\delta T_{\alpha\alpha})}{c^2}
       \right|_{\text{Ref.~\cite{Fearn2016}, Eq.~(93)}}
 &=\frac{1}{4\pi G}
       \left(\frac{\ddot m}{m}-2\frac{\dot m^{\,2}}{m^2}\right).
 \label{eq:fearn-three-diagonals}
\end{align}
The printed expression is neither the covariant trace nor the positive
sum of all four correctly evaluated diagonal components. The latter also
requires three spatial terms. Raising one index supplies the Lorentzian
signs; raising both indices does not.
Equation~(93) of Ref.~\cite{Fearn2016} is therefore not the trace of the tensor equation in that reference and does
not derive the stated mass-density fluctuation. The corrected scalar in
Eq.~\eqref{eq:fearn-effective-trace} remains a contraction of geometric
derivative terms.

The error is tensorial. With $x^0=ct$, all diagonal entries may have the same
units, and their positive sum is still not the trace. With $x^0=t$, the
coordinate factors also enter. Coordinate invariance of $\Box m/m$
likewise does not imply Weyl invariance of a measured force.

\subsubsection{The scalar field in the gravitational coupling}

Equation~(90) of Ref.~\cite{Fearn2016} replaces the material coefficient $6/m^2$
by $8\pi G$ while retaining derivatives of $m$. To test this approximation,
set $c=1$ and write
\begin{equation}
 m=m_0+m_s,\qquad T_{ab}=\overline T_{ab}+\delta T_{ab},\qquad
 \frac{|m_s|}{m_0}\ll1,
\end{equation}
where $m_0$ is constant. Expanding the complete product gives
\begin{equation}
 \frac{6T_{ab}}{m^2}
 =\frac6{m_0^2}\overline T_{ab}
  +\frac6{m_0^2}\delta T_{ab}
  -\frac{12m_s}{m_0^3}\overline T_{ab}
  +O(\delta^2).
 \label{eq:fearn-coupling-product}
\end{equation}
Here $O(\delta^2)$ denotes products of first-order perturbations, with their
spatial and temporal scales held fixed in the amplitude expansion. If the
background material tensor vanishes and $\delta T_{ab}$ is first order, the
correction to the coefficient multiplying $\delta T_{ab}$ is second order.
Replacing that coefficient by $6/m_0^2$ while retaining terms linear in
$\nabla_a\nabla_b m_s$ is then consistent. In a material or cosmological background
with $\overline T_{ab}\ne0$, the last explicitly displayed term is first order
and must be retained unless a further ordering justifies its omission.

Thus a constant coefficient and first-order scalar derivatives are consistent about a vacuum background, but a nonzero background stress generally requires the additional first-order term.

The two cases and the required coupled expansion are summarized in Fig.~\ref{fig:fearn_green_contrast}.
\begin{figure}[tbp]
\centering
\begin{tikzpicture}[font=\footnotesize,node distance=5mm,
 box/.style={draw,rounded corners,align=center,text width=6.35cm,inner sep=5pt},
 arr/.style={-Latex,thick}]
\node[box] (a1) {\textbf{A. A substitution to be tested}\\[3pt]
 Replace $6/m^2$ by $6/m_0^2$\\while retaining derivatives of $m_s$.};
\node[box,below=of a1] (a2) {Which background and perturbative order?\\
 $m=m_0(1+s)$,\quad $T_{ab}=\overline T_{ab}+\delta T_{ab}$};
\node[box,below=of a2] (a3) {If $\overline T_{ab}\ne0$, the omitted term\\
 $-12s\overline T_{ab}/m_0^2$ is first order.\\Its omission requires an additional ordering.};
\node[box,below=9mm of a3] (a4) {If $\overline T_{ab}=0$, the correction\\
 $s\,\delta T_{ab}$ is second order.\\The frozen prefactor can then be consistent.};
\draw[arr] (a1)--(a2);\draw[arr] (a2)--(a3);
\node[font=\scriptsize] at ($(a3.south)!0.5!(a4.north)$) {alternative background};
\node[box,right=9mm of a1] (b1) {\textbf{B. A controlled coupled expansion}\\[3pt]
 Specify background, matter action,\\orders and boundary data.};
\node[box,below=of b1] (b2) {Expand the complete material term:\\[2pt]
 $\displaystyle\frac{6T_{ab}}{m^2}=\frac6{m_0^2}
 [\overline T_{ab}+\delta T_{ab}-2s\overline T_{ab}]$\\
 $\displaystyle{}+O(\delta^2)$.};
\node[box,below=of b2] (b3) {Retain the coupled metric response,\\including $m_0\delta R/6$ in the scalar\\equation for constant background $m_0$.};
\node[box,below=of b3] (b4) {Determine the sourced field with its\\Green operator and boundary data;\\test the calibrated apparatus observable.};
\draw[arr] (b1)--(b2);\draw[arr] (b2)--(b3);\draw[arr] (b3)--(b4);
\end{tikzpicture}
\caption{Perturbation ordering for the coupling replacement in Eq.~(90) of Ref.~\cite{Fearn2016}. Expanding $6T_{ab}/m^2$ gives different retained terms for zero and nonzero background stress--energy. A constant background coupling is permitted only at the order justified by that product. The metric and material perturbations enter the same expansion.}
\label{fig:fearn_green_contrast}
\end{figure}

\subsubsection{The flat metric and the coupled perturbation}

Equations~(91)--(92) of Ref.~\cite{Fearn2016} evaluate the derivative terms with
a flat metric. The scalar Hessian is
\begin{equation}
 \nabla_a\nabla_bm=\partial_a\partial_bm
                    -\Gamma^{c}{}_{ab}\partial_cm.
 \label{eq:fearn-second-derivative}
\end{equation}
On a flat background with constant $m_0$, the connection multiplied by
$\partial_cm_s$ is second order. Partial derivatives therefore give the correct
first-order Hessian of $m_s$. This fact does not permit the curvature perturbation
in $Rm/6$ to be omitted, since $m_0\delta R/6$ is first order.

Linearizing Eq.~\eqref{eq:scalar} about a background $\overline g_{ab},\overline m$ gives
\begin{equation}
 \left(\overline\Box+\frac{\overline R}{6}
                    -12\alpha\overline m^{\,2}\right)\delta m
 =\delta\mathcal J
  -\left(\delta\Box+\frac{\delta R}{6}\right)\overline m.
 \label{eq:fearn-coupled-linearization}
\end{equation}
For constant $\overline m=m_0$, $\delta\Box\,m_0=0$, but $m_0\delta R/6$ remains first order. Consequently replacing the Hessian by partial derivatives is a valid first-order operation, whereas suppressing the perturbed curvature is an additional restriction. Small scalar amplitude alone does not supply that restriction.

\subsubsection{The material source and the conformal transformation}
Equation~\eqref{eq:ward}, $T=m\mathcal J$, relates the scalar source to the trace obtained from the same material action. The discussion following Eq.~(93) of Ref.~\cite{Fearn2016} recognizes the importance of stress. A polarized and strained solid requires that stress, the polarization and the energy transfers in both source equations. Keeping the dust source $\mathcal J=\lambda N$ while independently assigning electrical and elastic contributions to $T_{ab}$ violates this relation unless their variations reproduce the same $\mathcal J$.

Equations~(39)--(40) of Ref.~\cite{Fearn2015I} already give $\Omega=m/m_0$ and $\widehat g_{ab}=\Omega^2g_{ab}$. The interpretation in Ref.~\cite{Fearn2015I} of this transformation as analogous to a rest frame misses its action on the material standards. A comoving coordinate transformation does not change a body's invariant internal energy. A Weyl transformation changes the metric and the representation of those standards together. The effect on the measured force follows from the transformed action in Sect.~\ref{sec:equivalence}.

Equation~(90) of Ref.~\cite{Fearn2016} omits an explicit cosmological-constant term, whereas the paragraph following Eq.~(51) of Ref.~\cite{Fearn2015Mach0} requires the cosmological constant and negative pressure for its assumed accelerating background. Fearn invokes that acceleration and a cosmic horizon~\cite[pp.~90,106]{Fearn2016} to evade Hawking's objection, but does not derive Eqs.~(91)--(93) of Ref.~\cite{Fearn2016} by matching the laboratory perturbation to this cosmology and its absorber data. The cosmological argument therefore does not justify the mass-fluctuation formula. Woodward~\cite[pp.~362--364]{Woodward2017JBIS} likewise invokes $\Phi=c^2$ and accelerated expansion in the advanced/retarded interpretation, without deriving the laboratory source from a matched cosmological solution. The global requirements are stated in Sect.~\ref{sec:absorber}; the distinction from an automatic inference of divergence is recorded in Appendix~\ref{app:prior-corrections}.

\subsection{Propagation of waves for weak gravitational fields}
\label{sec:local}
\label{sec:HN-weakfield-waves}
For the weak metric
\begin{equation}
 g_{ab}=\eta_{ab}+\psi_{ab},\qquad
 \psi=\eta^{ab}\psi_{ab},\qquad |\psi_{ab}|\ll1.
\end{equation}
With the HN curvature convention, the linearized Ricci scalar is
\begin{equation}
 R^{(1)}=\Boxeta\psi-\partial_a\partial_b\psi^{ab}.
 \label{eq:HN-linear-R}
\end{equation}
The harmonic coordinate condition is the vanishing divergence of the trace-reversed perturbation,
\begin{equation}
 \partial_a\left(\psi^{ab}-\frac12\eta^{ab}\psi\right)=0.
\end{equation}
Substitution gives $R^{(1)}=\tfrac12\Boxeta\psi$.

For a local expansion about flat space, set $\Lambda=0$ and assign $T$,
$\mathcal J$, $\psi_{ab}$ and $s=m_s/m_0$ first order. The background has
$T_0=\mathcal J_0=R_0=0$. Restoring SI mass-density units, the simultaneous
trace and scalar equations are
\begin{align}
 \frac12\Boxeta\psi
 &=\frac{8\pi G_0}{c^2}T-6\Boxeta s,
 \nonumber\\
 m_0\Boxeta s+\frac{m_0}{12}\Boxeta\psi
 &=\mathcal J,
 \qquad T=m_0\mathcal J.
 \label{eq:HN-flat-coupled-system}
\end{align}
Here $\mathcal J=-(c\sqrt{-g})^{-1}\delta S_{\rm mat}/\delta m$ is the scalar source in the SI mass-density normalization. The factor $1/2$ follows from the harmonic coordinate condition for $g$.
Eliminating the curvature, $\delta R=\tfrac12\Boxeta\psi$, gives
\begin{equation}
 \Boxeta\psi=\frac{16\pi G_0}{c^2}(T-T_{\rm field}),
 \qquad
 T_{\rm field}=\frac{3c^2}{4\pi G_0}\Boxeta s
              =m_0^2\Boxeta s.
 \label{eq:trace-normalization}
\end{equation}
Substitution into the second line of
Eq.~\eqref{eq:HN-flat-coupled-system} cancels $\Boxeta s$ and returns
$\mathcal J=T/m_0$. The equations do not determine an additional scalar
mode: their dependence is required by Weyl symmetry. A Weyl gauge and the
mapped matter and boundary data complete the description.

Equation~\eqref{eq:trace-normalization} is the metric trace equation, not an
independent Newtonian-potential equation or an independent material source.

The full weak metric must also be retained. In the Newtonian limit of Einstein gravity, both the time-time perturbation and the spatial metric perturbations are of order $U/c^2$. Thus the trace cannot in general be replaced by the time-time component alone. The conformal change of variables introduces an additional gauge issue: if $\Omega=1+s$, then
\begin{equation}
 \widehat\psi_{ab}=\psi_{ab}+2s\eta_{ab},\qquad
 \widehat\psi=\psi+8s.
\end{equation}
Harmonic coordinates for $g$ need not be harmonic coordinates for $\widehat g$. In the former coordinates $\widehat R^{(1)}=\tfrac12\Boxeta\psi+6\Boxeta s$; substituting $\tfrac12\Boxeta\widehat\psi$ without a further coordinate adjustment would give an erroneous result.

Figure~\ref{fig:weakfield-reduction} summarizes the corrected reduction.
\begin{figure}[tbp]
\centering
\begin{tikzpicture}[font=\footnotesize,node distance=5mm,
 box/.style={draw,rounded corners,align=center,text width=10.5cm,inner sep=5pt},
 arr/.style={-Latex,thick}]
\node[box] (start) {Covariant trace identity, $\Lambda=0$\\
 $\displaystyle R=\frac6{m^2}T-\frac6m\Box m$};
\node[box,below=of start] (wf) {Local expansion about vacuum Minkowski space\\
 $g_{ab}=\eta_{ab}+\psi_{ab}$,\quad $m=m_0+m_s$,\quad
 $T,\psi_{ab},m_s/m_0=O(\delta)$};
\node[box,below=of wf] (Rgen) {Linearized Ricci scalar in the stated HN convention\\
 $R^{(1)}=\Box_\eta\psi-\partial_a\partial_b\psi^{ab}$};
\node[box,below=of Rgen] (gauge) {Harmonic coordinate condition\\
 $\partial_a(\psi^{ab}-\tfrac12\eta^{ab}\psi)=0$
 \quad$\Longrightarrow$\quad $R^{(1)}=\tfrac12\Box_\eta\psi$};
\node[box,below=of gauge] (wave) {Trace-sector wave equation\\[2pt]
 $\displaystyle\Box_\eta\psi=\frac{16\pi G_0}{c^2}(T-T_{\rm field})$,
 \qquad$\displaystyle T_{\rm field}=\frac{3c^2}{4\pi G_0}\frac{\Box_\eta m_s}{m_0}$};
\node[box,below=of wave] (observable) {To obtain a force: solve the remaining metric and material equations\\and transform the measuring apparatus and its boundary data consistently.};
\draw[arr] (start)--(wf);\draw[arr] (wf)--(Rgen);\draw[arr] (Rgen)--(gauge);
\draw[arr] (gauge)--(wave);\draw[arr] (wave)--(observable);
\end{tikzpicture}
\caption{Weak-field reduction of the HN trace equation. Harmonic gauge gives $\partial_a\partial_b\psi^{ab}=\tfrac12\Box_\eta\psi$ and the coefficient $16\pi G_0/c^2$. The trace $\psi$ includes both temporal and spatial metric perturbations; $T_{\rm field}$ denotes a rearranged geometric term.}
\label{fig:weakfield-reduction}
\end{figure}

\subsubsection{Near-zone ordering and the spatial wave term}
For a response varying on a characteristic source scale $L$ at angular frequency $\omega$, the ratio of characteristic derivative terms is
\begin{equation}
 \frac{|c^{-2}\partial_t^2m_s|_{\rm char}}
 {|\nabla^2m_s|_{\rm char}}=O\!\left[\left(\frac{\omega L}{c}\right)^2\right].
 \label{eq:HN-nearzone}
\end{equation}
This ratio is small for a centimeter-scale resonator at tens of kilohertz.
It compares characteristic derivatives of a source-scale response, rather
than their values at every point; it supplies no scalar-amplitude bound.
In this near-zone ordering the spatial wave term cannot be discarded.

The distant-wave zone does not justify the opposite operation of dropping the spatial wave term. For a spherical outgoing solution outside its source,
\begin{equation}
 m_s(t,r)=\frac{f_{\rm rad}(t-r/c)}r,\qquad
 \nabla^2m_s=\frac{f_{\rm rad}''(t-r/c)}{c^2r}
 =c^{-2}\partial_t^2m_s,\qquad r>0.
 \label{eq:spherical-wave}
\end{equation}
Here $f_{\rm rad}$ is a twice differentiable radial-wave profile and primes differentiate its argument. Both terms are necessary and their difference vanishes. Increasing the distance reduces their amplitudes together. A field dominated by distant matter need not be an outgoing spherical wave; its spatial derivatives must be determined from the global solution and its boundary data. Distance alone establishes neither $\nabla^2m_s=0$ nor $\Box m_s\simeq c^{-2}\partial_t^2m_s$.

For a specified metric perturbation, the scalar equation in
Eq.~\eqref{eq:HN-flat-coupled-system} admits the decomposition
\begin{equation}
 m_s=m_{s,\mathrm p}+m_{s,\mathrm h},\qquad
 \Boxeta m_{s,\mathrm p}=\mathcal J-\frac{m_0}{6}\delta R,
 \qquad \Boxeta m_{s,\mathrm h}=0.
 \label{eq:HN-particular-homogeneous}
\end{equation}
Equation~\eqref{eq:HN-nearzone} orders the source-scale particular response,
not independent homogeneous data. A well-posed wave problem has $m_{s,\mathrm h}=0$ if its homogeneous Cauchy value and normal
derivative vanish and its boundary data supply no incoming field. A spatially
constant homogeneous solution in this flat reduction has
$\ddot m_{s,\mathrm h}=0$; a nearly uniform finite-frequency wave retains
$\nabla^2m_{s,\mathrm h}=c^{-2}\ddot m_{s,\mathrm h}$. A cosmological
homogeneous component requires the corresponding curved operator and
global data. The coupled continuum theory also permits $s\mapsto s+\zeta_W$ and $\psi_{ab}\mapsto\psi_{ab}-2\zeta_W\eta_{ab}$, where $\zeta_W$ is an infinitesimal dimensionless change of scalar representative. These simultaneous changes leave $\widehat\psi_{ab}=\psi_{ab}+2s\eta_{ab}$ invariant. Such changes do not
supply an additional measured force under Proposition~\ref{prop:equivalence}.

\addtocontents{toc}{\protect\newpage}
\subsection{Review and analysis of Woodward's space-propulsion concept}
\label{sec:woodward}

Equation~(2) of Ref.~\cite{Fearn2013}, Eq.~(14) of Ref.~\cite{Fearn2014}, and Eq.~(7) of Ref.~\cite{Fearn2015II} identify the leading proposed mass fluctuation as
\begin{equation}
 \delta\mathcal M_W(t)\simeq
 \frac{1}{4\pi G\rho_{\rm ref}c^2}\frac{dP}{dt},
 \label{eq:woodward-proposed-mass}
\end{equation}
where $\rho_{\rm ref}$ denotes the reference density written as $\rho_0$ in Refs.~\cite{Fearn2013,Fearn2014,Fearn2015II}, and $P$ is its assigned rate of energy change. The preceding GR and HN calculations show why a geometric derivative term cannot be interpreted as this additional material source. In particular, dividing such a term by $G$ changes its normalization, not its physical status.

Equations~(13)--(21) of Ref.~\cite{Fearn2015II} combine piezoelectric and electrostrictive responses at the second harmonic. For $\delta\mathcal M_W=\Delta\mathcal M_2\cos(2\omega t+\theta_M)$ and $a_2=a_{2,\rm pk}\cos(2\omega t+\theta_a)$,
\begin{equation}
 \left\langle\delta\mathcal M_W a_2\right\rangle
 =\frac12\Delta\mathcal M_2a_{2,\rm pk}\cos(\theta_M-\theta_a).
 \label{eq:woodward-correlation}
\end{equation}
This correlation supplies neither a mass-fluctuation law nor the momentum transferred to the apparatus.

\subsubsection{The background potential and the local gravitational field}

Equation~(8) of Ref.~\cite{Fearn2014} expands the device density (denoting its delay distance by $R_{\rm sep}$) at $t\pm R_{\rm sep}/c$. Equations~(9)--(12) of Ref.~\cite{Fearn2014} retain the second time derivative and introduce $2\pi G\rho_{\rm ref}R_{\rm sep}^2=c^2$, although $\rho_{\rm ref}$ is the device density and $R_{\rm sep}$ is subsequently assigned a cosmological scale. The local PZT density cannot be substituted for the separately defined cosmological density. The time expansion fails independently: for a harmonic density component, the exact symmetric average is
\begin{equation}
 \frac{\rho_1\cos[\omega(t+R_{\rm sep}/c)]+\rho_1\cos[\omega(t-R_{\rm sep}/c)]}{2}
 =\rho_1\cos\omega t\cos(\omega R_{\rm sep}/c).
 \label{eq:woodward-delay}
\end{equation}
Replacing the last factor by $1-(\omega R_{\rm sep}/c)^2/2$ requires $|\omega R_{\rm sep}/c|\ll1$. A kilohertz source and a cosmological $R_{\rm sep}$ do not satisfy this condition. Equal advanced and retarded weights eliminate odd Taylor terms, not the higher even terms. The retained second derivative therefore does not approximate the stated delayed source.

Woodward's reply~\cite{Woodward2020Reply} invokes Brans' spectator-matter argument to assign the cosmic potential a value proportional to $c^2$ and the inertial interaction zeroth order. Denote that cosmic normalization by $\Phi$ and the local perturbation by $\phi$. If $\Phi$ is constant throughout the laboratory spacetime region,
\begin{equation}
 \partial_t^2(\Phi+\phi)=\partial_t^2\phi,
 \qquad
 \Box_{\bar g}(\Phi+\phi)=\Box_{\bar g}\phi,
 \label{eq:woodward-background-split}
\end{equation}
where $\bar g$ is the specified background. Adding $\Phi$ does not replace the local metric coefficient multiplying a derivative in the relaxed Einstein equation. Such a replacement changes the equation and its perturbation order. The distinction made in Sect.~3.6 of Ref.~\cite{Rodal2019} therefore survives the proposed cosmic normalization. A calculation about a cosmological background must derive the local coefficient from that background and its perturbation; it cannot insert $\Phi/c^2\sim1$ into a selected term of the local weak-field equation.

\subsubsection{Variable rest mass and the momentum geodesic}

Equations~(13)--(15) of Ref.~\cite{Fearn2015I} assume a zero covariant momentum derivative; Eq.~(18) of Ref.~\cite{Fearn2015I} subsequently introduces a varying mass. With $p^a=M_{\rm rest}U^a$, $U^aU_a=c^2$ and signature $(+---)$, the starting equation gives
\begin{equation}
 p^a\nabla_ap^b=M_{\rm rest}\frac{\dd M_{\rm rest}}{\dd\tau} U^b+M_{\rm rest}^2A^b=0,
 \qquad
 U_b p^a\nabla_ap^b=M_{\rm rest}\frac{\dd M_{\rm rest}}{\dd\tau} c^2=0,
 \label{eq:woodward-geodesic-mass}
\end{equation}
Here $U^a=\dd x^a/\dd\tau$ and $A^a=U^b\nabla_bU^a$; all mass derivatives in these equations are with respect to proper time. For $M_{\rm rest}>0$, it requires $\frac{\dd M_{\rm rest}}{\dd\tau}=0$. The variable-mass result in Ref.~\cite{Fearn2015I} is incompatible with its starting equation. The acknowledgement in Sect.~3.2 of Ref.~\cite{Fearn2015I} that the procedure is not fully Lorentz invariant does not resolve this contraction.

An open body's monopole momentum instead satisfies
\begin{align}
 \frac{Dp^a}{d\tau}&=\mathcal Q^a,\qquad
 \frac{\dd M_{\rm rest}}{\dd\tau}=\frac{U_a\mathcal Q^a}{c^2},\nonumber\\
 M_{\rm rest} A^a&=\left(\delta^a{}_b-\frac{U^aU_b}{c^2}\right)\mathcal Q^b,
 \label{eq:woodward-mass-exchange}
\end{align}
Here $\mathcal Q^a$ is the rate of four-momentum transfer by matter, radiation and applied stresses. Its parallel component changes the rest energy and its transverse component accelerates the body. A geodesic trajectory can therefore coexist with mass loss: isotropic emission in the instantaneous rest frame has $A^a=0$ and $\mathcal Q^a=\frac{\dd M_{\rm rest}}{\dd\tau} U^a\ne0$. What fails in Ref.~\cite{Fearn2015I} is the simultaneous use of varying $M_{\rm rest}$ and zero covariant momentum derivative. The radiation loss and Poynting--Robertson interpretation invoked after Eq.~(18) of Ref.~\cite{Fearn2015I} require a nonzero transfer law in both the temporal and spatial equations.

The argument following Eq.~(3.11) of Woodward~\cite[pp.~77--78]{WoodwardStarship} sets the velocity-dependent mass-transfer term to zero in the instantaneous rest frame. Equation~\eqref{eq:woodward-mass-exchange} shows that this does not set the four-momentum transfer to zero or establish its proposed exchange with cosmic matter.

For a scalar-dependent particle mass, variation of
$-\int M_{\rm p}(x)c\,\dd s_g$, with $U^aU_a=c^2$ and signature $(+---)$, gives
\[
 \frac{D(M_{\rm p}U^a)}{\dd\tau}=c^2\nabla^aM_{\rm p},\qquad
 M_{\rm p}A^a=(c^2g^{ab}-U^aU^b)\nabla_bM_{\rm p}.
\]
For $M_{\rm p}=\lambda m$, division by $M_{\rm p}c^2$ gives
Eq.~\eqref{eq:geodesicmap}. This is again a nonzero momentum derivative.
The scalar interaction and its source must follow from the same action.
The product rule alone supplies neither this interaction nor the exhaust law of a rocket.

Consequently, Eqs.~(19)--(21) of Ref.~\cite{Fearn2015I} do not derive a new gravitational source from the preceding momentum equation. The HN derivation in Ref.~\cite{Fearn2015I} also fails at the $00$ component: the $m\ddot m$ term in Eq.~(43) of Ref.~\cite{Fearn2015I}, and in Eq.~(5) of Ref.~\cite{Fearn2015II}, is absent from the correctly evaluated tensor expression, Eqs.~\eqref{eq:fearn-00-cancellation}--\eqref{eq:fearn-normalized-00}.

\subsubsection{Density variation, compression and spatial derivatives}

The dust identity $\rho_{\rm rest}=\lambda mN$ gives Eq.~\eqref{eq:densityidentity}, including the number-density derivative and the product of gradients. The compression in Eq.~\eqref{eq:numberstrain} changes $N$ while the mass per particle can remain fixed. Thus a density oscillation in the stack is not a measurement of the HN mass field. Ref.~\cite{Fearn2014} itself distinguishes energy stored in material bonds from elementary particle mass; that distinction does not justify substituting the specimen's energy density for the compensator.

An observer-dependent translational energy is likewise not the HN scalar. Internal kinetic, elastic and electromagnetic energies can contribute to a composite body's invariant mass, but that fact supplies no identification of the compensator with the moving element's energy divided by $c^2$.

The proposed reduction $\Box m_s\simeq c^{-2}\ddot m_s$ is not established for the locally sourced or outgoing-wave responses. In the laboratory near zone, Eq.~\eqref{eq:HN-nearzone} gives the source-scale ordering of the spatial and temporal derivatives. Outside a spherical source, Eq.~\eqref{eq:spherical-wave} instead gives $\nabla^2m_s=c^{-2}\partial_t^2m_s$ for an outgoing wave. A decreasing amplitude with distance does not remove the spatial phase derivative. Neither limit justifies the temporal-only expression. Independent homogeneous data must instead satisfy the conditions discussed after Eq.~\eqref{eq:HN-particular-homogeneous}. Figure~\ref{fig:woodward-flow} displays the two cases.
\begin{figure}[tbp]
\centering
\begin{tikzpicture}[font=\footnotesize,node distance=5mm,
 wide/.style={draw,rounded corners,align=center,text width=13.3cm,inner sep=5pt},
 half/.style={draw,rounded corners,align=center,text width=6.2cm,inner sep=5pt},
 arr/.style={-Latex,thick}]
\node[wide] (start) {The scalar contribution contains the full wave operator\\
 $\Box_\eta m_s=c^{-2}\partial_t^2m_s-\nabla^2m_s$.};
\node[half] (near) at (-3.5,-2.6)
 {\textbf{Local source-scale expansion}\\[2pt]
 $\displaystyle\frac{|c^{-2}\partial_t^2m_s|_{\rm char}}
 {|\nabla^2m_s|_{\rm char}}=O[(\omega L/c)^2]$\\[3pt]
 $\omega L/c\ll1$ for the laboratory device.\\This is characteristic ordering,\\not a pointwise inequality.};
\node[half] (far) at (3.5,-2.6)
 {\textbf{Outgoing auxiliary wave, $r>0$}\\[2pt]
 $\displaystyle m_s=\frac{f_{\rm rad}(t-r/c)}r$\\[3pt]
 $\nabla^2m_s=c^{-2}\partial_t^2m_s$\\Both terms remain; their difference\\vanishes outside the source.};
\coordinate (mid) at ($(near.south)!0.5!(far.south)$);
\node[wide,below=7mm of mid] (test)
 {Neither locality nor distance supplies the assumption $\Box_\eta m_s\simeq c^{-2}\partial_t^2m_s$.\\
 A time-only profile or another justified hierarchy must follow from the specified solution.};
\node[wide,below=of test] (notation)
 {Keep the cosmic normalization $\Phi$ distinct from the local perturbation $\phi$.\\
 A formal $\ddot m_s$ term is not a thrust prediction without the full source and momentum balance.};
\draw[arr] (start.south)--++(0,-.25)-|(near.north);
\draw[arr] (start.south)--++(0,-.25)-|(far.north);
\draw[arr] (near.south)--++(0,-.32)-|(test.north);
\draw[arr] (far.south)--++(0,-.32)-|(test.north);
\draw[arr] (test)--(notation);
\end{tikzpicture}
\caption{The spatial and temporal terms of the wave operator. The left branch gives the local source-scale ordering; the right gives an exact outgoing flat-space profile for which the two terms cancel. Neither calculation permits removal of the spatial derivatives merely because distant matter contributes to the background.}
\label{fig:woodward-flow}
\end{figure}

\subsubsection{Differentiation of mechanical power}

Equation~(7) of Ref.~\cite{Fearn2015II} and Eq.~(30) of Woodward's patent~\cite{WoodwardPatent2016} omit the second term in
\begin{equation}
 P=\mathbf F\cdot\mathbf v,
 \qquad
 \dot P=\mathbf F\cdot\mathbf a+\dot{\mathbf F}\cdot\mathbf v.
 \label{eq:woodward-power-rule}
\end{equation}
Equation~(15) of Ref.~\cite{Fearn2014} retains it. Woodward~\cite[Eqs.~(5.8)--(5.9), pp.~174--175]{WoodwardStarship} writes the full derivative and then discards $\mathbf v\cdot\dot{\mathbf F}$ as higher order. Similarly, Eq.~(19) of Ref.~\cite{Fearn2015I} neglects $\mathbf v\cdot\dot{\mathbf a}$. With constant modal mass $M_{\rm mod}$, $F=M_{\rm mod}\ddot x$ and $x=A\cos\omega t$, direct differentiation gives
\begin{equation}
 \dot P=M_{\rm mod}A^2\omega^4\cos2\omega t,
 \qquad
 Fa=\frac12M_{\rm mod}A^2\omega^4(1+\cos2\omega t).
 \label{eq:woodward-power-harmonics}
\end{equation}
The omitted term has the same order in $A$ and $\omega$: it cancels the constant part and doubles the second-harmonic coefficient of $Fa$ in $\dot P$, so $\langle\dot P\rangle=0$. Small $v/c$ does not justify its removal. For this single harmonic, both $\langle\dot P a\rangle$ and $\langle(Fa)a\rangle$ nevertheless vanish; the error in the power derivative is not itself a nonzero mean thrust.

Equation~\eqref{eq:woodward-power-harmonics} tests the stated constant-modal-mass approximation. The coupled electrical, elastic and dissipative power balance is given in Sect.~\ref{sec:laboratory}; an open body's momentum balance follows below.

\subsubsection{Open-system momentum balance and the variable-mass term}
\label{sec:open-momentum}

For a symmetric stress--energy tensor, the open-system balance starts from
\begin{equation}
 \widehat\nabla_a\widehat{\mathsf T}^{ab}
 =\widehat{\mathsf f}_{\rm ext}^{\,b},\qquad
 \widehat\nabla_a\big(\widehat{\mathsf T}^{ab}\chi_{{\rm bal}\,b}\big)
 =\widehat{\mathsf f}_{\rm ext}^{\,b}\chi_{{\rm bal}\,b}
  +\frac12\widehat{\mathsf T}^{ab}
          \mathcal L_{\chi_{\rm bal}}\widehat g_{ab}.
 \label{eq:covariant-momentum-current}
\end{equation}
Here $\widehat{\mathsf T}^{ab}$ is in energy-density units and contains all
retained material, electromagnetic and reservoir contributions.
The force density $\widehat{\mathsf f}_{\rm ext}^{\,b}$ describes interactions
with fields or matter omitted from that tensor; it is zero for a complete
closed material--electromagnetic system. In GR the hats merely identify the
physical metric and tensor. In the continuum HN sector they denote the
invariant variables of Proposition~\ref{prop:equivalence}. An untransformed
material tensor need not have zero covariant divergence: its exchange with
the scalar must then be included. Gravitational momentum is not an extra
local material tensor. The coordinate-dependent total gravitational balance
is instead the Landau--Lifshitz identity in Sect.~\ref{sec:GR}.

Let $\mathcal W_{\rm cv}$ have spacelike caps $\Sigma_1,\Sigma_2$ and a
timelike side $\mathcal B_{\rm cv}$. The smooth vector
$\chi_{\rm bal}^a$, with
$\chi_{{\rm bal}\,a}=\widehat g_{ab}\chi_{\rm bal}^b$, specifies the
laboratory's momentum component throughout the region. Define
\begin{equation}
 \mathfrak P[\chi_{\rm bal};\Sigma]
 =\frac1c\int_\Sigma
   \widehat{\mathsf T}^{ab}\chi_{{\rm bal}\,b}\,
   \dd\widehat\Sigma_a.
 \label{eq:covariant-momentum-functional}
\end{equation}
The cap normal in this definition is future directed. On the closed
boundary the past cap has the opposite orientation; on the timelike side,
$\dd\widehat\Sigma_a=-\widehat s_a\,\dd\widehat\Sigma_3$ for an outward unit
spacelike normal $\widehat s^a\widehat s_a=-1$, with $\dd\widehat\Sigma_3$ the positive proper three-volume on the side. The divergence theorem gives
\begin{align}
 \Delta\mathfrak P[\chi_{\rm bal}]
 ={}&-\frac1c\int_{\mathcal B_{\rm cv}}
       \widehat{\mathsf T}^{ab}\chi_{{\rm bal}\,b}\,
       \dd\widehat\Sigma_a\nonumber\\
 &+\frac1c\int_{\mathcal W_{\rm cv}}
 \left(\widehat{\mathsf f}_{\rm ext}^{\,b}\chi_{{\rm bal}\,b}
       +\frac12\widehat{\mathsf T}^{ab}
          \mathcal L_{\chi_{\rm bal}}\widehat g_{ab}\right)
       \dd\widehat V_4.
 \label{eq:covariant-worldtube-balance}
\end{align}
This is a scalar momentum functional, not a sum of vectors at different
spacetime points~\cite{Harte2012}. The metric term vanishes for a Killing
field; no such symmetry is assumed for the laboratory. Curvature and
acceleration or rotation of the selected observers remain in this term.
No nonrelativistic approximation has been made.

Split the side into the apparatus--support interface and the remaining
boundary. Denote the corresponding inward support impulse by
$\mathcal I_{\rm S\to A}[\chi_{\rm bal}]$, the remaining outward impulse
by $\mathcal I_{\rm out}[\chi_{\rm bal}]$, and the volume integral in
Eq.~\eqref{eq:covariant-worldtube-balance} by
$\mathcal I_{\rm bulk}[\chi_{\rm bal}]$. Then
\begin{equation}
 \mathcal I_{\rm read}[\chi_{\rm bal}]
 =-\mathcal I_{\rm S\to A}[\chi_{\rm bal}]
 =\mathcal I_{\rm bulk}[\chi_{\rm bal}]
  -\mathcal I_{\rm out}[\chi_{\rm bal}]
  -\Delta\mathfrak P_{\rm A}[\chi_{\rm bal}].
 \label{eq:covariant-support-impulse}
\end{equation}
Using the same $\chi_{\rm bal}^a$ on both sides, the interface integrals
cancel when the apparatus and support regions are added. Any stored
interface momentum is assigned once. Division by the elapsed proper time
of the specified laboratory clock gives the corresponding mean projected
force. The balance therefore retains radiation, cables, stresses,
external fields and changing stored momentum before any velocity limit.

For $x^0=ct$, choose $\chi_{{\rm bal}\,b}=\delta^i{}_b$ for one coordinate
momentum component. Equation~\eqref{eq:covariant-momentum-current} gives
the exact coordinate-density balance
\begin{align}
 \partial_t\left(\frac{\sqrt{-\widehat g}}c
                    \widehat{\mathsf T}^{0i}\right)
 +\partial_j\left(\sqrt{-\widehat g}\,
                    \widehat{\mathsf T}^{ji}\right)
 &=\sqrt{-\widehat g}\left(
       \widehat{\mathsf f}_{\rm ext}^{\,i}
       -\widehat\Gamma^i{}_{ab}\widehat{\mathsf T}^{ab}\right),
 \label{eq:covariant-coordinate-balance}\\
 \mathfrak p^i&=\frac{\sqrt{-\widehat g}}c
                   \widehat{\mathsf T}^{0i},\qquad
 \mathsf J_p^{ji}=\sqrt{-\widehat g}\,
                   \widehat{\mathsf T}^{ji}.
 \label{eq:covariant-laboratory-densities}
\end{align}
On a moving boundary its outward flux is
$(\mathsf J_p^{ji}-w^j\mathfrak p^i)n_j\,\dd A$, where $n_j\,\dd A$ is the outward coordinate area covector. In Minkowski coordinates,
this follows directly from
$\dd\Sigma_a=(-\mathbf w\cdot\mathbf n/c,n_j)c\,\dd t\,\dd A$.
Thus the transport term is already part of the relativistic side integral.

Only for the following Newtonian material specialization take
$|\mathbf v|/c\ll1$, weak gravity and small internal energy and stress
relative to $\rho c^2$. Then
$\widehat{\mathsf T}^{0i}/c=\rho v^i+O(c^{-2})$ and
$\widehat{\mathsf T}^{ji}=\rho v^jv^i+\Pi^{ji}+O(c^{-2})$, with
$\Pi^{ji}$ the positive-pressure stress. With the surface stress moved
to the applied force and separately conserved material mass, the result is
Eq.~\eqref{eq:open-control-volume}. Electromagnetic momentum, when retained,
must still use Eq.~\eqref{eq:covariant-coordinate-balance}; it is not
replaced by a mass current.

The force identification in Eq.~\eqref{eq:woodward-correlation} requires a momentum balance. Neither $\mathbf F=M_{\rm cv}\dot{\mathbf V}$ nor $\mathbf F=\dd(M_{\rm cv}\mathbf V)/\dd t$ is a universal equation for an open system~\cite{Tiersten1969,Sommerfeld1952,GantmacherLevin1950}. In that Newtonian specialization, let $\mathcal V(t)$ contain matter with locally conserved mass, $\mathbf v$ be its inertial velocity, $\mathbf w$ the boundary velocity and $\mathbf n$ the outward normal. The transport theorem gives
\begin{align}
 \mathbf P_{\mathcal V}&=\int_{\mathcal V(t)}\rho\mathbf v\,\dd^3x,
 \qquad M_{\rm cv}=\int_{\mathcal V(t)}\rho\,\dd^3x,\nonumber\\
 \dot M_{\rm cv}&=-\oint_{\partial\mathcal V}\rho(\mathbf v-\mathbf w)\cdot\mathbf n\,\dd A,\nonumber\\
 \frac{\dd P_{\mathcal V}^i}{\dd t}
 &=F_{\rm body}^i-\Phi_p^i-\Phi_\Pi^i
  =F_{\rm ext}^i-\Phi_p^i,\nonumber\\
 \Phi_p^i&=\oint_{\partial\mathcal V}
       \rho v^i(v^j-w^j)n_j\,\dd A,\qquad
 \Phi_\Pi^i=\oint_{\partial\mathcal V}\Pi^{ji}n_j\,\dd A,\nonumber\\
 F_{\rm ext}^i&=F_{\rm body}^i-\Phi_\Pi^i.
 \label{eq:open-control-volume}
\end{align}
Here $F_{\rm body}^i$ contains body forces. The tensile-positive Cauchy
stress is $\sigma^{ij}=-\Pi^{ji}$, so the surface traction contributes
$-\Phi_\Pi^i=\oint\sigma^{ij}n_j\,\dd A$ to $F_{\rm ext}^i$.
Moving this term between the flux and the applied force does not remove it.
On a material boundary, $(\mathbf v-\mathbf w)\cdot\mathbf n=0$
sets the advective flux to zero, but leaves the stress flux unchanged.
Elastic-wave transfer therefore remains present without mass escaping.
The transported momentum uses $\mathbf v$; the crossing rate uses
$\mathbf v-\mathbf w$. Stress at an exhaust cut is included once, through
this traction or an equivalent exhaust convention.

For a housing marker of velocity $\mathbf V$, define
\begin{equation}
 \mathbf P_{\mathcal V}=M_{\rm cv}\mathbf V+\mathbf p_{\rm rel},
 \qquad
 \mathbf p_{\rm rel}=\int_{\mathcal V}\rho(\mathbf v-\mathbf V)\,\dd^3x.
 \label{eq:open-internal-momentum}
\end{equation}
The internal momentum $\mathbf p_{\rm rel}$ need not vanish. For streams of inertial velocity $\mathbf v_i$, with $\dot M_{{\rm tr},i}$ positive for inflow and negative for outflow, Eq.~\eqref{eq:open-control-volume} becomes
\begin{align}
 \frac{\dd}{\dd t}(M_{\rm cv}\mathbf V+\mathbf p_{\rm rel})
 &=\mathbf F_{\rm ext}+\sum_i\dot M_{{\rm tr},i}\mathbf v_i,\nonumber\\
 M_{\rm cv}\dot{\mathbf V}
 &=\mathbf F_{\rm ext}
   +\sum_i\dot M_{{\rm tr},i}(\mathbf v_i-\mathbf V)-\dot{\mathbf p}_{\rm rel}.
 \label{eq:open-stream-balance}
\end{align}
For an ideal one-dimensional rocket, write $v_h$ for the axial housing velocity. With negligible $p_{\rm rel}$, $\dot M_{\rm cv}<0$ and exhaust velocity $v-v_h=-v_e$, the balance reduces to
\begin{equation}
 M_{\rm cv}\dot v_h=F_{\rm ext}-v_e\dot M_{\rm cv}.
 \label{eq:rocket-limit}
\end{equation}
A zero applied force thus permits acceleration through exhaust momentum. The same product rule without the flux term would give the wrong rocket equation. Related variable-mass models treat orbital motion and draining fluids~\cite{Berkovic1981,McDonaldLeakyTank,OttoMcDonald2018}.

For assigned Newtonian masses $m_i(t)$, positions $\mathbf x_i$, $M_{\rm sum}=\sum_i m_i$ and $\mathbf X=M_{\rm sum}^{-1}\sum_i m_i\mathbf x_i$, differentiation gives
\begin{equation}
 \sum_i m_i\dot{\mathbf x}_i
 =M_{\rm sum}\dot{\mathbf X}-\sum_i\dot m_i(\mathbf x_i-\mathbf X).
 \label{eq:rodal-centroid-momentum}
\end{equation}
Thus $M_{\rm sum}\ddot{\mathbf X}$ alone is not the force law for those selected degrees of freedom. A prescribed variation of these weights does not specify the mass and momentum transfer.

Equations~(1)--(2) and~(17) of Wanser~\cite{Wanser2013} give
\begin{equation}
 \sum_{i=1}^2m_i\ddot x_i=0,\qquad
 \dot p_{\rm pair}=\sum_{i=1}^2\dot m_i\dot x_i,\qquad
 p_{\rm pair}=\sum_{i=1}^2m_i\dot x_i.
 \label{eq:wanser-transfer}
\end{equation}
These Newtonian equations permit comoving mass transfer: zero relative transfer velocity removes the rocket term, not the transported inertial-frame momentum. The two particles are then an open subsystem. Wanser~\cite[pp.~126,128]{Wanser2013} acknowledges the required outside agent, but does not derive its proposed coupling to cosmic matter. Cyclic mass variation does not imply zero momentum transfer.

Equation~(14) of Ref.~\cite{Wanser2013}, with Eq.~\eqref{eq:wanser-transfer}, is exactly
\begin{equation}
 M_{\rm sum}\ddot X
 =2\sum_{i=1}^2\dot m_i(\dot x_i-\dot X)
  +\sum_{i=1}^2\ddot m_i(x_i-X).
 \label{eq:wanser-centroid}
\end{equation}
For the harmonic drive and small mass variations in Eqs.~(22)--(23) of Ref.~\cite{Wanser2013}, take the off-resonant periodic particular response at zeroth order, without free oscillations, and choose its centre of mass at rest at the origin. Let $m_i=m_{0i}+\delta m_i\cos(\omega t+\phi_i)$, $M_{\rm sum,0}=m_{01}+m_{02}$, $\omega_0^2=k(1/m_{01}+1/m_{02})$ and $J_W=\langle\sum_i\dot m_i\dot x_i^{(0)}\rangle$. Periodicity gives $\langle\sum_i\ddot m_i x_i^{(0)}\rangle=-J_W$. Hence, to first order in the fractional mass variations,
\begin{align}
 M_{\rm sum,0}\langle\ddot X\rangle
 &=2J_W-J_W=J_W=\langle\dot p_{\rm pair}\rangle,\nonumber\\
 \langle\ddot X\rangle
 &=\frac{\omega^2F_0}{2M_{\rm sum,0}(\omega_0^2-\omega^2)}
   \left(\frac{\delta m_1}{m_{01}}\cos\phi_1
              -\frac{\delta m_2}{m_{02}}\cos\phi_2\right).
 \label{eq:wanser-corrected-average}
\end{align}
Here $k$ is the spring stiffness and $F_0\cos\omega t$ the internal drive in Ref.~\cite{Wanser2013}. Equation~(21) of Ref.~\cite{Wanser2013} is twice this acceleration. Equation~(26) of Ref.~\cite{Wanser2013} agrees with $J_W$, so the factor $1/2$ in Eq.~(27) of Ref.~\cite{Wanser2013} is incorrect at this order. The Newtonian limit is not itself the objection. The prescribed mass variations and omitted agent do not supply a relativistic gravitational source or an HN momentum-transfer law.

If the stored momentum in Eq.~\eqref{eq:open-control-volume} is periodic, averaging gives
\begin{equation}
 \langle\mathbf F_{\rm ext}\rangle=\langle\boldsymbol\Phi_p\rangle.
 \label{eq:open-mean-force}
\end{equation}
It does not give zero mean force when the mean momentum flux is nonzero. Moreover, periodic internal driving does not establish periodicity of the overall translational momentum. Assuming that periodicity to exclude propulsion would be circular. A cancellation between $\langle M_{\rm cv}\dot{\mathbf V}\rangle$ and $\langle\dot M_{\rm cv}\mathbf V\rangle$ is therefore not a substitute for the transfer law.

For an electrically driven solid, $M_{\rm rest}=E_{\rm rest}/c^2$ describes rest-energy exchange, not necessarily material crossing of the boundary. Assigning an arbitrary exhaust velocity to $\dot E_{\rm rest}/c^2$ would misapply Eq.~\eqref{eq:open-stream-balance}. The momentum transferred by the electromagnetic field, elastic stresses, supply and supports follows from their stress--energy balance. The proposed $\langle\delta\mathcal M_W a\rangle$ must satisfy that balance; the correlation alone is not a derivation of thrust. Equation~\eqref{eq:apparatus-measured-mean} gives the measured mean force on the support: the mean external force minus the mean outward momentum flux and the change in stored apparatus momentum per period.

\paragraph{The force transmitted to the support.}
With the coordinate densities in Eq.~\eqref{eq:covariant-laboratory-densities}, the local balance is
\begin{equation}
 \partial_t\mathfrak p^i+\partial_j\mathsf J_p^{ji}=f_{\rm ext}^i,
 \qquad
 P_{\rm A}^i=\int_{\mathcal V_{\rm A}(t)}\mathfrak p^i\,\dd^3x.
 \label{eq:apparatus-local-momentum}
\end{equation}
Here $f_{\rm ext}^i$ is the right-hand side of Eq.~\eqref{eq:covariant-coordinate-balance}, including its geometric term for the chosen laboratory chart. The densities are per coordinate volume. In the Newtonian limit the geometric term includes the ordinary gravitational body force. An interaction is not counted again if its transfer is already represented by the boundary flux. For laboratory control volumes containing the apparatus and the supports, respectively, integration gives
\begin{align}
 \dot{\mathbf P}_{\rm A}
 &=\mathbf R_{\rm AS}+\mathbf F_{\rm A}^{\rm ext}-\boldsymbol\Phi_{\rm A},\nonumber\\
 \dot{\mathbf P}_{\rm S}
 &=\mathbf R_{\rm lab}-\mathbf R_{\rm AS}
   +\mathbf F_{\rm S}^{\rm ext}-\boldsymbol\Phi_{\rm S},\nonumber\\
 \frac{\dd}{\dd t}(\mathbf P_{\rm A}+\mathbf P_{\rm S})
 &=\mathbf R_{\rm lab}+\mathbf F_{\rm A}^{\rm ext}
   +\mathbf F_{\rm S}^{\rm ext}
   -\boldsymbol\Phi_{\rm A}-\boldsymbol\Phi_{\rm S}.
 \label{eq:apparatus-support-balance}
\end{align}
The force $\mathbf R_{\rm AS}$ is transmitted by the support to the apparatus, and $\mathbf R_{\rm lab}$ by the surrounding laboratory to the supports. The fluxes $\boldsymbol\Phi_{\rm A,S}$ cross the remaining boundaries, including electrical connections, escaping radiation and any material transport. The common-interface transfer is included in $\mathbf R_{\rm AS}$ once; any stored interface momentum is assigned to one side. On a boundary moving with velocity $\mathbf w$, the flux integrand is $(\mathsf J_p^{ji}-w^j\mathfrak p^i)n_j$. The mechanical contact carries no matter across the apparatus--support interface. This includes the transport term in Eq.~\eqref{eq:open-control-volume}.

For a material contact surface $\Gamma_{\rm AS}(t)$ with outward apparatus
normal $n_j$, the mechanical reaction is
\begin{equation}
 R_{\rm AS}^{i}(t)=\int_{\Gamma_{\rm AS}(t)}\sigma^{ij}n_j\,\dd A
       =-\int_{\Gamma_{\rm AS}(t)}\Pi^{ji}n_j\,\dd A.
 \label{eq:elastic-support-traction}
\end{equation}
The remaining stress flux belongs to $\boldsymbol\Phi_{\rm A}$, not to
$\mathbf R_{\rm AS}$ a second time. A bonded interface without its own surface
inertia has continuous velocity and opposite tractions on its two sides.
The support displacement and traction must satisfy the support's dynamic
boundary problem. For a single collocated support port with no independent support excitation,
its linear impedance $Z_s(\omega_n)$ gives
\begin{equation}
 \widetilde R_{{\rm AS},n}=-Z_s(\omega_n)\widetilde v_{\Gamma,n},\qquad
 \langle P_{{\rm A}\to{\rm S},n}\rangle
 =-\frac12\operatorname{Re}(\widetilde R_{{\rm AS},n}\widetilde v_{\Gamma,n}^{\,*})
 =\frac12\operatorname{Re}Z_s(\omega_n)|\widetilde v_{\Gamma,n}|^2,
 \label{eq:support-port-power}
\end{equation}
where $\omega_n=n\omega$ and $v_\Gamma$ is the port velocity.
For the spring--damper example below, $Z_s=b_s+k_s/(i\omega_n)$ and
$\widetilde v_{\Gamma,n}=i\omega_n\widetilde X_{h,n}$.
No particular impedance is inferred for the measured apparatus.
Equation~\eqref{eq:elastic-support-traction} uses the current surface,
normal and full stress; a second-order mean cannot be inferred by retaining
only their first-order perturbations.

Define the mechanical force on the support by $\mathbf F_{\rm read}=-\mathbf R_{\rm AS}$. Integration over one drive period $\mathcal P$, without assuming periodic total momentum, yields
\begin{equation}
 \langle\mathbf F_{\rm read}\rangle
 =\langle\mathbf F_{\rm A}^{\rm ext}\rangle
  -\langle\boldsymbol\Phi_{\rm A}\rangle
  -\frac{\mathbf P_{\rm A}(t+\mathcal P)-\mathbf P_{\rm A}(t)}{\mathcal P}.
 \label{eq:apparatus-measured-mean}
\end{equation}
This is the laboratory-coordinate form of Eq.~\eqref{eq:covariant-support-impulse}; physical calibration specifies the clock and projection.
A calibrated balance must also include its mechanical response. For example, a one-dimensional linear mount with displacement $X_h$ relative to its fixed base, stiffness $k_s$ and damping $b_s$ has $R_{\rm AS}=-k_sX_h-b_s\dot X_h$, and hence
\begin{equation}
 k_s\langle X_h\rangle+b_s\frac{X_h(t+\mathcal P)-X_h(t)}{\mathcal P}
 =\langle F_{\rm A}^{\rm ext}-\Phi_{\rm A}\rangle
  -\frac{P_{\rm A}(t+\mathcal P)-P_{\rm A}(t)}{\mathcal P}.
 \label{eq:soft-mount-mean}
\end{equation}
Only a bounded periodic steady state removes both difference terms. For a free apparatus $R_{\rm AS}=0$; its translational momentum must instead be solved from the first line of Eq.~\eqref{eq:apparatus-support-balance}. A prescribed internal harmonic is permissible when it is measured or obtained from the coupled apparatus--support equations, or in a stated clamped approximation. Its sustaining reaction cannot then be discarded. Equation~\eqref{eq:woodward-correlation} evaluates a product of specified harmonics; it does not replace Eq.~\eqref{eq:apparatus-measured-mean}.

\subsection{Constitutive and energetic requirements for laboratory tests}
\label{sec:laboratory}

The displacement and power formulas in Refs.~\cite{Fearn2013,Fearn2014,Fearn2015I,Fearn2015II} contain errors independent of the gravitational derivation. The reported devices and drive conditions differ~\cite{March2021}; their voltages, frequencies and measured displacements cannot be combined as one experiment.

\subsubsection{Electrical energy and mechanical work}
\label{sec:electrical-energy}

Reference~\cite{Fearn2015I} identifies $V_0^2/X_C$ as electrical power while specifying a peak voltage and a capacitive load. For its assumed ideal capacitor,
\begin{equation}
 U_C=\frac{C_{\rm el}V_0^2}{4}(1+\cos2\omega t),\qquad
 p_C=\dot U_C=-\frac{\omega C_{\rm el}V_0^2}{2}\sin2\omega t,
 \qquad \langle p_C\rangle=0.
 \label{eq:capacitor-power}
\end{equation}
The reactive-power magnitude is $V_0^2/(2X_C)$, not $V_0^2/X_C$; neither is the real power consumed. For measured periodic voltage and current, with a DC contribution added if present,
\begin{equation}
 \overline P=\frac1{\mathcal P}\int_0^{\mathcal P}V I_{\rm el}\,\dd t
 =\frac12\sum_{n\geq1}\operatorname{Re}
       (\widetilde V_n\widetilde I_{{\rm el},n}^{\,*}).
 \label{eq:real-power}
\end{equation}
Here $\mathcal P=2\pi/\omega$ and the tildes denote peak complex amplitudes with time dependence $e^{in\omega t}$. The terminal current includes dielectric loss and the motional current. Electrical input, elastic storage and kinetic-energy change are different quantities.

\subsubsection{Displacement coefficients and measured motion}
\label{sec:displacement-audit}

Equation~(5) of Ref.~\cite{Fearn2013} uses
\begin{equation}
 x=L(1+K_p^{(13)}V+K_e^{(13)}V^2),\qquad
 [K_p^{(13)}]=\mathrm{V}^{-1},\quad
 [K_e^{(13)}]=\mathrm{V}^{-2}.
 \label{eq:old-displacement}
\end{equation}
Its numerical substitutions instead use coefficients in metres per volt and metres per volt squared. The resulting force estimate is dimensionally inconsistent. Equations~(5.12) and~(5.19)--(5.20) of Woodward~\cite[pp.~176--177]{WoodwardStarship} make the same substitution: the fractional-displacement coefficient is replaced by the piezoelectric coefficient in metres per volt, without a length normalization. Equation~(15) of Ref.~\cite{Fearn2015II}, also used in Ref.~\cite{Fearn2014}, has the alternative form
\begin{equation}
 x=x_0+N_df_p(K_p^{(15)}V+K_e^{(15)}V^2),\qquad
 \frac{K_e^{(15)}}{K_p^{(15)}}=\frac{2A_2}{V_0A_1},
 \label{eq:new-displacement}
\end{equation}
where $N_d$ is the disc count, $f_p$ a dimensionless preload factor, and $A_1,A_2$ the signed cosine amplitudes for $V=V_0\cos\omega t$ in this instantaneous real-coefficient law. This form allows coefficients in metres per volt and metres per volt squared, but Ref.~\cite{Fearn2015II} substitutes the length $0.020$ m for $N_df_p$. Its further identification $K_e^{(15)}=0.22K_p^{(15)}$ from a harmonic ratio omits the required inverse-voltage factor. An electrical harmonic ratio also requires the mechanical transfer functions before it can determine $A_2/A_1$.

The additional factor $4\pi$ in the numerical estimates of Refs.~\cite{Fearn2013,Fearn2015II}, with the ordinary SI value of $G$ retained, is not a conversion between Gaussian and SI units. The end-cap excursion in the breadboard-mounted optical test of Ref.~\cite{Fearn2014} does not determine the second-harmonic displacement, phase or modal mass on the thrust balance. A dimensional repair supplies neither those data nor the rejected gravitational source in Eq.~\eqref{eq:woodward-proposed-mass}.

\subsubsection{Damped electroelastic response and quality factor}
\label{sec:damping-quality}

Equation~(25) of Ref.~\cite{Fearn2015II} includes the denominator $\omega_0^2-\omega^2+i\gamma_d\omega$. With $\gamma_d=b_{\rm mod}/M_{\rm mod}$, the isolated viscous mode obeys
\begin{align}
 M_{\rm mod}\ddot q+b_{\rm mod}\dot q+k_{\rm mod}q&=f_{\rm mod},
 \qquad\omega_0^2=k_{\rm mod}/M_{\rm mod},
 \quad Q_m=M_{\rm mod}\omega_0/b_{\rm mod},
 \label{eq:damped-modal-equation}\\
 \widetilde q_n&=\frac{\widetilde f_{{\rm mod},n}/k_{\rm mod}}
 {1-r_n^2+ir_n/Q_m},
 \qquad r_n=n\omega/\omega_0,
 \quad\widetilde q(\omega_0)=-iQ_m\widetilde f_{\rm mod}/k_{\rm mod}.
 \label{eq:quality-transfer}
\end{align}
Thus damping fixes both amplitude and phase. Reference~\cite{Fearn2015II} discusses damping and quality factor, but its force estimate and numerical substitution, Eqs.~(21) and~(27), do not consistently carry the complex responses at both harmonics into the force correlation. The viscoelastic calculation in Sect.~4 of Ref.~\cite{Rodal2016Electroelastic} retains these responses. The assembly quality factor depends on preload, electrical termination and restraints; it cannot simply be replaced by the ceramic supplier's value.

A heterogeneous stack need not have proportional damping or real normal modes~\cite{TisseurMeerbergen2001}. With real spatial interpolation and complex coordinate amplitudes, a reciprocal linear reduction gives, where the inverse exists,
\begin{align}
 \widetilde{\mathbf q}_n
 &=\big[\mathbf K_{\rm ve}(\omega_n)-\omega_n^2\mathbf M_{\rm mod}\big]^{-1}
       \big(\mathbf G_{{\rm pz},n}\widetilde V_n
                    +\widetilde{\mathbf f}_{{\rm ext},n}\big),
 \qquad \omega_n=n\omega,\nonumber\\
 \widetilde I_{{\rm el},n}
 &=i\omega_n\big(\mathbf G_{{\rm pz},n}^{\top}\widetilde{\mathbf q}_n
                         +C_{{\rm pz},n}\widetilde V_n\big),\qquad
 \widetilde V_{{\rm src},n}
 =\widetilde V_n+Z_{{\rm src},n}\widetilde I_{{\rm el},n}.
 \label{eq:causal-complex-stiffness}
\end{align}
The stiffness, piezoelectric coupling and capacitance can all be complex, with causal frequency dependence and jointly passive losses~\cite{Crandall1991,Sherrit1997}. Their imaginary parts cannot be inserted into an instantaneous real enthalpy. The motional current and source impedance determine the terminal voltage and mechanical response together. Nonlinear harmonic loads and currents must be derived consistently from the same constitutive law; the linear response alone does not generate a second harmonic. No simultaneous diagonalization or single real mode shape is assumed. A viscous equivalent at one frequency need not reproduce the response at twice that frequency.

Even a formal evaluation of the proposed correlation requires
\begin{equation}
 \langle\delta\mathcal M_{W,2}a_2\rangle
 =\frac12\operatorname{Re}
  (\widetilde{\delta\mathcal M}_{W,2}\widetilde a_2^{\,*})
 =\frac12|\widetilde{\delta\mathcal M}_{W,2}|\,|\widetilde a_2|
       \cos(\theta_M-\theta_a).
 \label{eq:damped-correlation}
\end{equation}
For the incremental mechanical input $P_{\rm mech,in}=f_{\rm mod}\dot q$, its fundamental-product contribution satisfies
\begin{equation}
 \widetilde{\dot P}_{{\rm mech,in},2}
 =-\omega^2\widetilde f_{{\rm mod},1}\widetilde q_1.
 \label{eq:viscoelastic-input-harmonic}
\end{equation}
This is the full derivative, including $\dot f_{\rm mod}\dot q$; it retains the phases of the drive, coupling and response. Spatially distributed correlations likewise require the local complex amplitudes before integration. A real efficiency factor cannot replace these phases or the distinct responses at $\omega$ and $2\omega$. Equation~\eqref{eq:damped-correlation} remains a formal correlation inside the rejected mass law, not the force on the support in Eq.~\eqref{eq:apparatus-measured-mean}.

\subsubsection{Energy transfer, dissipation and entropy}
\label{sec:damping-entropy}

For a conservative linear single-coordinate piezoelectric reduction, the electric enthalpy and its conjugate variables obey
\begin{equation}
 \mathcal H_{\rm pz}=\tfrac12k_{\rm mod}q^2-g_{\rm pz}qV
                    -\tfrac12C_{\rm pz}V^2,\qquad
 F_{\rm int}=\partial_q\mathcal H_{\rm pz},\quad
 Q_{\rm el}=-\partial_V\mathcal H_{\rm pz}.
 \label{eq:piezoelectric-reciprocity}
\end{equation}
Here $g_{\rm pz}$ is real and $Q_{\rm el}$ is electrode charge. With mechanical loss force $F_{\rm loss}$, loss current $I_{\rm loss}$, and external mechanical load $f_{\rm ext,mod}$, the mechanical equation and terminal current give
\begin{align}
 M_{\rm mod}\ddot q+F_{\rm int}+F_{\rm loss}&=f_{\rm ext,mod},
 &I_{\rm el}&=\dot Q_{\rm el}+I_{\rm loss},\nonumber\\
 VI_{\rm el}+f_{\rm ext,mod}\dot q
 &=\frac{\dd}{\dd t}
  \left[\tfrac12M_{\rm mod}\dot q^2+\mathcal H_{\rm pz}+VQ_{\rm el}\right]
     +F_{\rm loss}\dot q+VI_{\rm loss}.
 \label{eq:coupled-piezoelectric-power}
\end{align}
The term $g_{\rm pz}\dot q$ in $\dot Q_{\rm el}$ is the electrical back-action of mechanical motion. A dispersive material retains its internal-variable storage and reservoirs in this balance; its instantaneous dissipation cannot be inferred from a complex stiffness alone.

For the viscous mode, $F_{\rm loss}=b_{\rm mod}\dot q$ and periodic motion gives
\begin{equation}
 \langle P_{\rm mode,loss}\rangle
 =\frac{b_{\rm mod}}2\sum_{n\geq1}(n\omega)^2|\widetilde q_n|^2,
 \qquad
 2\pi\frac{E_{\rm mech}}{\Delta E_{\rm loss}}=Q_m
 \quad(\omega=\omega_0).
 \label{eq:damped-mean-power}
\end{equation}
The last equality uses the forced single-frequency solution, for which $E_{\rm mech}=k_{\rm mod}|\widetilde q|^2/2$ and $\Delta E_{\rm loss}$ is the energy removed per cycle. Modal loss carried into supports is a boundary transfer; only the part thermalized within the selected material region contributes to its local heating.

To leading nonrelativistic order, a material region without matter or radiation crossing its boundary obeys the local-equilibrium entropy balance
\begin{equation}
 \frac{\dd S_{\rm app}}{\dd t}
 +\int_{\partial\mathcal V_{\rm mat}}\frac{\mathbf j_h\cdot\mathbf n}{\Theta}\,\dd A
 =\int_{\mathcal V_{\rm mat}}\sigma_{\rm irr}\,\dd^3x\geq0.
 \label{eq:material-entropy-balance}
\end{equation}
Here $\Theta>0$ is temperature and $\mathbf j_h$ is heat flux relative to the material. The material-frame, low-frequency conduction closure can retain the relativistic acceleration term~\cite{LopezMonsalvoAndersson2011}:
\begin{equation}
 \mathbf X_{\rm th}=\boldsymbol\nabla_\perp\Theta+\Theta\mathbf A/c^2,
 \quad \mathbf j_h=-\mathbf k_T\mathbf X_{\rm th},\qquad
 \sigma_{\rm irr}=\frac{\mathcal D_{\rm irr}}{\Theta}
     +\frac{\mathbf X_{\rm th}^{\top}\mathbf k_T\mathbf X_{\rm th}}{\Theta^2}\geq0.
 \label{eq:local-entropy-production}
\end{equation}
Here $\boldsymbol\nabla_\perp$ is the material rest-space gradient. The thermal conductivity $\mathbf k_T$ is positive semidefinite, $\mathcal D_{\rm irr}\geq0$ excludes reversible storage, and $\mathbf A$ is the material proper acceleration in its orthonormal rest frame. This closure requires resolved thermal periods and lengths large compared with the eliminated relaxation times and transport lengths, in every material direction; it is not asserted to hold at arbitrary frequency. In the leading nonrelativistic limit $\mathbf X_{\rm th}=\nabla\Theta$. Other transfer mechanisms require their entropy fluxes.

In a periodic thermal state, heat exports the generated entropy. The measured $Q_m$, heating and temperature-dependent response must satisfy the same energy balance. Dissipation transfers energy to reservoirs; it supplies neither a new HN scalar source nor net momentum to a closed apparatus.

\subsubsection{Tajmar's mechanical model and its experimental comparison}
\label{sec:tajmar-mechanics}

Equation~(25) of Ref.~\cite{Tajmar2017} assigns the dynamic mechanical power
\begin{equation}
 P_{\rm Dyn}=\left(\frac{N_dm_P}{3}-m_b\right)av.
 \label{eq:tajmar-dynamic-power}
\end{equation}
Here $m_P$ is the mass of one PZT disc and $m_b$ the brass end-cap mass. The subtraction is attributed to opposite brass-end-cap motion. But $v_b=-v$ and $a_b=-a$ give $m_ba_bv_b=+m_bav$. Summing signed inertial forces and multiplying by one velocity is not summing mechanical powers. For $x_i=\chi_{{\rm sh},i}q$, the kinetic energy is instead
\begin{equation}
 E_{\rm kin}=\tfrac12M_{\rm mod}\dot q^2,\qquad
 M_{\rm mod}=\sum_i m_i\chi_{{\rm sh},i}^{\,2}>0.
 \label{eq:tajmar-modal-energy}
\end{equation}
Opposite mode-shape signs do not cancel kinetic energy.

Equation~(26) of Ref.~\cite{Tajmar2017} fails independently even if its signed-power definition is retained. For the prescribed motion in Eq.~(11) of Ref.~\cite{Tajmar2017}, write $x=A_T\sin\omega t+B_T\sin^2\omega t$, with $v=\dot x$, $a=\dot v$ and $j=\dot a$. Periodic integration gives
\begin{equation}
 \langle(a^2+vj)a\rangle=\tfrac12\langle a^3\rangle
 =-\tfrac34A_T^2B_T\omega^6.
 \label{eq:tajmar-average-identity}
\end{equation}
Consequently its definition $F=-\dot P_{\rm Dyn}a/(4\pi Gc^2\rho_{\rm avg})$ yields
\begin{align}
 \overline F_{\rm literal}&=\mathcal K_T(N_dm_P-3m_b),
 &\mathcal K_T&=\frac{A_T^2B_T\omega^6}{16\pi Gc^2\rho_{\rm avg}},
 \label{eq:tajmar-literal-force}\\
 \overline F_{\rm printed}&=-\mathcal K_T(N_dm_P+3m_b),
 \label{eq:tajmar-printed-force}
\end{align}
where $\rho_{\rm avg}$ is the density prescribed in Ref.~\cite{Tajmar2017}. The printed PZT term has the wrong sign. The full derivative, including the jerk, is retained here; Eq.~(17) of Ref.~\cite{Tajmar2017} likewise retains the full quasistatic power derivative. The sign error, signed-power error and invalid gravitational-source assignment are separate. Neither expression above predicts or bounds a physical residual force. The conserved support force remains Eq.~\eqref{eq:apparatus-measured-mean}.

Figure~5 in Tajmar's paper~\cite{Tajmar2017} compares switching transient peaks with a periodic mean-force formula; their relation requires the drive envelope and balance response. Figure~6 in Tajmar's paper~\cite{Tajmar2017} uses Buldrini's existing data~\cite{Buldrini2016}, with only two nonzero voltage points and no plotted uncertainties, rather than a new independent experiment. Resonances are excluded before Eq.~(27) of Ref.~\cite{Tajmar2017}; its fixed-coefficient frequency powers are not general laws near resonance.

Equation~(1) of Fearn and Woodward~\cite{FearnWoodward2017Voltage} assumes the source in Eq.~\eqref{eq:woodward-proposed-mass}; their quartic voltage scaling does not derive it. Their Figs.~12 and~13 fit the switch-off transient measurements defined in their Fig.~10, with exponents $3.97$ and $3.4$ for the reversed and forward orientations. Both captions disclose an artificial near-zero point introduced to constrain the fit and increase the coefficient of determination; neither gives an uncertainty on the exponent. These fits do not independently establish the gravitational mechanism. Removal of electrical input does not remove stored energy or the balance response.

Monette, K\"o{\ss}ling and Tajmar~\cite{Monette2021} found signals consistent with thermal and vibration-related artifacts in their tested configurations. Ciomperlik's translating-actuator control~\cite{Ciomperlik2020}, also discussed in Ref.~\cite{Monette2021}, produced reversals and switching transients without the proposed mass mechanism, but supplied no calibrated quantitative explanation of Woodward's apparatus. Such signal features alone do not establish propulsion. Alternative rotational testbeds include magnetic levitation~\cite{MierHicks2017}; their readings still require the corresponding momentum-transfer and instrument-response equations.

\subsection{Scalar charge and cosmological boundary conditions}
\label{sec:HN-local-suppression}
\label{sec:absorber}
An unweighted charge ratio is not a bound on the local scalar field. Even a quasistatic $1/r$ Green function gives the characteristic scaling
\begin{equation}
 m_{\rm dev}\sim\frac{\lambda Q_{\rm dev}}{L_{\rm src}},\qquad
 m_0\sim\frac{\lambda Q_{\rm cosmic}}{R_{\rm cos}},\qquad
 \frac{m_{\rm dev}}{m_0}\sim
 \frac{Q_{\rm dev}}{Q_{\rm cosmic}}\frac{R_{\rm cos}}{L_{\rm src}}.
 \label{eq:weighted}
\end{equation}
Here $Q_{\rm dev}$ and $Q_{\rm cosmic}$ are integrated particle-number sources in the dust model with common coupling $\lambda$; $L_{\rm src}$ and $R_{\rm cos}$ are characteristic local and cosmological source distances. The factor $R_{\rm cos}/L_{\rm src}$ invalidates the unweighted comparison. A cosmological calculation changes the kernel and includes its homogeneous contribution; Eq.~\eqref{eq:weighted} exhibits the missing distance weighting; it is not a replacement bound. The force conclusion below follows from the action instead.

Advanced and retarded prescriptions enter the boundary problem of an action-at-a-distance theory~\cite{Hoyle1974,Hawking1965,WheelerFeynman1945,FeynmanWheeler1949}. For the conformal operator $\mathcal O_g=\Box_g+R/6$ and a smooth test scalar $\chi_{\rm test}$,
\begin{equation}
 \mathcal O_{\widehat g}(\Omega^{-1}\chi_{\rm test})=\Omega^{-3}\mathcal O_g\chi_{\rm test}.
 \label{eq:operatorcovariance}
\end{equation}
Normalize a Green function in its first argument by
\begin{equation}
 \mathcal O_{g,x}\mathcal G(x,y)=\delta_g(x,y),\qquad
 \delta_g(x,y)=\frac{\delta^{(4)}(x-y)}{\sqrt{-g(x)}}.
 \label{eq:green-normalization}
\end{equation}
If this Green function exists with the specified boundary conditions, an admissible conformal transformation gives
\begin{equation}
 \widehat{\mathcal G}(x,y)=\Omega^{-1}(x)\Omega^{-1}(y)\mathcal G(x,y).
 \label{eq:greenmap}
\end{equation}
Indeed, Eq.~\eqref{eq:operatorcovariance} gives
\begin{align}
 \mathcal O_{\widehat g,x}\widehat{\mathcal G}(x,y)
 &=\Omega^{-3}(x)\Omega^{-1}(y)\delta_g(x,y)\nonumber\\
 &=\Omega^{-4}(x)\delta_g(x,y)=\delta_{\widehat g}(x,y).
 \label{eq:green-normalization-map}
\end{align}
The second equality uses the support $x=y$ of the delta distribution; no reference-mass factor enters its normalization. A smooth positive conformal factor preserves null cones~\cite{Wald1984,Penrose2004}, and hence the advanced or retarded support of the mapped kernel. Conditional on existence and convergence of the chosen Green representation,
\begin{equation}
 m(x)=\int_{\mathcal X}\mathcal G(x,y)\mathcal J_{\rm tot}(y)\,\dd V_{4,g}(y)+m_{\rm h}(x),
 \qquad \mathcal O_gm_{\rm h}=0,
 \label{eq:HN-integral-restriction}
\end{equation}
where $\mathcal J_{\rm tot}=\mathcal J+4\alpha m^3$ and $\dd V_{4,g}=\sqrt{-g}\,\dd^4y$. With the same prescription and mapped sources, $\widehat{\mathcal J}_{\rm tot}=\Omega^{-3}\mathcal J_{\rm tot}$ and
\begin{align}
 \int_{\mathcal X}\widehat{\mathcal G}(x,y)\widehat{\mathcal J}_{\rm tot}(y)\,\dd V_{4,\widehat g}(y)
 &=\Omega^{-1}(x)\int_{\mathcal X}\mathcal G(x,y)\mathcal J_{\rm tot}(y)\,\dd V_{4,g}(y),\nonumber\\
 \widehat m_{\rm h}&=\Omega^{-1}m_{\rm h}.
 \label{eq:HN-integral-map}
\end{align}
In the Einstein gauge, $\widehat m=m_0$, the scalar equation reduces to $m_0\widehat R/6=\widehat{\mathcal J}+4\alpha m_0^3$, the Einstein trace together with the material Ward identity. It supplies no independent scalar evolution law. The integral representation still requires its prescribed boundary and homogeneous data. A homogeneous-free condition $m_{\rm h}=0$, when required by the selected direct-action prescription~\cite{Hawking1965}, is preserved. It is not implied by the differential equation. The cosmological geometry, absorber response and laboratory matching must satisfy the coupled equations and the integral restriction together. The transformation selects neither the Green prescription nor its convergence, and does not prove the existence of a global absorber. Every globally admitted solution satisfying Proposition~\ref{prop:equivalence} also satisfies the local observable identity~\eqref{eq:noextra}. Additional source-integral or absorber restrictions select among these solutions; they cannot restore Eq.~\eqref{eq:woodward-proposed-mass} as an independent material source or an additional measured force.

Equation~(19) of Ref.~\cite{Fearn2015Mach0} retains distinct advanced and
retarded mass fields in the metric variation, including their product and
the product of their gradients. Proposition~\ref{prop:equivalence} does
not identify every particle-level action-at-a-distance formulation with
Eq.~\eqref{eq:fullaction}. A global restriction on solutions of that action
preserves Eq.~\eqref{eq:noextra}; an additional nonlocal variational term
not contained in that action requires a separate derivation. The latter
has neither been specified as a conserved apparatus model nor shown here
to restore Eq.~\eqref{eq:woodward-proposed-mass}.

\subsection{Observable gravitational couplings and frequency}
\label{sec:astro-observations}
Equation~\eqref{eq:Galgebraic} defines an algebraic coefficient $G_{\rm alg}\propto m^{-2}$. A Weyl transformation changes that coefficient together with the metric and material standards. It therefore does not, by itself, change a measured dimensionless gravitational coupling. Applying an observational bound to $G_{\rm alg}$ would first require identifying it with the coupling measured by the experiment. The invariant variables and their equations are derived in Sect.~\ref{sec:equivalence}.

Astronomical timing and ephemeris constraints on secular variation~\cite{ShapiroEtAl1971,Gillies1997,BiesiadaMalec2004,PitjevaPitjev2013,Genova2018} cannot be transferred directly to a kilohertz modulation by inserting the drive frequency into a bound on $\dot G_{\rm obs}/G_{\rm obs}$, where $G_{\rm obs}$ is the physical coupling inferred from those measurements. That inference requires the theory's physical coupling and the measurement's temporal response. Frequency-specific laboratory searches likewise constrain specified scalar couplings~\cite{Deshpande2024}. These observational questions arise only after a consistent observable has been derived; they cannot correct the tensor and source errors demonstrated above.

\subsection{Ordinary momentum exchange}
\label{sec:ordinary}
For freely escaping electromagnetic radiation, the momentum carried through the boundary gives
\begin{equation}
 \mathbf F_\gamma=-\frac1c\int\mathbf e_\gamma\frac{\dd P}{\dd\Omega_s}\,\dd\Omega_s,
 \qquad |\mathbf F_\gamma|\leq\frac{P_{\rm escaping}}c.
 \label{eq:photonrecoil}
\end{equation}
Here $\mathbf e_\gamma$ is the unit propagation direction of the escaping radiation, $\dd\Omega_s$ is the solid-angle element and $\dd P/\dd\Omega_s$ is the radiated power per unit solid angle. The bound applies to emitted radiation; forces transmitted through cables, supports or external electromagnetic fields require their own momentum fluxes. External gravitational fields can also exert forces and torques. The conformal correspondence derived next preserves these interactions for corresponding data. It concerns the additional force attributed to the HN scalar, not the removal of ordinary momentum exchange.

\subsection{Physical observables and equivalence with the corresponding Einstein problem}
\label{sec:equivalence}
In this subsection the action and its variational identities again use $c=1$. For the action in Eq.~\eqref{eq:fullaction}, set
\begin{equation}
 \Omega=\frac{m}{m_0},\qquad
 \widehat g_{ab}=\Omega^2g_{ab},\qquad m_0>0.
\end{equation}
The transformation applies to the material action and boundary data as well as to the gravitational action. Its physical consequence is the following.

\begin{proposition}
\label{prop:equivalence}
Consider the classical continuum theory defined by Eq.~\eqref{eq:fullaction} on an open spacetime domain $\mathcal X$, including the boundary terms required for its variational problem. Assume that:
\begin{enumerate}
 \item $m$ is smooth and strictly positive on $\mathcal X$, and the conformal transformation and its inverse are regular and admissible at every boundary or asymptotic end used in the problem;
 \item every material system, including the source, measuring instrument, thermal reservoirs and environment, satisfies the universal coupling in Eq.~\eqref{eq:universal};
 \item the Cauchy, boundary, matching and external driving data are transformed with the fields, and any additional source-integral, absorber or other global restrictions are imposed on both representations.
\end{enumerate}
Then the solutions of this local field theory on $\mathcal X$, modulo the admissible Weyl freedom, correspond to Einstein solutions with the same physical matter model and mapped data. Corresponding solutions give identical physical observables. A global cosmological correspondence holds only for solutions satisfying the stipulated global restrictions and regularity conditions on the whole cosmological domain.
\end{proposition}

Positivity gives a positive lower bound on each compact subset of $\mathcal X$; it need not give a uniform lower bound on a noncompact domain. The proposition asserts no extension through $m=0$ and does not establish existence of a globally admissible absorber solution.

\subsubsection{Transformation of the gravitational action}
With our curvature convention,
\begin{equation}
 \widehat R=\Omega^{-2}\big[R+6\Omega^{-1}\Box\Omega\big].
\end{equation}
Multiplication by $\sqrt{-\widehat g}=\Omega^4\sqrt{-g}$ and integration by parts give
\begin{equation}
 \sqrt{-\widehat g}\,\widehat R
 =\sqrt{-g}\big[\Omega^2R-6(\nabla\Omega)^2\big]
 +6\pd_a\big(\sqrt{-g}\,\Omega\nabla^a\Omega\big).
 \label{eq:bulkidentity}
\end{equation}
Since $n_a=g_{ab}n^b$, one has $n_a\nabla^a\Omega=n^a\nabla_a\Omega$. With the outward normal of Sect.~\ref{sec:action}, the integrated divergence is
\begin{equation}
 6\int_{\mathcal X}\partial_a\big(\sqrt{-g}\,\Omega\nabla^a\Omega\big)\,\dd^4x
 =6\int_{\partial\mathcal X}\epsilon_n\sqrt{|h|}\,
       \Omega n^a\nabla_a\Omega\,\dd^3y.
 \label{eq:conformal-divergence-flux}
\end{equation}
The induced metric, unit normal and extrinsic curvature transform as
\begin{align}
 \widehat h_{ij}&=\Omega^2h_{ij},&
 \widehat n^a&=\Omega^{-1}n^a,\nonumber\\
 \widehat K_{ij}&=\Omega\big(K_{ij}+h_{ij}n^a\nabla_a\ln\Omega\big),&
 \widehat K&=\Omega^{-1}\big(K+3n^a\nabla_a\ln\Omega\big).
 \label{eq:boundarymap}
\end{align}
In particular,
\begin{equation}
 \epsilon_n\sqrt{|\widehat h|}\,\widehat K
 =\epsilon_n\sqrt{|h|}\big(\Omega^2K+3\Omega n^a\nabla_a\Omega\big).
 \label{eq:conformal-boundary-density}
\end{equation}
The two additional boundary contributions are respectively
$+(3/\kappa)\int_{\partial\mathcal X}\epsilon_n\sqrt{|h|}\,\Omega n^a\nabla_a\Omega\,\dd^3y$
from the Einstein bulk term and its negative from the GHY term. They cancel. Equation~\eqref{eq:fullaction} therefore becomes
\begin{align}
 S={}&\frac1{2\kappa}\int_{\mathcal X}\sqrt{-\widehat g}\,(\widehat R-2\Lambda)\,\dd^4x
 -\frac1\kappa\int_{\partial\mathcal X}\epsilon_n\sqrt{|\widehat h|}\,\widehat K\,\dd^3y
 \nonumber\\
 &+\widehat S_{\rm mat}[\widehat g,\widehat\Psi].
 \label{eq:einsteinaction}
\end{align}
This is the Einstein action with its non-null boundary term. Corners, null boundaries and asymptotic subtractions, when present, require the corresponding terms in both representations. The variables held fixed at the boundary are part of the variational problem~\cite{Wald1984,Padmanabhan2010}.

\subsubsection{Cancellation in the tensor equation}
The Einstein tensor transforms according to
\begin{align}
 \widehat G_{ab}={}&G_{ab}+\frac2m\nabla_a\nabla_bm
 -\frac4{m^2}\nabla_am\nabla_bm\nonumber\\
 &+g_{ab}\left[-\frac2m\Box m+\frac{(\nabla m)^2}{m^2}\right].
\end{align}
Substitution into Eq.~\eqref{eq:metriceq} cancels the second derivatives of $m$ and the products of its first derivatives. The remaining equation is
\begin{equation}
 \widehat G_{ab}+\Lambda\widehat g_{ab}=-\kappa\widehat T_{ab}.
 \label{eq:einsteineq}
\end{equation}
The cancellation is exact. No small-amplitude assumption for $m_s=m-m_0$ has been used.

\subsubsection{Clocks, trajectories and electromagnetic measurements}
\label{sec:observables}
Here proper lengths and times again use $c=1$, as in the action. Under a Weyl transformation with smooth dimensionless parameter $\chi$,
\begin{equation}
 g_{ab}\longmapsto e^{2\chi}g_{ab},\qquad
 m\longmapsto e^{-\chi}m,
\end{equation}
$\widehat g_{ab}$ is unchanged. For a material field of Weyl weight $w$, $\Psi\mapsto e^{w\chi}\Psi$, the representative $\widehat\Psi=\Omega^w\Psi$ is also unchanged. Physical readings are functions of these invariant variables and of the experimental worldlines and surfaces. Coordinate invariance of $R$ or $\Box m/m$ alone does not make either expression a Weyl-invariant observable.

The particle action transforms as
\begin{equation}
 -\lambda\int m\,\dd s_g=-\lambda m_0\int\dd s_{\widehat g}.
\end{equation}
Thus a freely falling particle follows a constant-mass geodesic of $\widehat g$.
The conformal connection is
\[
 \widehat\Gamma^a{}_{bc}=\Gamma^a{}_{bc}
 +\delta^a_b\nabla_c\ln\Omega+\delta^a_c\nabla_b\ln\Omega
 -g_{bc}g^{ad}\nabla_d\ln\Omega.
\]
With $u^au_a=+1$ in the $(+---)$ convention and
$a_g^a=u^b\nabla_bu^a=A^a/c^2$ in the $g$ representation,
\begin{align}
 a_g^a&=(g^{ab}-u^au^b)\nabla_b\ln\Omega,\nonumber\\
 \widehat a^a&=\Omega^{-2}\left[a_g^a-(g^{ab}-u^au^b)\nabla_b\ln\Omega\right]=0.
 \label{eq:geodesicmap}
\end{align}
The first line's apparent variable-mass acceleration cancels in the acceleration measured with the transformed clocks and rods.

Their proper time, local length and frequency satisfy
\begin{equation}
 \widehat\tau=\int\Omega\,\dd\tau_g,\qquad
 \dd\widehat\ell=\Omega\,\dd\ell_g,\qquad
 \widehat\omega=\Omega^{-1}\omega_g.
\end{equation}
These relations map the material standards; the invariant quantities are the calibrated readings and dimensionless ratios formed with them.

In four dimensions the electromagnetic potential $A_a^{\rm EM}$ and field $F_{ab}=\nabla_aA_b^{\rm EM}-\nabla_bA_a^{\rm EM}$ have zero Weyl weight, and the Maxwell action is unchanged. The electric field measured by an observer also involves its normalized velocity and tetrad. An applied voltage must consequently be specified as $\widehat V(\widehat\tau)$ with the transformed leads and impedances. Holding a coordinate waveform $V(t)$ fixed while changing the clock or material parameters can change the physical experiment.

Number density, velocity and number current satisfy
\begin{equation}
 \widehat N=\Omega^{-3}N,\qquad
 \widehat u^a=\Omega^{-1}u^a,\qquad
 \widehat J^a=\Omega^{-4}J^a,\quad J^a=Nu^a.
\end{equation}
Since the directed hypersurface element $\dd\Sigma_a$ has Weyl weight $4$, the particle number $Q_N=\int_\Sigma J^a\dd\Sigma_a$ is invariant. The same transformation of the instrument and its calibration preserves a force-balance reading.

\subsubsection{The physical drive waveform}
\label{sec:drive-clock}
\begin{equation}
 \widehat\alpha_{\rm clk}(t)=\frac{\dd\widehat\tau}{\dd t}
 =\Omega(t)\frac{\dd\tau_g}{\dd t},\qquad
 \widehat V(\widehat\tau(t))=\Omega^{-1}(t)V_g(t).
 \label{eq:drive-lapse}
\end{equation}
The voltage relation assumes a lumped terminal voltage and a spatially uniform $\Omega$ across the source and voltage-reference leads. For a nonuniform factor, the work per charge must instead be transformed along the specified lead path. A physical sinusoidal drive is represented by
\begin{equation}
 \widehat V(\widehat\tau)=V_{\rm pk}\cos(\omega\widehat\tau),\qquad
 V_g(t)=\Omega(t)V_{\rm pk}
 \cos\!\left(\omega\int_0^t\widehat\alpha_{\rm clk}(t')\,\dd t'\right).
 \label{eq:drive-mapped}
\end{equation}
The clock-rate function $\widehat\alpha_{\rm clk}$ is Weyl invariant and coordinate dependent; choosing $t=\widehat\tau$ sets it to one. A Weyl change of representation at fixed coordinates does not change the physical clock or voltage. In particular, the zero-weight coordinate covector $A_a^{\rm EM}(t)$ can remain fixed in the same physical experiment. Holding an untransformed local voltage $V_g(t)$ fixed while changing its material standard or clock relation can instead change $\widehat V(\widehat\tau)$. Any resulting difference in a balance reading is the ordinary response to different physical drive data. With the mapped waveform and the other data of Proposition~\ref{prop:equivalence}, the additional force remains identically zero.

For universally coupled clocks, voltage and length references and calibration masses,
\begin{align}
 \int\omega_{{\rm ref},g}\,\dd\tau_g
 &=\int\widehat\omega_{\rm ref}\,\dd\widehat\tau,
 &\frac{V_g}{V_{{\rm ref},g}}
 &=\frac{\widehat V}{\widehat V_{\rm ref}},\nonumber\\
 \frac{\ell_g}{\ell_{{\rm ref},g}}
 &=\frac{\widehat\ell}{\widehat\ell_{\rm ref}},
 &\frac{M_{{\rm rest},g}}{M_{{\rm ref},g}}
 &=\frac{\widehat M_{\rm rest}}{\widehat M_{\rm ref}}.
 \label{eq:fixed-physical-standards}
\end{align}
The first equality preserves the clock phase, hence its cycle count; the other ratios
preserve calibrated voltage, length and mass at the same event. Here $M_{{\rm rest},g}=\Omega\widehat M_{\rm rest}$ and $M_{{\rm ref},g}=\Omega\widehat M_{\rm ref}$ are the represented local rest masses at the comparison event. 

For an extended body, Eq.~\eqref{eq:ward} gives, on corresponding spacelike
slices $\Sigma$,
\begin{equation}
 \widehat T_{ab}=\Omega^{-2}T_{ab},\qquad
 \widehat u_{\rm obs}^a=\Omega^{-1}u_{\rm obs}^a,\qquad
 \widehat\nu_\Sigma^a=\Omega^{-1}\nu_\Sigma^a,
 \qquad \dd\widehat V_\Sigma=\Omega^3\dd V_{\Sigma,g}.
 \label{eq:extended-calibration-weights}
\end{equation}
Here $u_{\rm obs}^a$ specifies the observers and $\nu_\Sigma^a$ is the future-directed unit normal to $\Sigma$; both vectors have norm $+1$. In the units $c=1$ used in this subsection, define their energy
element by $\dd M_{\Sigma,g}=T_{ab}u_{\rm obs}^a\nu_\Sigma^b\dd V_{\Sigma,g}$, and define
$\dd\widehat M_\Sigma$ analogously. Then
\begin{equation}
 \dd M_{\Sigma,g}=\Omega(x)\dd\widehat M_\Sigma,\qquad
 M_{\Sigma,g}=\int_\Sigma\Omega(x)\dd\widehat M_\Sigma,\qquad
 \widehat M_\Sigma=\int_\Sigma\Omega^{-1}(x)\dd M_{\Sigma,g}.
 \label{eq:extended-calibration-integral}
\end{equation}
These identities hold on each mapped slice for a time-dependent,
spatially nonuniform $\Omega$. For body and reference slices
$\Sigma_B$ and $\Sigma_R$, the corresponding calibrated energy ratio is
\begin{equation}
 \frac{\displaystyle\int_{\Sigma_B}\Omega^{-1}\dd M_{\Sigma,g}}
      {\displaystyle\int_{\Sigma_R}\Omega^{-1}\dd M_{\Sigma,g}}
 =\frac{\widehat M_{\Sigma_B}}{\widehat M_{\Sigma_R}}.
 \label{eq:extended-calibration-ratio}
\end{equation}
The unweighted integrals do not in general share a single conformal
factor. These are specified observer-energy integrals, not a definition
of total gravitating mass or of a support-force reading. Remote energy
comparisons include their mapped transport and redshift; a force reading
uses the mapped stress flux and support reactions. No spatial gradient
of $\Omega$ has been set to zero. Proposition~\ref{prop:equivalence}
already permits the same nonuniform factor in the material action,
internal stresses and boundary data.

Holding these physical standards fixed is compatible with mapped data;
their being external to the device does not break universal coupling.
The conformal factor may equal one at an external reference while varying
inside the domain. Holding instead an un-hatted dimensionful field fixed
while changing its relation to the reference can change the physical
protocol. A nonuniversally coupled standard or a prescribed boundary that
forbids the transformation lies outside Proposition~\ref{prop:equivalence};
an externally calibrated experiment does not do so merely by definition.

\subsubsection{Elastic, dielectric and source contributions}
For an elastic solid, take invariant material coordinates $X^I$ and define
\begin{equation}
 \mathcal C^{IJ}=-\widehat g^{ab}\pd_aX^I\pd_bX^J.
\end{equation}
The sign makes $\mathcal C^{IJ}$ positive on spatial material gradients for our metric signature. A constitutive action can depend on these strain variables, polarization, electromagnetic fields and entropy variables. The material action has contributions
\begin{equation}
 \widehat S_{\rm mat}=S_{\rm EM}+S_{\rm solid+polarization}
 +S_{\rm supply}+S_{\rm reservoir}+S_{\rm supports}+S_{\rm environment}.
 \label{eq:mattercompletion}
\end{equation}
Reservoir variables, or a consistent effective reduction of them, account for dissipation. An external voltage source prescribes boundary data and transfers energy and momentum through its connections. Equation~\eqref{eq:universal} transforms all these terms together, with the same constitutive coefficients and measured driving data in the invariant representation.

For a local electromagnetic and polarization action with algebraic dependence on $F_{ab}$, define its excitation at fixed metric, scalar and other material variables by
\begin{equation}
 \delta S_{{\rm EM}+{\rm polarization}}
 =-\frac12\int_{\mathcal X}\sqrt{-g}\,
       \mathcal D_{\rm em}^{ab}\delta F_{ab}\,\dd^4x.
 \label{eq:dielectric-excitation}
\end{equation}
The same variation in invariant variables has the same $F_{ab}$ and $\delta F_{ab}$. Thus
\begin{equation}
 \sqrt{-\widehat g}\,\widehat{\mathcal D}_{\rm em}^{ab}
 =\sqrt{-g}\,\mathcal D_{\rm em}^{ab},\qquad
 \widehat{\mathcal D}_{\rm em}^{ab}=\Omega^{-4}\mathcal D_{\rm em}^{ab}.
 \label{eq:dielectric-weyl}
\end{equation}
The contravariant excitation has weight $-4$; its densitized form has weight zero. The piezoelectric, electrostrictive, dielectric and elastic constitutive functions are defined with respect to $\widehat g_{ab}$, $\mathcal C^{IJ}$ and the invariant material variables. Their un-hatted representatives follow by substitution in the same action. Vacuum Maxwell invariance alone does not establish this constitutive property. Holding a phenomenological permittivity or mechanical modulus fixed independently in both representations need not satisfy Eq.~\eqref{eq:universal} and is outside Proposition~\ref{prop:equivalence}.

The damping in Eq.~\eqref{eq:damped-modal-equation} describes a reduced material subsystem. It is not an additional conservative force obtained by inserting $b_{\rm mod}\dot q$ into an ordinary elastic potential. Eliminating reservoir variables must retain the energy and entropy balances in Sect.~\ref{sec:damping-entropy}; the full matter source includes the energy transferred to those reservoirs.

Under Proposition~\ref{prop:equivalence}, local temperature and the total coarse-grained entropy current transform as
\begin{equation}
 \widehat\Theta=\Omega^{-1}\Theta,\qquad
 \widehat{\mathcal S}^{\,a}=\Omega^{-4}\mathcal S^a,\qquad
 \widehat\nabla_a\widehat{\mathcal S}^{\,a}
 =\Omega^{-4}\nabla_a\mathcal S^a.
 \label{eq:entropy-weyl-map}
\end{equation}
Entropy per particle, in units of Boltzmann's constant, is invariant; its proper density has weight $-3$. The current includes conductive entropy transport. The last equality follows from $\sqrt{-\widehat g}=\Omega^4\sqrt{-g}$, and preserves the entropy-production inequality. Likewise,
\begin{equation}
 \widehat S_\Sigma
 =\int_\Sigma\widehat{\mathcal S}^{\,a}\,\dd\widehat\Sigma_a
 =\int_\Sigma\mathcal S^a\,\dd\Sigma_a=S_\Sigma.
 \label{eq:entropy-invariant}
\end{equation}
The reservoir state, heat boundary data and constitutive transport laws must be mapped together. An unchanged Fourier law is not to be assumed after a position-dependent conformal rescaling: the temperature gradient also acquires derivatives of $\Omega$.
If additional nonequilibrium thermal variables are retained, their
constitutive laws and their contributions to the total entropy current
must be mapped as well; the current-density identity in
Eq.~\eqref{eq:entropy-weyl-map} does not depend on the Fourier closure.

The quality factor measured from the same mode with the mapped clocks and material standards is a dimensionless observable,
\begin{equation}
 \widehat Q_m^{\,\rm obs}=Q_m^{\rm obs}.
 \label{eq:quality-invariant}
\end{equation}
For nonconstant $\Omega$, a reduced oscillator written in the un-hatted variables need not retain constant coefficients. Equation~\eqref{eq:quality-invariant} compares corresponding measurements; it does not follow by independently rescaling $M_{\rm mod}$, $b_{\rm mod}$ and $k_{\rm mod}$ while holding the drive fixed. Any finite damping and positive entropy production in the apparatus are preserved in the corresponding Einstein problem. They supply no additional compensator force.

Equation~(7) of Ref.~\cite{Tajmar2017},
\begin{equation}
 \Delta M_{\rm T}=-\frac{\dot P_{\rm mech}}{4\pi Gc^2\rho_{\rm ref}},
 \label{eq:tajmar-mass-law}
\end{equation}
selects the time derivative of mechanical power as the source of a new gravitational mass fluctuation. Equation~\eqref{eq:mattercompletion} gives no such selective coupling: electrical and elastic energies, their stresses and their exchanges with the supply belong to the same action. Equation~\eqref{eq:tajmar-mass-law} therefore does not follow from the universal theory. A nonuniversal coupling would define a different theory and would require its own action.

\subsubsection{Boundary and matching conditions}
The Einstein Cauchy data are $\widehat h_{ij}$, $\widehat K_{ij}$ and the invariant material data and conjugate momenta, subject to the constraints. Equation~\eqref{eq:boundarymap} maps the geometrical data; the transformed action determines the material momenta. On each non-null spacetime boundary component, $h_{ij}$ denotes the full induced three-metric. Its Dirichlet data obey
\begin{equation}
 \delta\widehat h_{ij}=0
 \quad\Longleftrightarrow\quad
 \delta h_{ij}+2h_{ij}\delta\ln\Omega=0.
 \label{eq:dirichlet}
\end{equation}
Fixing $h_{ij}$ and $m$ separately is a choice of representation and can impose more data than this physical Dirichlet condition.

On a timelike component, $h_{ij}$ includes the temporal and mixed
components. With a fixed boundary embedding and foliation, write
\[
 h_{ij}\,\dd y^i\dd y^j
 =\mathfrak n_\partial^2(\dd y^0)^2
 -\mathfrak s_{AB}(\dd y^A+\beta_\partial^A\dd y^0)
                  (\dd y^B+\beta_\partial^B\dd y^0),
 \qquad A,B=1,2,
\]
where $\mathfrak n_\partial>0$ is the boundary lapse,
$\beta_\partial^A$ the boundary shift and $\mathfrak s_{AB}$ the
positive-definite spatial two-metric. The Weyl map gives
\[
 \widehat{\mathfrak n}_\partial=\Omega\mathfrak n_\partial,
 \qquad\widehat\beta_\partial^A=\beta_\partial^A,
 \qquad\widehat{\mathfrak s}_{AB}=\Omega^2\mathfrak s_{AB}.
\]
Consequently Eq.~\eqref{eq:dirichlet} already includes
\[
 \delta\mathfrak n_\partial+\mathfrak n_\partial\delta\ln\Omega=0,
 \qquad\delta\beta_\partial^A=0,
 \qquad\delta\mathfrak s_{AB}
       +2\mathfrak s_{AB}\delta\ln\Omega=0.
\]
These are components of the same condition, not additional independent
data. On a spacelike Cauchy cap, the spacetime lapse and shift specify
the normal coordinate extension and are not extra intrinsic Dirichlet
data required to cancel the gravitational boundary variation. This
variational statement does not establish a well-posed initial-boundary
value problem for arbitrary data; the evolution gauge, constraints
and corner compatibility must also be satisfied.

Without a surface layer, continuity of $\widehat h_{ij}$ and $\widehat K_{ij}$ gives the matching conditions. In $g,m$ variables the latter contains the normal derivative of $\Omega$ in Eq.~\eqref{eq:boundarymap}. A surface stress requires the corresponding surface action. Prescribing untransformed metric junction conditions and an independent scalar jump would generally specify a different problem.

For a boundary embedding $x^a=\iota^a(y)$ define its tangent basis and
dual tangential projector by
\begin{equation}
 \mathsf e_i{}^a=\frac{\partial\iota^a}{\partial y^i},\qquad
 h_{ij}=g_{ab}\mathsf e_i{}^a\mathsf e_j{}^b,\qquad
 \mathsf e^i{}_b=h^{ij}g_{bc}\mathsf e_j{}^c.
 \label{eq:boundary-tangent-projector}
\end{equation}
Here $a,b$ are spacetime indices and $i,j$ label the three boundary
coordinates. For the same embedding, $\widehat{\mathsf e}_i{}^a=
\mathsf e_i{}^a$ and $\widehat{\mathsf e}^i{}_b=\mathsf e^i{}_b$.
Antisymmetry gives $n_bn_a\mathcal D_{\rm em}^{ab}=0$.
The boundary vector density therefore transforms as
\begin{equation}
 \sqrt{|\widehat h|}\,\widehat n_a
 \widehat{\mathcal D}_{\rm em}^{ab}\mathsf e^i{}_b
 =\sqrt{|h|}\,n_a\mathcal D_{\rm em}^{ab}\mathsf e^i{}_b,
 \label{eq:dielectric-boundary-map}
\end{equation}
because $\widehat n_a=\widehat g_{ab}\widehat n^b=\Omega n_a$
and the weights sum to $3+1-4=0$. Its oriented contribution to the
action variation is
\begin{equation}
 \left.\delta S_{{\rm EM}+{\rm polarization}}\right|_{\partial\mathcal X}
 =-\int_{\partial\mathcal X}\epsilon_n\sqrt{|h|}\,
 n_a\mathcal D_{\rm em}^{ab}\mathsf e^i{}_b\,
 \delta A_i^{\rm EM}\,\dd^3y,
 \qquad A_i^{\rm EM}=\mathsf e_i{}^a A_a^{\rm EM}.
 \label{eq:electromagnetic-boundary-variation}
\end{equation}
The same formula holds for vacuum Maxwell theory with the corresponding
vacuum excitation. In boundary-adapted coordinates the projector selects the components tangent to the boundary. Boundary indices are not restricted to spacelike directions. On a timelike worldtube, projection onto a tangent unit observer encodes the normal electric displacement, with its sign fixed by the outward orientation; the two spatial tangential projections encode magnetic boundary data. This
three-dimensional boundary vector density is not a scalar electric flux
through a spatial two-surface.

Surface polarization or free surface charge, when present, requires the corresponding mapped surface action and data. Mechanical tractions, heat flux, incident radiation and source connections transform with the material data. An asymptotic normalization such as $\Omega\to1$ restricts the admissible transformations in Proposition~\ref{prop:equivalence}.

\subsubsection{The scalar equation and the measured force}
Let $\widehat S$ denote the complete Einstein action in
Eq.~\eqref{eq:einsteinaction}. At fixed invariant material fields,
the bulk functional chain rule gives
\begin{align}
 \left.\frac{\delta S}{\delta g_{ab}}\right|_{m,\widehat\Psi}
 &=\Omega^2\left.\frac{\delta\widehat S}{\delta\widehat g_{ab}}
              \right|_{\widehat\Psi},\nonumber\\
 \left.\frac{\delta S}{\delta m}\right|_{g,\widehat\Psi}
 &=\frac2m\widehat g_{ab}
       \left.\frac{\delta\widehat S}{\delta\widehat g_{ab}}
              \right|_{\widehat\Psi}.
 \label{eq:invariant-chain-rule}
\end{align}
These variational identities use $c=1$. The functional derivatives here
and in the following Ward identities use the covariant metric $g_{ab}$;
the definition of $T_{ab}$ uses $g^{ab}$. Since
$\delta g^{cd}=-g^{ca}g^{db}\delta g_{ab}$,
\[
 \frac{\delta S_{\rm mat}}{\delta g_{ab}}
 =-g^{ac}g^{bd}\frac{\delta S_{\rm mat}}{\delta g^{cd}}
 =-\frac{\sqrt{-g}}2T^{ab}.
\]
This derivative is a contravariant tensor density.
Thus
\begin{equation}
 2g_{ab}\left.\frac{\delta S}{\delta g_{ab}}\right|_{m,\widehat\Psi}
 -m\left.\frac{\delta S}{\delta m}\right|_{g,\widehat\Psi}=0.
 \label{eq:fixed-invariant-Ward}
\end{equation}
For the material action its left-hand side is
$\sqrt{-g}(-T+m\mathcal J)$, recovering Eq.~\eqref{eq:ward}.
This is an off-shell identity. With noninvariant fields
$\widehat\Psi_r=\Omega^{w_r}\Psi_r$ instead held independent,
\begin{equation}
 2g_{ab}\left.\frac{\delta S}{\delta g_{ab}}\right|_{m,\Psi}
 -m\left.\frac{\delta S}{\delta m}\right|_{g,\Psi}
 +\sum_r w_r\Psi_r\frac{\delta S}{\delta\Psi_r}=0.
 \label{eq:raw-matter-Ward}
\end{equation}
The matter Euler derivatives disappear only on the matter equations.
The scalar equation therefore supplies no independent propagating
scalar mode. With the mapped data, the remaining metric and material
equations are those of the Einstein problem.

Conversely, an Einstein solution generates $g_{ab}=\Omega^{-2}\widehat g_{ab}$ and $m=\Omega m_0$ on the domain of every smooth admissible positive $\Omega$. It solves a specified global HN integral problem only when the mapped fields also satisfy that problem's source-integral and boundary restrictions. Together with the transformed boundary terms and Eq.~\eqref{eq:dirichlet}, this proves Proposition~\ref{prop:equivalence}.

For corresponding force-balance measurements it follows that
\begin{equation}
 F_{\rm additional}^{\rm obs}
 \equiv F_{\rm conformal}^{\rm obs}-F_{\rm GR}^{\rm obs}=0.
 \label{eq:noextra}
\end{equation}
Ordinary momentum fluxes through radiation and cables, support reactions and external-field forces occur in both corresponding solutions for mapped data and cancel in this difference. Equation~\eqref{eq:noextra} is exact for every solution satisfying Proposition~\ref{prop:equivalence}, independently of scalar amplitude; it does not set either total force reading to zero. Consistently mapped absorber conditions restrict the admitted solutions but cannot restore Eq.~\eqref{eq:woodward-proposed-mass} as an additional material source. Additional physical scalar fields, creation fields, nonuniversal couplings and singular transformations lie outside the proposition and require a separate dynamical theory.

\section{Conclusion}
\label{sec:conclusion}
For nonzero proper density and constant physical speed of light, Woodward's energy-density identities~\cite{Woodward2004}, when imposed exactly on the same fields, fix the potential at a constant value. His subsequent treatment of that potential as an independent variable is mathematically inconsistent. For positive gravitational coupling and the un-subtracted source in Ref.~\cite{Woodward2004}, the exact identities also contradict the source equation for every smooth strictly positive periodic density. A background-only equality is a different assumption and changes the quadratic terms; it does not preserve his derivation. A consistently subtracted source has no positive-density integral contradiction.

The exact GR calculation places the second time derivative in the quasilinear gravitational operator. Its coefficient is fixed by the metric. An order-unity coordinate lapse is not a new mass source; local normalization transforms the clocks and rods with the operator. The local Friedmann expansion retains cosmological curvature and tidal terms. Replacing one weak-field coefficient by a cosmic potential while retaining the other local equations changes the problem. Moving the derivative to the other side does not convert it into a material source. Woodward's proposed mass fluctuation~\cite{Woodward2004,WoodwardStarship} is not a source derived from Einstein's equations.

The linear second time derivative retained in Refs.~\cite{Fearn2015I,Fearn2015II} cancels from the time--time component of the HN derivative tensor. The squared-gradient terms retained in Fearn's diagonal sum~\cite{Fearn2016} cancel from the covariant trace. Their mass-fluctuation formulas therefore result from incorrect tensor calculations. Neither an efficiency factor nor a numerical adjustment repairs these errors.

The mechanical substitutions in Refs.~\cite{Fearn2015I,Fearn2015II} introduce separate failures. Reference~\cite{Fearn2015I} assumes zero covariant momentum transfer and subsequently introduces varying rest mass. These assumptions are incompatible. Differentiating mechanical power retains the force derivative as well as the acceleration term. Tajmar's signed-force construction~\cite{Tajmar2017} does not equal the sum of the kinetic powers, and the printed dynamic average in Ref.~\cite{Tajmar2017} has a separate algebraic inconsistency. A mass--acceleration correlation cannot replace boundary flux and internal momentum in the force balance.

A quantitative resonant prediction requires the assembled stack's quality factors and complex responses at both harmonics. Reference~\cite{Fearn2015II} includes a damped oscillator but does not carry its complex harmonic responses consistently into its thrust estimate. The mechanical power balance retains elastic storage and modal energy loss. A periodic thermal state exports the entropy generated by dissipation. These energy and entropy balances supply no HN mass-fluctuation law.

Under Proposition~\ref{prop:equivalence}, a smooth strictly positive scalar, an admissible Weyl transformation, universal material coupling and consistently mapped matter, driving and boundary data give the corresponding Einstein problem. In this sector the HN scalar is a conformal compensator. For the same physical matter model and mapped data, the additional measured force relative to that Einstein solution is exactly zero. Ordinary momentum exchange and support reactions remain in both readings. Global source-integral and absorber conditions compatible with these hypotheses select solutions and preserve the equality of their observables.

Woodward and Fearn's mass-fluctuation and propulsion formulas do not follow from their stated GR or continuum conformal HN field equations~\cite{Woodward2004,Fearn2015I,Fearn2015II,Fearn2016}. Their derivations contain the explicit differentiation, component and contraction errors demonstrated above. The claimed source and additional propulsion force are therefore not predictions of the stated GR equations or of the local continuum conformal HN theory under Proposition~\ref{prop:equivalence}. Nonlocal variational dynamics not contained in that continuum action require a separate force calculation; they are neither validated nor excluded by this equivalence result.

\clearpage
\appendix
\section{Corrections to prior analyses}
\label{app:prior-corrections}
Table~\ref{tab:prior-corrections} identifies the corrections and changed interpretations of Refs.~\cite{Rodal2019,Rodal2025v1,Rodal2016Electroelastic}. The derivations are given at the indicated locations and are not repeated here.

\begin{table}[!ht]
\centering
\small
\renewcommand{\arraystretch}{1.18}
\begin{tabular}{@{}>{\raggedright\arraybackslash}p{0.24\linewidth}p{0.51\linewidth}>{\raggedright\arraybackslash}p{0.19\linewidth}@{}}
\toprule
Earlier result & Correction or qualification & Present result\\
\midrule
Derivative objection in Ref.~\cite{Rodal2025v1}
& The product rule is valid for exact $E_0=\rho_0\varphi$. The simultaneous exact identities fix $\varphi=c^2$; background equality alone does not imply $\dot E_0=c^2\dot\rho_0$.
& Eqs.~\eqref{eq:intro-exact-constraint}--\eqref{eq:intro-source-defects}\\[4pt]

Ref.~\cite{Rodal2019}, Eq.~(23) and GR order; Ref.~\cite{Rodal2025v1}, Eqs.~(72), (76)
& Harmonic gauge applies to the trace reversal, giving $R^{(1)}=\tfrac12\Boxeta\psi$ and $16\pi$, not $8\pi$. The selected GR term is 2PN relative to the Newtonian $00$ equation, despite its absolute $c^{-6}$ order.
& Eqs.~\eqref{eq:gr-linear-harmonic}, \eqref{eq:gr-pn-count}, \eqref{eq:trace-normalization}\\[4pt]

Ref.~\cite{Rodal2019}, Eq.~(18), $m_s=0$
& With the retained HN convention and $\Lambda>0$ denoting positive vacuum energy, the constant-scalar trace is $R-4\Lambda=6T/m_0^2$, rather than $R+4\Lambda=6T/m_0^2$.
& Eqs.~\eqref{eq:curvature-definition}, \eqref{eq:constant-m-trace}\\[4pt]

Ref.~\cite{Rodal2019}, scalar reduction and differentiation after Eq.~(17)
& Retain $m_0\delta R/6$ and the product-gradient and number-density terms in $\Box(\lambda mN)$. Proposition~\ref{prop:equivalence} replaces the additional suppressed scalar-effect interpretation by the exact corresponding-force identity.
& Eqs.~\eqref{eq:HN-coupled-linear}, \eqref{eq:densityidentity}, \eqref{eq:noextra}\\[4pt]

Trace criticism in Ref.~\cite{Rodal2019}; constant-coupling objection in Ref.~\cite{Rodal2025v1}
& Common component units for $x^0=ct$ do not make the positive diagonal sum a Lorentzian trace. The coupling correction is first order for nonzero background stress, but second order for a wholly first-order material source.
& Eqs.~\eqref{eq:fearn-explicit-contraction}, \eqref{eq:fearn-coupling-product}\\[4pt]

Ref.~\cite{Rodal2019}, Sect.~3.4
& The cosmology invoked by Fearn~\cite{Fearn2015Mach0,Fearn2016} has not been matched to the laboratory propulsion derivation. This objection is retained; omission of an explicit local $\Lambda$ alone does not establish Hawking's divergence.
& Sect.~\ref{sec:absorber}\\[4pt]

Cosmic-charge comparison in Ref.~\cite{Rodal2025v1}
& An unweighted charge ratio is not a local-field bound. Even the quasistatic $1/r$ kernel introduces the source-distance ratio.
& Eq.~\eqref{eq:weighted}\\[4pt]

Ref.~\cite{Rodal2016Electroelastic}, Eqs.~(3)--(4), Sect.~5
& The identification $\lambda m=\gamma(v)M_{p0}$, with rest mass $M_{p0}$, equates the HN scalar with observer-dependent energy divided by $c^2$. A variable-mass centroid also requires the momentum-transfer term. Equation~(9) of that reference correctly retained the jerk contribution.
& Eqs.~\eqref{eq:woodward-power-rule}, \eqref{eq:rodal-centroid-momentum}\\[4pt]

Ref.~\cite{Rodal2016Electroelastic}, Sect.~7
& Excitation multipliers fitted the force agreement; it was not an independent determination of the gravitational source. The dependence on stiffness, damping, phase and end conditions remains necessary for the mechanical response.
& Sects.~\ref{sec:damping-quality}--\ref{sec:damping-entropy}\\
\bottomrule
\end{tabular}
\caption{Corrections and qualifications of the cited earlier analyses.}
\label{tab:prior-corrections}
\end{table}

The cosmological sign follows directly from Eq.~\eqref{eq:curvature-definition}: for $\dd s^2=(\dd x^0)^2-e^{2H_{\rm cos}x^0/c}\dd\mathbf x^2$, with constant $H_{\rm cos}$,
\begin{equation}
 G_{00}=-\frac{3H_{\rm cos}^2}{c^2},\qquad
 R=\frac{12H_{\rm cos}^2}{c^2},\qquad
 G_{00}+\Lambda=0\ \Longrightarrow\ R=4\Lambda.
 \label{eq:de-sitter-sign-check}
\end{equation}
Consistent changes of signature or Ricci contraction preserve both derivative cancellations in Fearn's tensor. Equation~\eqref{eq:noextra} follows from the exact action and observable transformation under Proposition~\ref{prop:equivalence}, independently of the intermediate expansions and numerical estimates corrected above.

\bibliographystyle{unsrt-initials}
\bibliography{references,additional_references}
\end{document}